\documentclass[ aps,twocolumn, showkeys, notitlepage, superscriptaddress]{revtex4-2}

\usepackage{natbib}
\usepackage{lipsum}
\usepackage{graphicx}
\usepackage{dcolumn}
\usepackage{bm}
\usepackage{amsmath}
\usepackage{float}
\usepackage{multirow}
\usepackage{slashed}
\usepackage{xcolor}
\usepackage{physics}
\usepackage{multirow}
\usepackage{gensymb}
\usepackage[colorlinks=true, pdfstartview=FitV, bookmarks=true, bookmarksnumbered=true, breaklinks]{hyperref}
\usepackage{mathtools,braket}
\usepackage{lipsum}  
\usepackage{color}
\definecolor{blue}{rgb}{0.0, 0.0, 1.0}
\definecolor{red}{rgb}{1.0, 0.0, 0.0}
\definecolor{royalblue}{rgb}{0.0, 0.14, 0.4}
\hypersetup{linkcolor=blue, citecolor=blue, urlcolor=blue}

\def\orcid#1{\kern .08em\href{https://orcid.org/#1}{\includegraphics[keepaspectratio,width=0.7em]{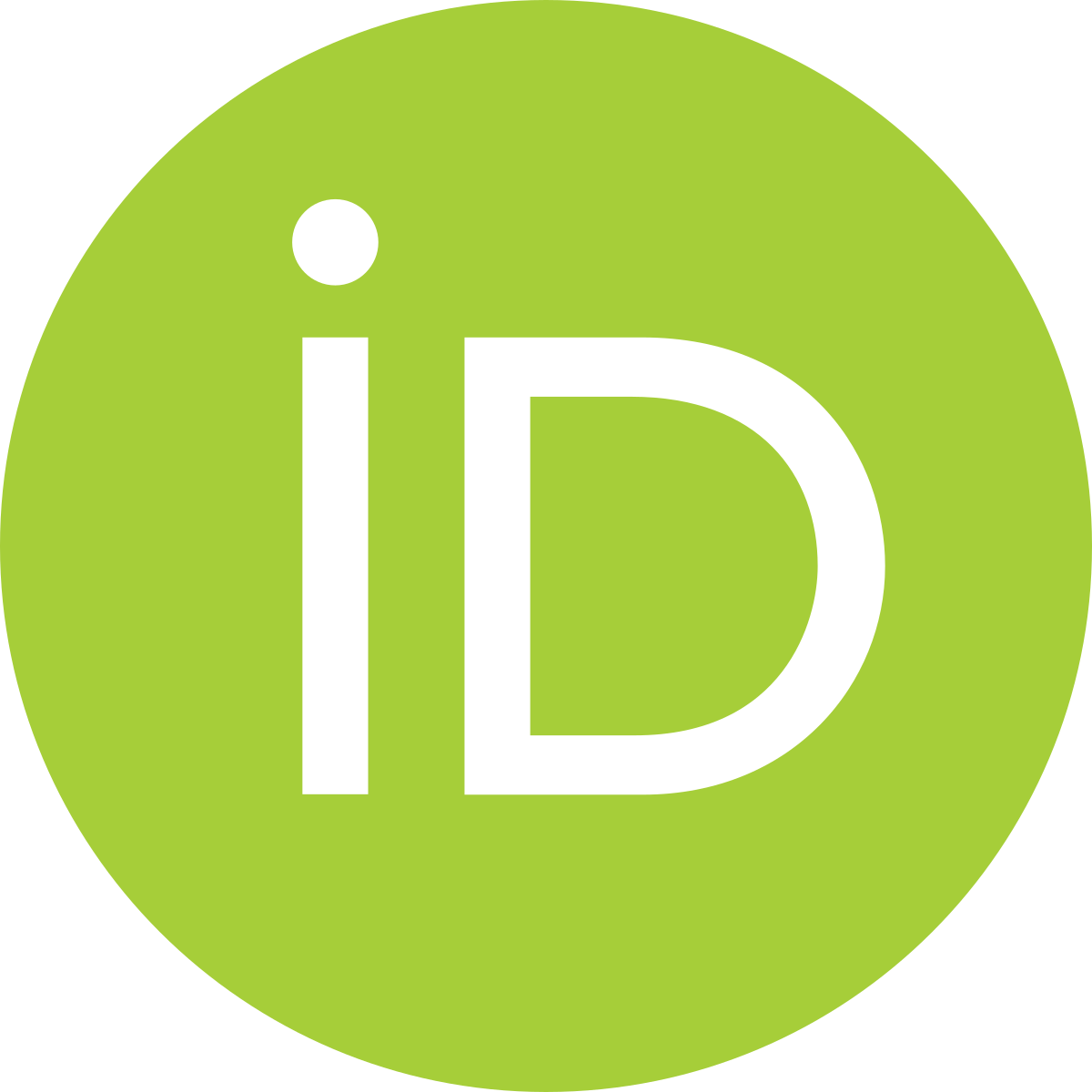}}}

\begin{document}

\title{Hadronic exceptional points}

\author{Ahmad Jafar Arifi\orcid{0000-0002-9530-8993}}
\email{aj.arifi01@gmail.com}
\affiliation{Advanced Science Research Center, Japan Atomic Energy Agency, Tokai, Ibaraki 319-1195, Japan}
\affiliation{Research Center for Nuclear Physics, The University of Osaka, Ibaraki, Osaka 567-0047, Japan}

\author{Kei Suzuki\orcid{0000-0002-8746-4064}}
\email{k.suzuki.2010@th.phys.titech.ac.jp}
\affiliation{Advanced Science Research Center, Japan Atomic Energy Agency, Tokai, Ibaraki 319-1195, Japan}

\date{\today}

\begin{abstract} 
Exceptional points, where eigenvalues and eigenvectors coalesce, are a defining feature of non-Hermitian systems and have been extensively observed in photonic, atomic, and condensed matter systems.
However, they have received little attention 
in quantum chromodynamics (QCD), which is the fundamental theory of quarks, gluons, and hadrons.
We propose that imaginary magnetic fields provide a simple realization of non-Hermitian dynamics in hadronic systems. 
Based on two theoretical approaches, a hadronic effective Lagrangian and a constituent quark model, 
we compute mass spectra of neutral mesons and find exceptional points separating the real-spectrum and complex-eigenvalue regimes. 
In small fields, the real spectrum exhibits level attraction between hadronic states, whereas in larger fields, hadrons are deconfined, which is a signature of a field-induced inverted potential.
Our findings open a new avenue for studying QCD dynamics in non-Hermitian environments. 
\end{abstract}

\maketitle


\textit{Introduction.}---Non-Hermitian quantum systems have attracted significant attention in the study of dissipation, amplification, and instability phenomena~\cite{Bender:2007nj, El-Ganainy:2018ksn, Bergholtz:2019deh, Ashida:2020dkc,Ding:2022juv}. 
A hallmark of such systems is the presence of \emph{exceptional points} (EPs)~\cite{Kato1966}, where both eigenvalues and eigenvectors coalesce.
EPs have been established also in experiments~\cite{Dembowski:2001zz,Guo:2009yqd,Ruter:2010qjb}.
While these phenomena are extensively studied in quantum optics, condensed matter, and atomic systems, their realization in quark and hadronic systems remains open questions.

Non-Hermitian concepts have historically played an important role in the description of resonances in scattering problems~\cite{Gamow:1928zz, Breit:1936zzb,Siegert:1939zz, Fano:1961zz,Feshbach:1958nx}.
In the low-energy regime of quantum chromodynamics (QCD), effective degrees of freedom are hadrons, including bound states and resonances.
Non-Hermitian structures naturally emerge in the description of unstable hadron resonances, or they are introduced as mathematical tools~\cite{Aguilar:1971ve,Ho:1983lwa,Moiseyev:1998gjp,Nawa:2011pz}\footnote{Other related discussions of exceptional points and hadron physics can be found in Ref.~\cite{Ghodrati:2025fah}.} for describing resonance phenomena. 
In contrast, it is still missing to study non-Hermitian dynamics of hadrons including bound states through external fields or other tunable physical parameters.

\begin{figure}[b]
    \centering
    \includegraphics[width=0.99\columnwidth]{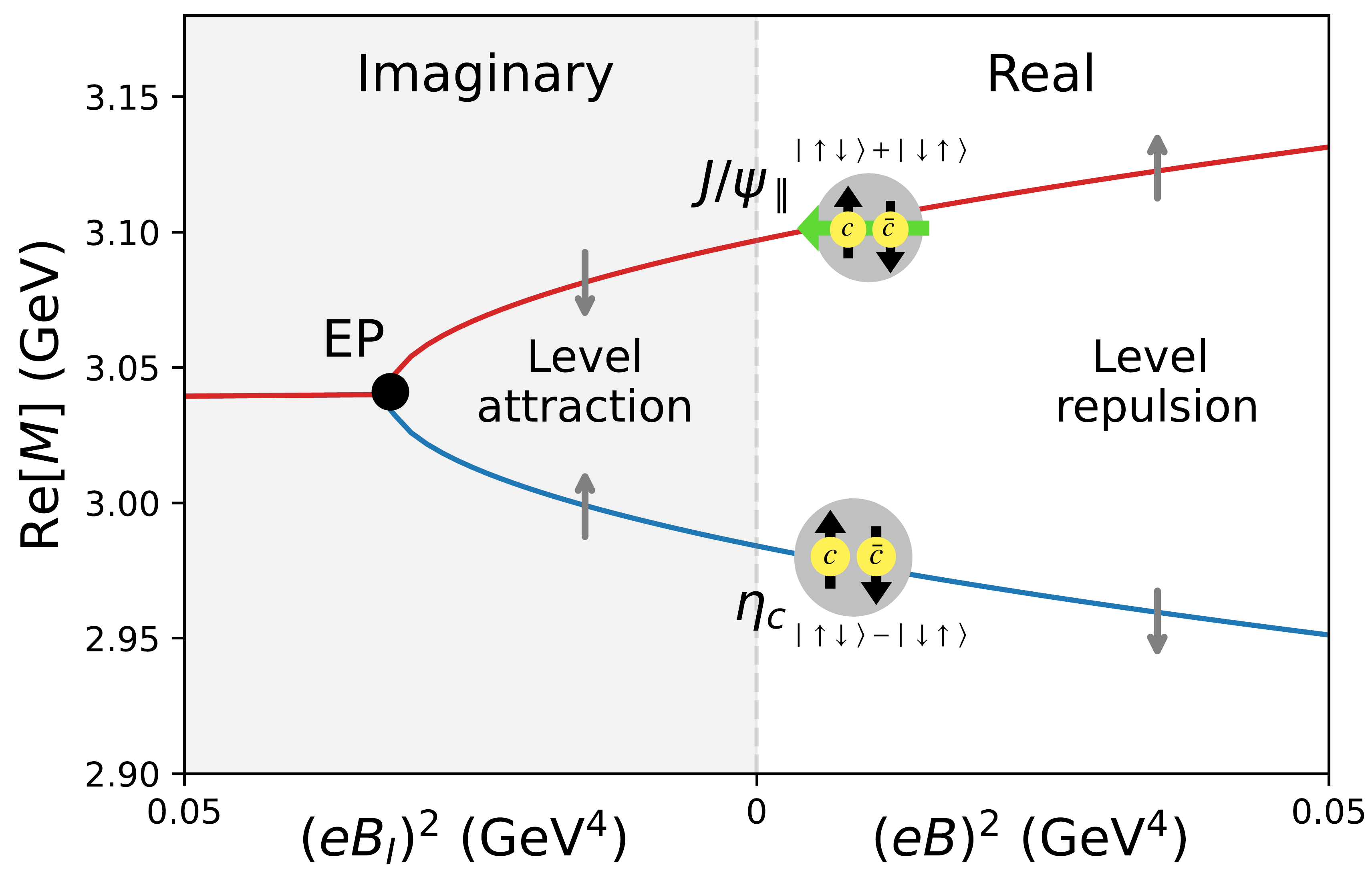}  
    \caption{Schematic illustration of the charmonium mass evolution as a function of a magnetic field. Level attraction in the imaginary-magnetic-field $(eB_I)$ region gives rise to an EP, while level repulsion occurs in the real-magnetic-field $(eB)$ region.}
    \label{fig:real-imaginary}
\end{figure}

In this letter, we propose \emph{imaginary magnetic fields} as a simple route to tunably realize EPs in hadron mass spectra and to explore non-Hermitian phenomena in hadronic systems.
A typical behavior of charmonium mass spectrum is illustrated in Fig.~\ref{fig:real-imaginary}. 
Based on an effective Lagrangian and quark model approach, we find that (i) adjacent energy levels exhibit \emph{level attraction}, (ii) EPs emerge at moderate field strengths, (iii) the spectrum has \emph{real eigenvalue} below the EPs and becomes \emph{complex eigenvalue} above the EPs, 
and (iv) at stronger fields, the system undergoes \emph{deconfinement}. 
These findings establish a concrete connection between non-Hermitian physics and hadron spectroscopy and open a new direction for exploring QCD in non-Hermitian environments.

Lattice QCD simulations, i.e, the most reliable first-principles approach of QCD, have extensively explored hadrons in real magnetic fields~\cite{Bali:2011qj,Luschevskaya:2012xd,Hidaka:2012mz,Luschevskaya:2014lga,Bonati:2014ksa,Luschevskaya:2015cko,Bonati:2016kxj,Bali:2017ian,Bali:2018sey,Luschevskaya:2018chr,Hattori:2019ijy,Endrodi:2019whh,Bignell:2019vpy,Bignell:2020dze,Ding:2020hxw,DElia:2021tfb, Ding:2022tqn,Ding:2025pbu,Ding:2026qzu} (see Ref.~\cite{Endrodi:2024cqn} for a review).
On the other hand, the study in real electric fields is limited due to the sign problem in the Monte Carlo method. \emph{Imaginary electric fields} were originally introduced as a way to circumvent the sign problem~\cite{Fiebig:1988en,Christensen:2004ca,Engelhardt:2007ub,Detmold:2009dx,Detmold:2010ts,DElia:2012ifm,Lujan:2014kia,Freeman:2014kka,Lujan:2016ffj,Niyazi:2021jrz,Yang:2022zob,Endrodi:2023wwf,Yang:2026tzt}~\footnote{As other techniques for real electric fields, there are methods with partially quenched real electric fields~\cite{Shintani:2006xr,Shintani:2008nt} and isospin electric fields~\cite{Yamamoto:2012bd}.}, but they can also serve as a platform for studing non-Hermitian QCD phenomena. As shown in this letter, imaginary magnetic fields~\footnote{Imaginary magnetic fields have historically been introduced as an artificial but useful parameter, for example, in the context of Lee-Yang zeros~\cite{Yang:1952be,Lee:1952ig}.
More recently, a few phenomenological studies~\cite{Ozawa:2023oqc,Montag:2026cqh} have investigated physical consequences of imaginary magnetic fields in non-Hermitian environments.} provide one of the simplest realizations of non-Hermitian QCD phenomena. Therefore, non-Hermitian signatures, such as hadronic EPs, offer a natural benchmark for future developments of QCD simulation methods, including non-Monte-Carlo approaches~\cite{Banuls:2019rao} and improved Monte-Carlo techniques.

\label{sec:eff-lagrangian}

\textit{Approach 1: Effective Lagrangian approach.}---We first formulate an effective Lagrangian for hadron fields in the presence of an imaginary magnetic field.
As an illustrative example, here we focus on charmonia, i.e., meson states composed of charm and anticharm quarks.
The advantages of this system are that (i) not only relativistic frameworks but also nonrelativistic descriptions work well, and
(ii) the properties in real magnetic fields have been extensively investigated in phenomenological studies
~\cite{Marasinghe:2011bt,Tuchin:2011cg,Yang:2011cz,Tuchin:2013ie,Machado:2013rta,Alford:2013jva,Cho:2014exa,Cho:2014loa,Dudal:2014jfa,Bonati:2015dka,Guo:2015nsa,Sadofyev:2015hxa,Suzuki:2016kcs,Yoshida:2016xgm,Suzuki:2016fof,Hasan:2017fmf,Dutta:2017pya,Singh:2017nfa,Hoelck:2017dby,Braga:2018zlu,Iwasaki:2018pby,Hasan:2018kvx,Amal:2018qln,Bagchi:2018olp,Iwasaki:2018czv,Braga:2019yeh,Hasan:2020iwa,Chen:2020xsr,Zhou:2020ssi,Braga:2020hhs,Iwasaki:2021kms,Braga:2021fey,Jena:2022nzw,Hu:2022ofv,Ghosh:2022sxi,Parui:2022msu,Sebastian:2023tlw,Nilima:2024nvd,Jena:2024cqs,Shukla:2024qlf,Wen:2025dwy,Jena:2025xcf,Arifi:2025ivt,Yan:2025tlx,Dominguez:2025nar,Arifi:2025atv, Arifi:2026asz} 
(see Refs.~\cite{Hattori:2016emy,Zhao:2020jqu,Iwasaki:2021nrz} for reviews), while many of the phenomena predicted here also appear in other hadronic systems (see e.g., Refs.~\cite{Gusynin:1995gt,Gusynin:1995nb,Shushpanov:1997sf,Agasian:2001ym,Chernodub:2010qx,Chernodub:2011mc,Callebaut:2011ab,Ammon:2011je,Chernodub:2011gs,Andersen:2012dz,Andersen:2012zc,Kojo:2012js,Andreichikov:2013zba,Orlovsky:2013gha,Fayazbakhsh:2013cha,Frasca:2013kka,Andreichikov:2013pga,Kamikado:2013pya,Liu:2014uwa,Haber:2014ula, Kamikado:2014bua,Taya:2014nha,He:2015zca,Avancini:2015ady,Kawaguchi:2015gpt,Hattori:2015aki,Zhang:2016qrl,Avancini:2016fgq,He:2016oqk,Li:2016gfn,Andreichikov:2016ayj,Kawaguchi:2016gbf,Ghosh:2016evc,Mao:2017wmq,Ghosh:2017rjo,GomezDumm:2017jij,Aguirre:2017dht,Wang:2017vtn,Liu:2018zag,Coppola:2018vkw,Andreichikov:2018wrc,Mao:2018dqe,Ayala:2018zat,He:2018vfc,Coppola:2018ygv,Avancini:2018svs,Ghosh:2019fet,Yakhshiev:2019gvb,Das:2019ehv,GomezDumm:2020bxj,Li:2020hlp,Coppola:2020mon,Sheng:2020hge,Ayala:2020dxs,Kojo:2021gvm,Moreira:2022dwo,Carlomagno:2022inu,Sheng:2022ssp,Carlomagno:2022arc,Mei:2022dkd,Braghin:2022uih,Chen:2023gws,Chen:2023jbq,Li:2023rsy,Ayala:2023llp,Coppola:2023mmq,Mei:2024rjg,Coppola:2024uvz,Hernandez:2025inu,Coppola:2025nus,Mei:2026xlj} for light hadrons in real magnetic fields).

By analytically continuing the magnetic field as $B \to iB_I$ in the effective Lagrangian for charmonia in real magnetic fields~\cite{Cho:2014exa,Cho:2014loa}, the interaction Lagrangian is
\begin{align}
\mathcal{L}_{{VP}\gamma } = \frac{g_{{VP}\gamma}}{m_0} \, i e \,\tilde{F}_{\mu\nu} (\partial^\mu P) V^\nu,
\end{align}
where $P$ and $V^\mu=(V_0,\bm{V}_\perp,V_\parallel)$ denote pseudoscalar and vector fields, respectively, $g_{{VP}\gamma}$ is a dimensionless coupling, and $m_0\equiv(m_{P}+m_{V})/2$ is the average mass of the pseudoscalar and vector states. Here, $\tilde{F}_{\mu\nu}$ is the dual electromagnetic field strength tensor incorporating an external field, and $e$ is the elementary electric charge. 
The coupling constant $g_{{VP}\gamma}$ is determined as $g_{{VP}\gamma} = \sqrt{12\pi m_0^2 e^{-2} k_\gamma^{-3}\Gamma_{\mathrm{M1}}}$, where $\Gamma_{\mathrm{M1}}$ is an M1 decay width for ${V}\to {P}+\gamma$, and $k_\gamma=(m_{{V}}^2 - m_{{P}}^2)/(2m_{{V}})$ is the outgoing photon momentum.
The experimental values of $\Gamma_{\mathrm{M1}}$ are taken from Particle Data Group~\cite{ParticleDataGroup:2024cfk}. 

When we assume an imaginary magnetic field along the $z$ direction $\tilde{F}_{03}=B_I$ and a vanishing spatial momentum $p^\mu=(\omega,0,0,0)$, the equations of motion reduce to a set of coupled algebraic equations for the pseudoscalar field $P$ and the longitudinal component $V_\parallel$ of vector fields, and $\bm{V}_\perp$ are independent of other fields.
The resulting Klein-Gordon equation incorporating the coupled channel of $P$-$V_\parallel$ can be written as
$\mathcal{M}\Psi=\omega^2\Psi$, where $\Psi$ is the eigenfunction of a matrix operator $\mathcal{M}$.
As an example, for $P=\eta_c$ and $V_\parallel=J/\psi_\parallel$,
\begin{align}
\begin{pmatrix}
 m_{\eta_c}^2 - \omega^2 & \dfrac{g_{J/\psi\eta_c\gamma}}{m_0} \omega eB_I \\
- \dfrac{g_{J/\psi\eta_c\gamma}}{m_0} \omega eB_I & m_{J/\psi}^2 -\omega^2 
\end{pmatrix}
\begin{pmatrix}
\eta_c \\[4pt] J/\psi_\parallel
\end{pmatrix}
= 0, \label{eq:EFT-eigen}
\end{align}
where $m_0=(m_{\eta_c}+m_{J/\psi})/2$. 
Due to the off-diagonal components, this matrix is manifestly non-Hermitian, $\mathcal{M}_{12}=-\mathcal{M}_{21}$, and does not satisfy $\mathcal{M}^\dagger=\mathcal{M}$.
Thus, the relativistic Klein-Gordon equation yields a quadratic eigenvalue problem in $\omega$. 
We emphasize that this matrix is pseudo-Hermitian~\cite{Mostafazadeh:2001jk,Mostafazadeh:2001nr,Mostafazadeh:2002id}, which allows either real or complex-conjugate pair eigenvalues (see Supplemental Material S1).

By solving the condition $\det (\mathcal{M}-\omega^2 I)= 0$, the analytic solutions of mass formulas are obtained as
\begin{align} 
\omega^2_{\pm} = \frac{1}{2} \left( m_{\eta_c}^2 + m_{J/\psi}^2 - \gamma^2 \pm \sqrt{\Delta
} \right), \label{eq:eigenvalues}
\end{align} 
where we define 
\begin{align}
\Delta=(m_{J/\psi}^2 - m_{\eta_c}^2)^2 - 2\gamma^2 (m_{\eta_c}^2 + m_{J/\psi}^2) + \gamma^4, 
\end{align} 
with $\gamma^2=g_{J/\psi\eta_c\gamma}^2(eB_I)^2/m_0^2$. The location of the EP is analytically determined by $\Delta=0$~\footnote{Formally, for the discriminant in Eq.~\eqref{eq:eigenvalues}, there is another solution but located at much stronger field region at $eB_I^{\mathrm{EP}} = {(m_{J/\psi} + m_{\eta_c})^2}/{2g_{J/\psi\eta_c\gamma}}$.}:
\begin{align}
    eB_I^{\mathrm{EP}} = \frac{m_{J/\psi}^2 - m_{\eta_c}^2}{2g_{J/\psi\eta_c\gamma}}. \label{eq:EPlocation}
\end{align}
Using the experimental charmonium masses~\cite{ParticleDataGroup:2024cfk}, $m_{\eta_c}=2984.1$ MeV and $m_{J/\psi}=3096.9$ MeV, and $g_{J/\psi\eta_c\gamma} = 1.912$, we obtain $eB_I^{\mathrm{EP}} \approx 0.18~\mathrm{GeV}^2$. We emphasize that the location of the EP is solely determined by the mass gap and its coupling strength of the M1 transition between them.
Furthermore, this two coupled-channel system can also be similarly applied for $\eta_c(2S)$ and $\psi(2S)$.

\begin{figure}[t]
    \centering
    \includegraphics[width=0.49\columnwidth]{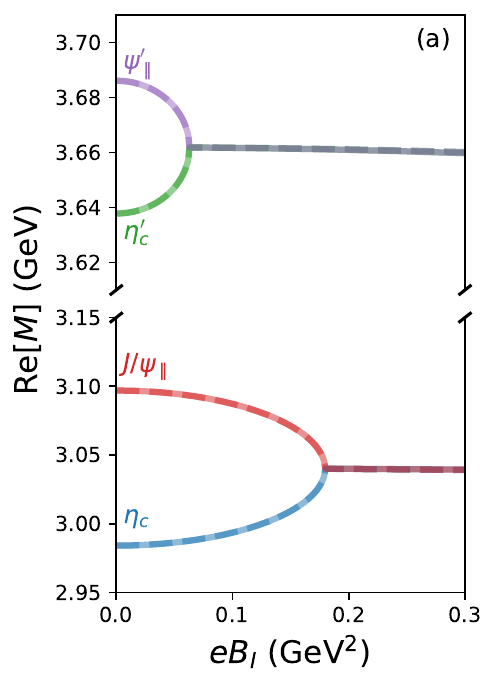} 
    \includegraphics[width=0.49\columnwidth]{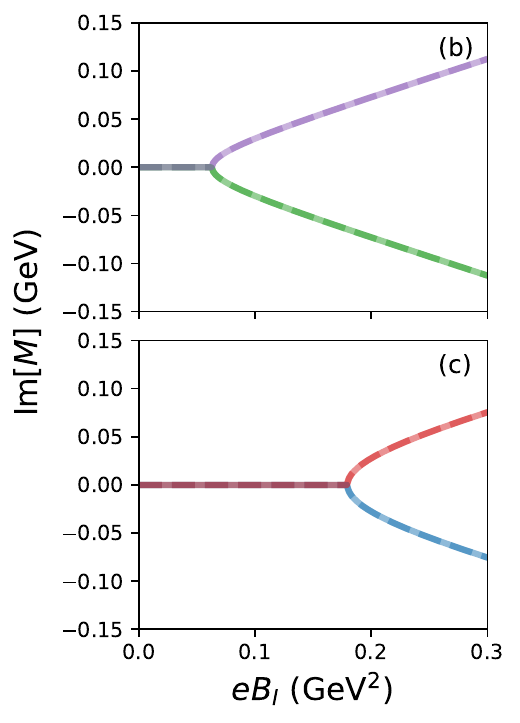}
\caption{Eigenvalues of the longitudinal charmonium states in imaginary magnetic fields, computed within the effective Lagrangian approach, where we plot
(a) the real part and (b,c) the imaginary part. 
Dashed lines: the results from the two-coupled-channel model.
Solid lines: the four-coupled-channel model, demonstrating the robustness of the results against the inclusion of additional channels.}
    \label{fig:charm-eft}
\end{figure}

Figure~\ref{fig:charm-eft} shows that the two mass eigenvalues undergo level attraction with increasing $eB_I$.
This is in clear contrast to the level repulsion typically found for real magnetic fields $eB$~~\cite{Cho:2014exa,Cho:2014loa}. 
We further find that the masses are smoothly connected across the real- and imaginary-magnetic-field regions when plotted as functions of $(eB)^2$, as illustrated in Fig.~\ref{fig:real-imaginary}.
This behavior can be derived from the small $(eB)^2$ or $(eB_I)^2$ expansion of corresponding mass formulas. 
The two eigenvalues and eigenvectors coalesce at the EP, which occurs at $eB_I^{\mathrm{EP}} \approx 0.18~\mathrm{GeV}^2$. 
In addition, the EPs for higher excited states tend to occur at smaller values of $eB_I$, due to the smaller mass gap and the larger coupling constant as expected in Eq.~\eqref{eq:EPlocation}.
For $eB_I>eB_I^\mathrm{EP}$, the eigenvalues split into a complex conjugate pair as shown in the right panels of Fig.~\ref{fig:charm-eft}, implying that the eigenvalues become complex: the real part corresponds to the energy, while the imaginary part encodes the decay or amplification rate, characteristic of a non-Hermitian system. 

The extension to a four-coupled-channel system is straightforward and naturally incorporates additional mixing among the states. 
We find that the results remain qualitatively robust in the small $eB_I$ region, as shown in Fig.~\ref{fig:charm-eft}.
However, noticeable differences from the two-coupled-channel results emerge only at larger magnetic fields, typically for $eB_I \gtrsim 6~\mathrm{GeV}^2$, where the additional channel couplings become increasingly important (see Supplemental Material S2).

Within the effective Lagrangian approach, bound states are introduced as explicit hadronic degrees of freedom and thus remain formally well-defined even in the strong-field regime. 
In the present work, however, we restrict our analysis to moderate values of $eB_I$, where the EPs already emerge.
This is because, as will be discussed later, at a sufficiently large field, quarkonia are expected to dissociate due to deconfinement.
Furthermore, in the current approach, we can study only the $eB_I$ dependence of the longitudinal charmonia whereas the transverse states are independent of $eB_I$. Because of this, the more microscopic model is required to get a more complete description.

\label{sec:quark-model}

\textit{Approach 2: Quark model.}---To obtain a more microscopic picture, we investigate charmonium spectra within a nonrelativistic quark model that explicitly incorporates the underlying quark structure in the presence of an imaginary magnetic field. The Hamiltonian for two-body systems consisting of a charm quark $c$ and an anticharm quark $\bar{c}$ under a real magnetic field is given by~\cite{Alford:2013jva}
\begin{equation}
H_{c\bar{c}} = \sum_{i=c,\bar{c}} \left[
    m_i + \frac{(\bm{p}_i - q_i \bm{A})^2}{2m_i}
    - \bm{\mu}_i \cdot \bm{B}
\right] + V(r),
\end{equation}
with the quark mass $m_i$, the momentum $\bm{p}_i$, the electric charge $q_i$, the vector potential $\bm{A}=\frac12 \bm{B}\times \bm{r}$ in the symmetric gauge, and the quark magnetic moment $\bm{\mu}_i = g q_i \bm{S}_i/(2m_i)$, where the quark spin $\bm{S}_i$ and the Land{\'e g-}factor $g=2$.
The interquark potential is
\begin{equation}
V(r)= C + \sigma r - \frac{\alpha}{r}
+ \beta(\bm{S}_c\!\cdot\!\bm{S}_{\bar c}) e^{-\Lambda r^2}.
\end{equation}
The parameters $\sigma$, $\alpha$, and $\beta$ characterize the confining, Coulomb, and spin-spin interactions, and $\Lambda$ controls Gaussian smearing of the spin-spin potential~\cite{Barnes:2005pb}.  
The constant $C$ fixes the overall mass shift.  
For the model parameters, we use the same set as Refs.~\cite{Suzuki:2016kcs,Yoshida:2016xgm}:  
$m_c=1.784\,{\rm GeV}$, $\alpha=0.713$, $\sqrt{\sigma}=0.402\,{\rm GeV}$,  
$\beta=0.4778\,{\rm GeV}$, $\Lambda=1.020\,{\rm GeV}^2$, $C=-0.5693\,{\rm GeV}$, and $q_c= -q_{\bar{c}}=(2/3)e$ with the elementary charge $e$.

The Hamiltonian in relative coordinates is well-known~\cite{Machado:2013rta,Alford:2013jva,Andreichikov:2013zba}.
For systems with vanishing pseudo-momentum and the analytic continuation $B \to i B_I$,
\begin{equation}
H_{c\bar{c}} = -\frac{\bm{\nabla}^2}{2\mu}
- \frac{q^2 B_I^2}{8\mu} r_\perp^2
+ V(r)
- i \sum_{i=c,\bar c} \bm{\mu}_i \cdot \bm{B}_I. \label{eq:H_rel}
\end{equation}

The presence of an imaginary magnetic field in the quark model induces two important effects.
(i) The transverse harmonic-oscillator proportional to $+B^2 r_\perp^2$ known in a real magnetic field, which leads to Landau levels of quarks, is transformed into an \emph{inverted potential} proportional to $-B_I^2 r_\perp^2$ in Eq.~(\ref{eq:H_rel}).
As a result, this term induces deconfinement of quarks at a sufficiently large magnetic field.
(ii) The interaction between the spin and the field, the fourth term in Eq.~(\ref{eq:H_rel}), becomes purely imaginary, which corresponds to the imaginary Zeeman effect of quarks.
Similar to the effective Lagrangian approach, this term gives rise to an spin-state mixing between the spin eigenstates of charmonia, the spin-singlet state $\ket{00}$ and the longitudinal spin-triplet state $\ket{10}$, where we labeled $\ket{SS_z}$ with the total spin $S$ and its $z$ component $S_z$: the off-diagonal matrix element is written as
\begin{equation}
\bra{10} -i(\bm{\mu}_c+\bm{\mu}_{\bar{c}} )\cdot\bm{B}_I \ket{00}
= -i\frac{g q B_I}{4\mu}.
\end{equation}
On the other hand, the transverse spin-triplet states, $S_z=\pm1$, are not mixed with  other states and are not affected by the imaginary Zeeman interaction due to an exact cancellation between the quark and antiquark.

Thus, the Schr\"odinger equation for $S_z=0$ is a non-Hermitian coupled-channel equation, while those for $S_z=\pm1$ are Hermitian single-channel equations.
To numerically solve these equations and precisely extract both the real and imaginary parts of the spectrum, we apply the cylindrical Gaussian expansion method (CGEM)~\cite{Suzuki:2016kcs,Yoshida:2016xgm}.
Although this method was originally developed for systems in real magnetic fields, we find that this method can be straightforwardly extended to a complex Hamiltonian. 
Note that if the Hamiltonian is reduced to a $2\times2$ coupled-channel Hamiltonian, we can discuss the the parity-time (PT) symmetry~\cite{Bender:1998ke,Bender:2002vv} (see Supplemental Material S1).

\begin{figure}[t]
    \centering
    \includegraphics[width=0.49\columnwidth]{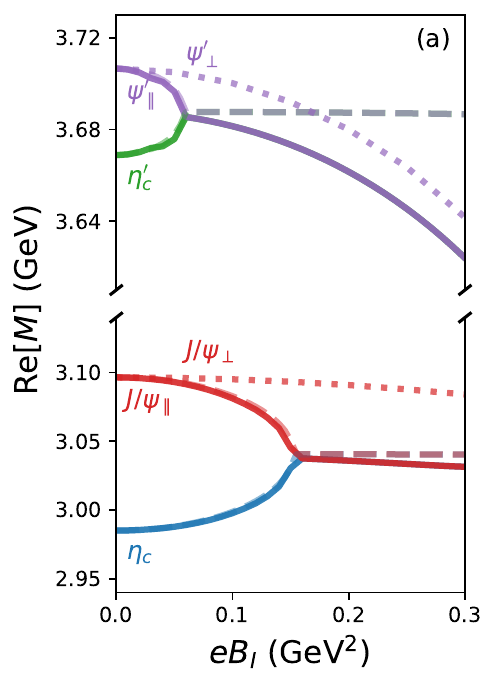} 
    \includegraphics[width=0.49\columnwidth]{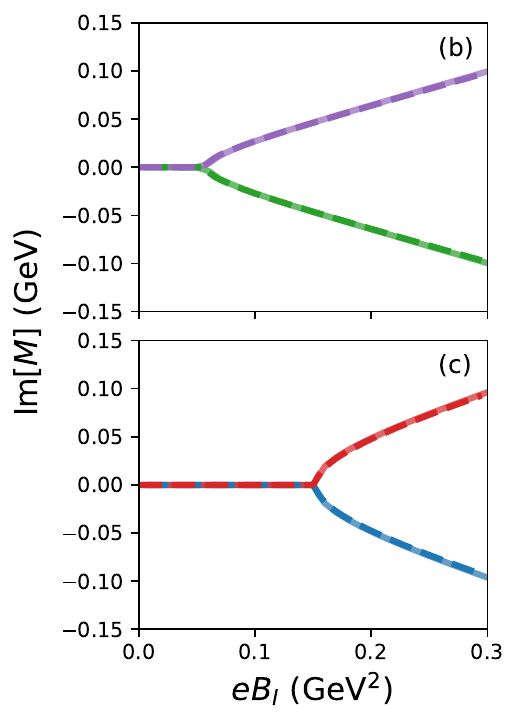} 
\caption{
Eigenvalues of the charmonium states in imaginary magnetic fields, computed using the constituent quark model, where we plot (a) the real part and (b,c) the imaginary part.
Solid lines: the full results for the longitudinal charmonia.
Dashed lines: the longitudinal charmonia {\it without} the inverted potential.
Dotted lines: the transverse charmonia.}
    \label{fig:charm-qm}
\end{figure}

For longitudinal charmonia, the mixing between spin-singlet and spin-triplet states induces level attraction and gives rise to EPs, as shown in Fig.~\ref{fig:charm-qm}. Within the present parameter set, the EP appears at $eB \sim 0.15~\mathrm{GeV}^2$ for the ground state and at a smaller field strength for the excited state. Beyond the EPs, the eigenvalues become complex. We find that the locations of the EPs are insensitive to the inclusion of the inverted potential, indicating that the EP structure is governed primarily by the imaginary Zeeman interaction, whose strength is controlled by the quark magnetic moment.

Unlike the effective Lagrangian approach based on the point-particle picture, in the quark model, the internal structure of charmonium is reflected through an inverted potential in a magnetic field. Figure~\ref{fig:charm-qm} compares the mass spectra obtained with (solid lines) and without (dashed lines) this contribution. The inverted potential primarily affects the real part of the eigenvalues: we find a reduction of the masses after the EPs, while the imaginary part is nearly unchanged.

In the transverse channel, there is no EP structure due to the absence of mixing, and the inverted potential becomes the only magnetic-field effect. Consequently, the masses decrease monotonically as the magnetic field increases, as shown in Fig.~\ref{fig:charm-qm}.
We find that the mass shift for the excited state is larger than that for the ground state. This is because the broader spatial wave function is strongly affected by the inverted potential.

The inverted potential weakens the confinement and eventually leads to the dissolution of bound states. This situation might be similar to the thermal deconfinement (i.e., meson melting at finite temperature~\cite{Matsui:1986dk,Hashimoto:1986nn}), but its mechanism originates from the non-Hermiticity.
Then, when the magnitude of the field exceeds a certain value, the system becomes unbound.
In the current parameter sets, the numerical solutions corresponding to bound states become unstable at around $eB \sim 0.30~\mathrm{GeV}^2$.
This is regarded as a numerical signal of the onset of deconfinement. 
As a numerical check, if we neglect the inverted potential term (like the dashed lines in Fig.~\ref{fig:charm-qm}), the solutions for bound states remain stable, and the deconfinement does not occur, which is similar to the situation in the effective Lagrangian approach.

\label{sec:discussion}

\textit{Discussion and conclusion.}—In this work, we have demonstrated that imaginary magnetic fields induce non-Hermitian dynamics in charmonium systems. Using the two approaches, we consistently find the emergence of level attraction, EPs, and complex eigenvalues.

\paragraph{Location and robustness of EPs.}
One of the central findings of this work is the location and robustness of the EPs studied within the two approaches. Taking into account parameter sensitivities, we find that the EP for the $J/\psi$-$\eta_c$ system most likely appears at $eB_I \sim 0.10$--$0.20~\mathrm{GeV}^2$. Because this scale is primarily governed by the charmonium mass splitting and the corresponding coupling constant, which is similar to the case in real magnetic fields, the EP location is expected to be relatively model-independent.
Importantly, the EP is expected to emerge before the onset of the deconfinement regime: physical phenomena around the EP can be observed without being hindered by deconfinement.
To quantitatively constrain the EP location, further studies using more refined theoretical models are expected. 

\paragraph{EPs and avoided crossings.}
A situation analogous to EPs in imaginary magnetic fields is the avoided level crossing known to occur in real magnetic fields (e.g., see Ref.~\cite{Suzuki:2016kcs,Yoshida:2016xgm}).
The avoided level crossing is an interesting phenomenon, while it requires the consideration of radially excited states, such as $\eta_c(2S)$ and $\psi(2S)$.
In this sense, its theoretical or experimental treatment becomes somewhat more complicated.
In contrast, the EP positions can be reliably captured within a minimal two-channel model composed of only the ground states. For the charmonium system, where higher states are well separated, the inclusion of additional channels does not substantially modify the EP location. This indicates that EP predictions are more robust and tightly constrained than those of avoided crossings.

\paragraph{EPs in other neutral mesons.} 
Here, we focused on the emergence of EPs in charmonium spectra, since quarkonia are relatively more stable against strong interaction compared to light mesons. 
More generally, EPs can emerge systematically in mesons with different quark flavor content, although their precise locations depend sensitively on the mass splitting and couplings (see Supplemental Material S2).

\label{sec:outlook}

\textit{Outlook.}---In this work, we focused on only neutral S-wave mesons as a simple system.
For future topics, EPs may emerge even in (i) higher partial waves such as P-waves, (ii) baryons, and (iii) charged hadrons, where EPs are induced by similar spin-eigenstate mixing.
In lattice QCD simulations, the signals of S-wave mesons are clearer than those of (i) higher partial waves and (ii) baryons.
(iii) Charged hadrons are also interesting for comparison with neutral ones, but purely real spectra before EPs are not realized due to the imaginary Zeeman effect in the hadronic level.
In addition, (iv) imaginary electric fields are also platforms of EPs, where opposite-parity eigenstates of hadrons, e.g., S-waves and P-waves, are mixed, and EPs between such states can appear.
In this sense, neutral S-wave mesons in imaginary magnetic fields may be one of the best choices for investigating hadronic EPs.

For future lattice QCD studies, finding hadronic EPs remains an important and open challenge.
Monte Carlo simulations of lattice QCD with imaginary magnetic fields will require further methodological developments for mitigating the sign problem, while recent progress in sign-problem-free algorithms~\cite {Banuls:2019rao} provides a promising direction. Our results motivate these approaches.
We also comment on analytic continuation.
To study physics in imaginary electric fields in Minkowski spacetime, simulations are performed in the corresponding Wick-rotated background in Euclidean spacetime. Then, the challenge is to relate the Euclidean results to the corresponding Minkowski physics. Since unlike electric fields, magnetic fields are invariant under Wick rotation, imaginary magnetic fields provide a more direct setting for investigating non-Hermitian QCD.

The key ingredient underlying the emergence of EPs in the present study is the imaginary spin-eigenstate mixing induced by the imaginary Zeeman coupling $- i(\bm{\mu}_c + \bm{\mu}_{\bar{c}}) \cdot \bm{B}_I$, while such imaginary mixing itself does not necessarily require the imaginary magnetic field $\bm{B}_I$.
A more general condition is an imaginary spin operator proportional to $- i (\gamma_1 S_{z,1}+ \gamma_2 S_{z,2})$ where $\gamma_1,\gamma_2$ are arbitrary real parameters satisfying $\gamma_1\gamma_2<0$: when two particles with nonzero spins show imaginary shifts with opposite signs, their bound states can exhibit EPs in a parameter region.
Such a situation can arise in environments with spin-dependent gain and loss, which can be engineered experimentally in a standard manner in atomic and phtonic systems, for example, using spontaneous emission in the spin-up state together with external pumping of the spin-down state, and optical loss in one polarization state and optical gain in the opposite polarization state provided by a pumped active medium.
Bound-state EPs in such systems are regarded as effective analogs to hadronic EPs induced by imaginary spin-state mixing.
Exploring these possibilities is an interesting direction for future research.

\textit{Acknowledgments.}---A. J. A. was supported by the JAEA Postdoctoral Fellowship Program, the RCNP Collaboration Research Network Program (COREnet 057), and the PUTI Q1 Grant, University of Indonesia (PKS-206/UN2.RST/HKP.05.00/2025), as well as Japan Society for the Promotion of Science (JSPS) KAKENHI (Grant No. 26K17163). K. S. was supported by JSPS KAKENHI (Grant No. JP24K07034).

\section*{Supplemental Materials}
\label{sec:supplemental}

\setcounter{section}{0}
\setcounter{equation}{0}
\setcounter{figure}{0}
\renewcommand{\theequation}{S\arabic{equation}}
\renewcommand{\thesection}{S\arabic{section}}
\renewcommand{\thefigure}{S\arabic{figure}}

\section{PT symmetry}

Here, we discuss the PT symmetry~\cite{Bender:1998ke,Bender:2002vv} and the pseudo-Hermiticity~\cite{Mostafazadeh:2001jk,Mostafazadeh:2001nr,Mostafazadeh:2002id} of the non-Hermitian hadron effective Lagrangian and the non-Hermitian quark-model Hamiltonian discussed in the main text.

\subsection{Quark-model Hamiltonian}
In the quark-model Hamiltonian under an imaginary magnetic field, there is mixing between a spin-singlet state and a longitudinal component of spin-triplet state.
When we focus on a single mixing pair, the original infinite-dimensional Hamiltonian can be reduced to an effective $2\times2$ Hamiltonian.
Then, the Shr\"odinger equation is reduced to a two-component coupled-channel equation.
A minimal form of this $2\times2$ Hamiltonian is written as
\begin{align}
H=
\begin{pmatrix}
a & ic \\
ic & b
\end{pmatrix},
\end{align}
characterized by real constants, $a$, $b$, and $c$.
This form belongs to a known general form of PT-symmetric $2\times2$ Hamiltonian~\cite{Bender:2003gu,Mostafazadeh:2003gz}.
We define a parity operator as
\begin{align}
P \equiv\sigma_z=
\begin{pmatrix}
1 & 0\\
0 & -1
\end{pmatrix}.
\end{align}
Note that this operator is not the conventional spatial parity operator but instead acts on an internal two-level space.
The time-reversal operator $T$ in this case acts as complex conjugation, $i\to-i$.
We confirm that the Hamiltonian satisfies
\begin{align}
PT\,H\,(PT)^{-1}=P H^\ast P=H.
\end{align}
Then, the PT symmetry guarantees that the eigenvalues $\omega$ are either real or a complex-conjugate pair.

The eigenvalues are obtained from $\det(H-\omega I)=0$:
\begin{align}
\omega_\pm=\frac{1}{2}\left(a+b\pm\sqrt{(a-b)^2-4c^2}\right).
\end{align}
The spectral behavior is governed by the discriminant
\begin{align}
\Delta=(a-b)^2-4c^2.
\end{align}
Three regimes can be identified:
\begin{itemize}
\item $|a-b|>2|c|$ ($\Delta>0$): the two eigenvalues are real.
This region is called the PT-symmetric phase.
\item $|a-b|=2|c|$ ($\Delta=0$): the two eigenvalues and eigenvectors coalesce at an exceptional point (EP), which separates the PT-symmetric and PT-broken phases.
\item $|a-b|<2|c|$ ($\Delta<0$): the eigenvalues form a complex-conjugate pair,
\begin{align}
\omega_\pm=\frac{a+b}{2}\pm \frac{i}{2}\sqrt{4c^2-(a-b)^2}.
\end{align}
This region is called the PT-broken phase.
\end{itemize}
Thus, the EP marking the transition from a real spectrum to a complex-conjugate pair occurs when the level splitting $|a-b|$ exactly balances the non-Hermitian coupling strength $2|c|$.

\subsection{Effective Lagrangian}

In the effective Lagrangian approach, the interaction Lagrangian results in a coupled-channel Klein-Gordon equation, $\mathcal{M}\Psi=\omega^2\Psi$.
A minimal form of the $2\times2$ matrix operator $\mathcal{M}$ is written as
\begin{align} 
\mathcal{M}=
\begin{pmatrix}
         a & + \gamma \omega \\
        - \gamma \omega&   b \\
    \end{pmatrix},
\end{align}
with real constants $a$, $b$, and $\gamma$.
This operator does not conserve the PT symmetry in the conventional manner, while we can derive the pseudo-Hermiticity~\cite{Mostafazadeh:2001jk,Mostafazadeh:2001nr,Mostafazadeh:2002id}.
Note that, strictly speaking, the pseudo-Hermiticity should be understood at the operator level by replacing $\omega$ with the time-derivative operator $i\partial_t$.
The pseudo-Hermiticity relation is~\cite{Mostafazadeh:2001jk}
\begin{align}
\mathcal{M}^\dagger=\eta\,\mathcal{M}\,\eta^{-1},
\end{align}
where $\eta$ is an invertible Hermitian operator.
In the present case, when we choose $\eta \equiv \sigma_z$, we find that $\mathcal M$ is pseudo-Hermitian.
Then, the pseudo-Hermiticity guarantees that the eigenvalues $\omega^2$ are either real or a complex-conjugate pair.

The eigenvalues are determined by solving $\det(\mathcal{M}-\omega^2 I)=0$:
\begin{align}
\omega_\pm^2 =\frac{1}{2}\left( a+b-\gamma^2  \pm 
\sqrt{\Delta}\right),
\end{align}
where
\begin{align}
\Delta=(a-b)^2+\gamma^4-2\gamma^2(a+b)
\end{align}
determines the spectral behavior.
Although the matrix $\mathcal{M}$ differs in structure from the non-relativistic quark-model Hamiltonian, its eigenvalues exhibit analogous behavior: they are real for $\Delta>0$, coalesce at an EP for $\Delta=0$, and form a complex-conjugate pair for $\Delta<0$.
However, in this case, since the left-handed side is $\omega^2$ instead of $\omega$, the spectrum contains four eigenvalues, corresponding to two pairs of positive- and negative-frequency solutions of the Klein-Gordon equation.
As a result, there are two types of EPs. While the first EP originates from mixing between the two hadron states, the second corresponds to the coalescence of positive- and negative-frequency branches.

\section{Effective Lagrangian approach}

Here, we discuss the construction of the effective Lagrangian approach for up to six coupled channel systems. 

\subsection{Coupled-channel systems} 

To consider imaginary magnetic fields, we use the replacement of the dual field strength tensor
\begin{align}\label{eq:tensor-imag}
    \tilde{F}_{\mu\nu} \to&~ \tilde{F}_{\mu\nu}^{\mathrm{[Re]}} + i\tilde{F}_{\mu\nu}^{\mathrm{[Im]}},
\end{align}
corresponding to
\begin{align}
    B \to&~ B + i B_I, \\
    E \to&~ E + i E_I.
\end{align}
This procedure corresponds to an analytic continuation of the external electromagnetic background. 

The total effective Lagrangian is written as $\mathcal{L} = \mathcal{L}_{\mathrm{free}} +\mathcal{L}_{VP\gamma},$
where the free Lagrangian for the pseudoscalar and vector meson fields is given by
\begin{align}
\mathcal{L}_{\mathrm{free}}
=&~ \frac{1}{2}(\partial_\mu P)(\partial^\mu P)
-\frac{1}{2}m_{P}^2P^2\nonumber\\
&~ + \frac{1}{2}(\partial_\mu V_\nu)(\partial^\mu V^\nu)
- \frac{1}{2}m_{V}^2V_\mu V^\mu.
\end{align}
For a purely imaginary magnetic field $eB_I$, the interaction Lagrangian becomes
\begin{align}
\mathcal{L}_{VP\gamma} = \frac{g_{VP\gamma}}{m_0}
\,ie\,\tilde{F}_{\mu\nu}^{[\mathrm{Im}]}
(\partial^\mu P)V^\nu.
\end{align}
This interaction is non-Hermitian due to the imaginary unit $i$, while the free part is Hermitian.
Here, $P$ and $V^\mu=(V_0, \bm{V}_\perp, V_\parallel)$ denote the pseudoscalar and vector fields, respectively, and the coupling constant is represented by $g_{{PV}\gamma}$, while $m_0 = (m_{P} + m_{V})/2$ is the average mass of the pseudoscalar and vector states.
$\tilde{F}_{\mu\nu}$ is the dual electromagnetic field-strength tensor in the presence of an external field, and $e$ denotes the elementary electric charge.

\paragraph{Two coupled-channel model}

From the total Lagrangian, the equations of motion are given by
\begin{align}
    (\partial^2 + m_{P}^2)P + \frac{g_{{VP}\gamma}}{m_0}ie \tilde{F}_{\mu\nu}\partial^\mu V^\nu =&~ 0, \label{eq:KG1} \\
    (\partial^2 + m_{V}^2)V_\nu - \frac{g_{{VP}\gamma}}{m_0}ie \tilde{F}_{\mu\nu}(\partial^\mu P) =&~ 0. \label{eq:KG2}
\end{align}
We consider an imaginary magnetic field along the $z$ direction, $\tilde{F}_{03}=B_I$, and vanishing spatial momentum $p^\mu=(\omega,0,0,0)$. Consequently, only the temporal derivative contributes, $\partial_0 \to -i\omega$.
In this setup, only the longitudinal component $V_\parallel$ couples to the pseudoscalar field $P$ and $\bm{V}_\perp$ remains decoupled.

The coupled-channel Klein-Gordon equations in Eqs.~\eqref{eq:KG1} and~\eqref{eq:KG2} can be written as $\mathcal{M}\Psi=\omega^2\Psi$, where $\Psi$ is the eigenfunction of an operator $\mathcal{M}$. For the lowest two pseudoscalar-vector pairs, $2\times 2$ forms are
\begin{align} 
    \begin{pmatrix}
        -\omega^2 + m_{{P}_{1}}^2 & + \dfrac{g_{11}}{m_{11}} \omega eB_I \\
        -\dfrac{g_{11}}{m_{11}} \omega eB_I & -\omega^2 + m_{{V}_{1}}^2 \\
    \end{pmatrix} 
    \begin{pmatrix}
      P_{1} \\[8pt] V_{1}^\parallel
    \end{pmatrix} = 0,
\end{align}
and
\begin{align}
    \begin{pmatrix}
        -\omega^2 + m_{{P}_{2}}^2 & + \dfrac{g_{22}}{m_{22}} \omega eB_I \\
        - \dfrac{g_{22}}{m_{22}} \omega eB_I & -\omega^2 + m_{{V}_{2}}^2 \\
    \end{pmatrix} 
    \begin{pmatrix}
        {P}_{2} \\[8pt] {V}_{2}^\parallel
    \end{pmatrix} = 0.
\end{align}
Here, ${P}_i$ and ${V}_i^\parallel$ ($i={1},{2}$) denote general pseudoscalar and longitudinal vector states, respectively. 
The off-diagonal terms encode the field-induced mixing and scale as $\omega eB_I$.
Here, $(g_{11},g_{22}) \equiv (g_{{V}_1{P}_1\gamma},g_{{V}_2{P}_2\gamma})$ are shorthand notations for the corresponding coupling constants and, $m_{11}=(m_{{P}_{1}}+m_{{V}_{1}})/2$ and $m_{22}=(m_{{P}_{2}}+m_{{V}_{2}})/2$ are the averaged mass of the pseudoscalar and vector states. To get the mass of these states as a function of imaginary magnetic field, we should solve  $\det (\mathcal{M}-\omega^2 I)= 0$ that give us the second-order eigenvalue equation for $\omega$.

A crucial feature is the relative minus sign between the off-diagonal elements, $\mathcal{M}_{12}=-\mathcal{M}_{12}$, which differs from the case of real magnetic fields. This sign structure renders the effective interaction non-Hermitian and leads to qualitatively different spectral behavior, including the level attraction and the emergence of EPs.

\paragraph{Four coupled-channel model}

When combining the two pseudoscalar-vector pairs, i.e., $({P}_{1},{V}^\parallel_{1})$ and $({P}_{2},{V}^\parallel_{2})$, the full $4\times 4$ system can be written as
\begin{widetext}
\begin{align}
    \begin{pmatrix}
        -\omega^2 + m_{{P}_{1}}^2 
        & \dfrac{g_{11}}{m_{11}} \omega eB_I 
        & 0 
        & \dfrac{g_{12}}{m_{12}} \omega eB_I \\
        - \dfrac{g_{11}}{m_{11}} \omega eB_I 
        & -\omega^2 + m_{{V}_{1}}^2 
        & \dfrac{g_{21}}{m_{21}} \omega eB_I 
        & 0 \\
        0 
        & - \dfrac{g_{21}}{m_{21}} \omega eB_I 
        & -\omega^2 + m_{{P}_{2}}^2 
        & \dfrac{g_{22}}{m_{22}} \omega eB_I \\
        - \dfrac{g_{12}}{m_{12}} \omega eB_I  
        & 0
        & -\dfrac{g_{22}}{m_{22}} \omega eB_I 
        & -\omega^2 + m_{{V}_{2}}^2 
    \end{pmatrix} 
    \begin{pmatrix}
        {P}_{1} \\[8pt]
        {V}_{1}^\parallel \\[8pt]
        {P}_{2} \\[8pt]
        {V}_{2}^\parallel
    \end{pmatrix}
    = 0,
\end{align}
\end{widetext}
where the coupling constants between the pseudoscalar-vector pairs are denoted by $g_{12}$ and $g_{21}$. 
The corresponding average masses are also different, 
$m_{12}=(m_{{P}_{1}}+m_{{V}_{2}})/2$ and $m_{21}=(m_{{P}_{2}}+m_{{V}_{1}})/2$.
In contrast to the two-channel case, the system now includes both intra-level transitions $({P}_i \leftrightarrow {V}_i)$ between the same principle quantum number and inter-level transitions $({P}_i \leftrightarrow {V}_j,\ i \neq j)$, leading to a richer coupled-channel structure. The zero elements reflects the selection rules of the underlying interaction: ${P}_i \to {P}_i \gamma$ or ${V}_i \to {V}_j \gamma$ is forbidden.

\paragraph{Six coupled-channel model}
For completeness, we extend the model to three pseudoscalar-vector pairs, $({P}_i, {V}_i)$ with $i={1},{2},{3}$. 
This extension is particularly relevant for systems such as light mesons, where nearby states can play a non-negligible role.

The resulting $6\times 6$ system can be written as
\begin{widetext}
\begin{align}
\begin{pmatrix}
-\omega^2 + m_{{P}_{1}}^2 
& \dfrac{g_{11}}{m_{11}} \omega eB_I 
& 0 
& \dfrac{g_{12}}{m_{12}} \omega eB_I 
& 0 
& \dfrac{g_{13}}{m_{13}} \omega eB_I \\
- \dfrac{g_{11}}{m_{11}} \omega eB_I 
& -\omega^2 + m_{{V}_{1}}^2 
& \dfrac{g_{21}}{m_{21}} \omega eB_I 
& 0 
& \dfrac{g_{31}}{m_{31}} \omega eB_I 
& 0 \\
0 
& - \dfrac{g_{21}}{m_{21}} \omega eB_I 
& -\omega^2 + m_{{P}_{2}}^2 
& \dfrac{g_{22}}{m_{22}} \omega eB_I 
& 0 
& \dfrac{g_{23}}{m_{23}} \omega eB_I \\
- \dfrac{g_{12}}{m_{12}} \omega eB_I 
& 0 
& - \dfrac{g_{22}}{m_{22}} \omega eB_I 
& -\omega^2 + m_{{V}_{2}}^2 
& \dfrac{g_{32}}{m_{32}} \omega eB_I 
& 0 \\
0 
& - \dfrac{g_{31}}{m_{31}} \omega eB_I 
& 0 
& - \dfrac{g_{32}}{m_{32}} \omega eB_I 
& -\omega^2 + m_{{P}_{3}}^2 
& \dfrac{g_{33}}{m_{33}} \omega eB_I \\
- \dfrac{g_{13}}{m_{13}} \omega eB_I 
& 0 
& - \dfrac{g_{23}}{m_{23}} \omega eB_I 
& 0 
& - \dfrac{g_{33}}{m_{33}} \omega eB_I 
& -\omega^2 + m_{{V}_{3}}^2
\end{pmatrix}
\begin{pmatrix}
{P}_{1} \\[6pt]
{V}_{1}^\parallel \\[6pt]
{P}_{2} \\[6pt]
{V}_{2}^\parallel \\[6pt]
{P}_{3} \\[6pt]
{V}_{3}^\parallel
\end{pmatrix}
= 0.
\end{align}
\end{widetext}

\subsection{Coupling constant determination}

The coupling constants $g_{ij}$, which parametrize the M1 transitions 
${P}_i \leftrightarrow {V}_j$, can be determined from the corresponding M1 decay widths. 
In general, they are extracted using
\begin{align} \label{eq:coupling}
    g_{ij}= \sqrt{ \frac{(2J_i+1)4\pi m_0^2 \Gamma_{\mathrm{M1}}}{e^2 k_\gamma^3}},
\end{align}
where $e=\sqrt{4\pi\alpha_\mathrm{em}}$ is evaluated with the fine-structure constant $\alpha_\mathrm{em}$, and 
$k_\gamma=(m_{{V}_j}^2 - m_{{P}_i}^2)/(2m_{{V}_j})$ is the center-of-mass momentum of the final state photon.
The spin factor (2$J_i+1)$ with $J_{V}=1$ and $J_{P}=0$ is important for ${V}\to {P}\gamma$ and ${P}\to {V}\gamma$ transitions.
In the following, we demonstrate the determination of coupling constants for physical mesonic systems.

\paragraph{Light-meson sector}

For the M1 decay width $\Gamma_{\mathrm{M1}}$ in Eq.~\eqref{eq:coupling}, we input experimental (Exp) data  of total widths and the corresponding branching ratios from Particle Data Group (PDG)~\cite{ParticleDataGroup:2024cfk}. 
In the unflavored sector, we consider the lowest six neutral mesons: $\pi^0,\eta,\rho^0,\omega,\eta^\prime$, and $\phi$. 
The coupling constants are determined as
\begin{align}
 {\rho^0 \pi^0 \gamma}:& \quad  g= 0.333,  \quad  \Gamma = \phantom{0} 69.278~\mathrm{keV}, \quad \mathrm{(Exp)}\\
 {\rho^0 \eta \gamma}:&  \quad  g= 1.044,  \quad  \Gamma = \phantom{0}44.220~\mathrm{keV}, \quad \mathrm{(Exp)}\\
 {\omega \pi^0 \gamma}:& \quad  g= 1.069,  \quad  \Gamma = 723.044~\mathrm{keV}, 
 \quad \mathrm{(Exp)}\\
 {\omega \eta \gamma}:& \quad    g= 0.299,  \quad  \Gamma = \phantom{00}3.906~\mathrm{keV}, \quad \mathrm{(Exp)}
\end{align}
and
\begin{align} 
{\eta^\prime\rho\gamma}:& \quad  g= 1.125,  
\quad \Gamma = \phantom{0}55.422~\mathrm{keV},\quad \mathrm{(Exp)}\\
{\eta^\prime\omega\gamma}:& \quad  g=0.349, 
\quad \Gamma = \phantom{00}4.738~\mathrm{keV}, \quad \mathrm{(Exp)}\\
 {\phi \pi \gamma}:& \quad     g= 0.079,  \quad  \Gamma = \phantom{00}5.651~\mathrm{keV}, \quad \mathrm{(Exp)}\\
 {\phi \eta \gamma}:& \quad     g= 0.542,  \quad  \Gamma = \phantom{0}55.492~\mathrm{keV}, \quad \mathrm{(Exp)}\\
 {\phi\eta^\prime\gamma}:&\quad g= 0.705,  \quad  \Gamma = \phantom{00}0.265~\mathrm{keV}, \quad \mathrm{(Exp)}
\end{align}
where we have used the experimental masses~\cite{ParticleDataGroup:2024cfk}: $m_\pi=134.9768,\ m_\eta=547.862,\ m_\rho=775.26,\ m_\omega=782.66,\ m_{\eta^\prime}=957.78,$ and $m_\phi=1019.46$ MeV.  

For the strange sector, we consider the neutral $K^0$ and $K^{*0}$ mesons.
The coupling constant is determined as
\begin{align}
  {K^{*} K \gamma}:&\quad  g= 0.885, \quad \Gamma=116.358~\mathrm{keV}, \quad \mathrm{(Exp)}
\end{align}
where we have used the neutral kaon masses, $m_K=497.611$ and $m_{K^*}=895.55$ MeV~\cite{ParticleDataGroup:2024cfk}.
Overall, the couplings in the light-meson sector exhibit a relatively broad 
range of magnitudes.

\paragraph{Heavy-light meson sector}

For heavy-light mesons, we focus on ground-state transitions between pseudoscalar and vector mesons in the neutral channels. 

For the charm sector, we consider
\begin{align}
   {D^{*0} D^0 \gamma}: ~ g = 3.439, ~ \Gamma=19.732~\mathrm{keV},~ \mathrm{(Exp)}
\end{align}
with the neutral $D$ and $D^*$ meson masses, $m_{D^0}=1864.84$ and $m_{D^{*0}}=2006.85$ MeV.
Because the total width of the $D^{*0}$ meson is not experimentally known and only an upper bound exists~\cite{ParticleDataGroup:2024cfk}, 
we estimate it from the $D^{*+}$ width~\cite{Gubler:2015qok}.

For the bottom sector, there are no available experimental or lattice QCD data for the M1 decay. Accordingly, we determine the coupling constants using the M1 decay width estimated from a quark model (QM)~\cite{Choi:2007se}:
\begin{align}
    {B^{*0} B^0 \gamma}: & \quad g = 4.070, \quad \Gamma=0.13~\mathrm{keV}, \quad \mathrm{(QM)}
\end{align}  
with the neutral $B^0$ meson masses, $m_B=5279.63$ and $m_{B^*}=5324.75$ MeV~\cite{ParticleDataGroup:2024cfk}. 

\paragraph{Heavy-quarkonium sector}

For the charmonium sector, we consider the lowest four S-wave states below the $D\bar{D}$ threshold: $\eta_c,J/\psi,\eta_c^\prime$, and $\psi^\prime$.
The coupling constants are
\begin{align}
  {\psi \eta_c \gamma}:& \quad  g= 1.912, \quad \Gamma=1.306~\mathrm{keV}, \quad \mathrm{(Exp)} \\
  {\psi^\prime \eta_c \gamma}:& \quad g= 0.137,  \quad \Gamma=1.055~\mathrm{keV}, \quad \mathrm{(Exp)}\\
  {\psi^\prime\eta_c^\prime \gamma}:& \quad g = 2.810,  \quad \Gamma=0.158~\mathrm{keV}, \quad \mathrm{(Exp)}\\
  {\eta_c^\prime \psi \gamma}:& \quad g=0.441,
  \quad \Gamma=15.70~\mathrm{keV}, \quad \mathrm{(Lat)}
\end{align}
with the charmonium masses, $m_{\eta_c}=2984.1$, $m_{J/\psi}=3096.9$, $m_{\eta_c^\prime}=3637.8$, and $m_{\psi^\prime}=3686.097$ MeV~\cite{ParticleDataGroup:2024cfk}.
Here, the couplings are slightly different from those estimated in Ref.~\cite{Yoshida:2016xgm} because of the updated experimental values.
For the M1 decay width of $\eta_c^\prime \to J/\psi \gamma$, we used the lattice QCD data in Ref.~\cite{Becirevic:2014rda}.

For the bottomonium sector, transitions involving states up to the $3S$ level are, in principle, accessible since they lie below the $B\bar{B}$ threshold. However, the $\eta_b^{\prime\prime}$ state has not yet been experimentally observed. We therefore restrict our analysis to transitions involving states up to the $2S$ level. Owing to the limited experimental information available in this sector, the M1 decay widths except $\Upsilon^\prime \to \eta_b \gamma$ are determined from quark-model estimates~\cite{Ridwan:2024ngc}
\begin{align}
 {\Upsilon\eta_b \gamma}:&\quad  g= 1.208, \quad \Gamma=\phantom{0}9.29~\mathrm{eV},\quad \mathrm{(QM)} \\
{\eta_b^\prime \Upsilon \gamma}:&\quad  g= 0.048, \quad \Gamma= 25.30~\mathrm{eV},\quad \mathrm{(QM)} \\
 {\Upsilon^\prime \eta_b \gamma}:&\quad  g= 0.055, \quad \Gamma= 17.60~\mathrm{eV},\quad \mathrm{(Exp)} \\
 {\Upsilon^\prime \eta_b^\prime \gamma}:&\quad  g= 1.274, \quad \Gamma=\phantom{0}0.57~\mathrm{eV},\quad \mathrm{(QM)}
\end{align} 
with bottomonium masses, $m_{\eta_b}=9398.7$, $m_{\Upsilon}=9460.4$, $m_{\eta_b^\prime}=9999$, and $m_{\Upsilon^\prime}=10023.4$ MeV~\cite{ParticleDataGroup:2024cfk}. 

A clear hierarchy is observed in both sectors: intra-level transitions (e.g., $nS \leftrightarrow nS$) are significantly larger than inter-level ones ($nS \leftrightarrow mS$, $n\neq m$). This hierarchy becomes more pronounced in the bottomonium sector due to the larger heavy-quark mass.
As a result, the intra-level mixing dominates the coupled-channel dynamics, and inter-level mixing provides only subleading corrections. 
This feature underlies the robustness of the spectral structures, such as the location of EPs, discussed in this work.

\subsection{EPs in various hadronic spectra}

\begin{figure*}[t]
    \centering
    \includegraphics[width=1.85\columnwidth]{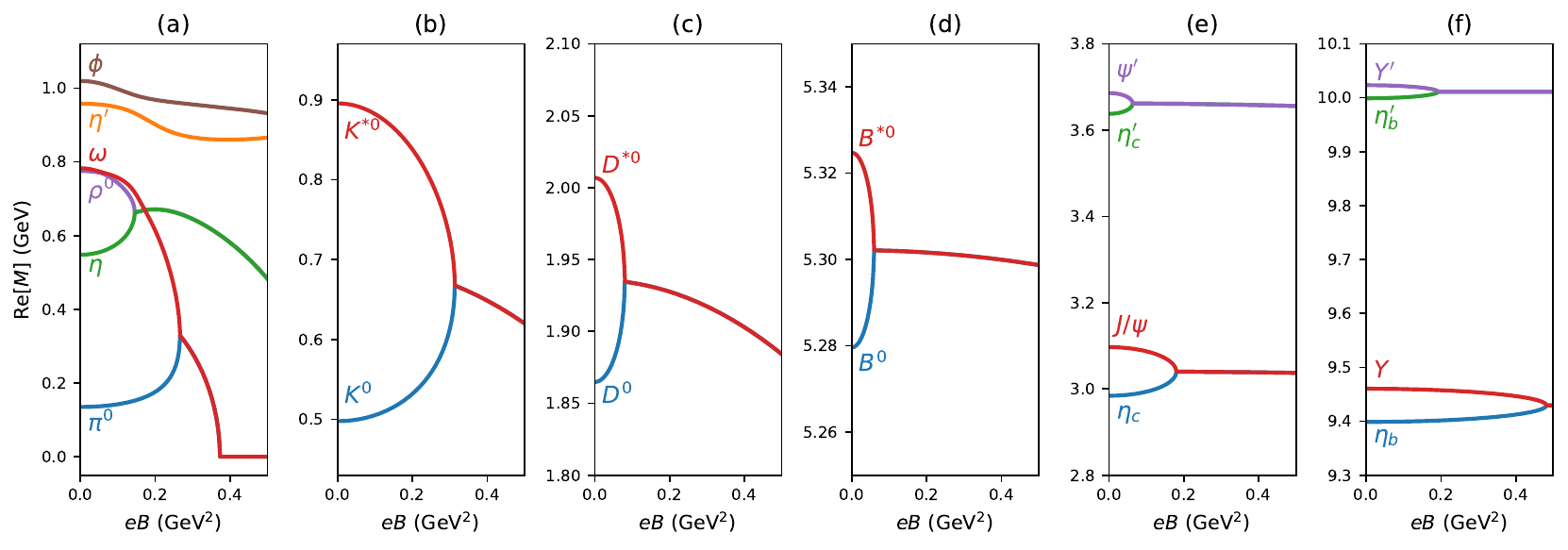}
\caption{Real parts of eigenvalues spectra of (a) light-light mesons started from $\pi^0$-$\eta$-$\rho^0$-$\omega$-$\eta^\prime$-$\phi$, (b) strange-light mesons started from $K^0$-$K^{\ast0}$, (c,d) heavy-light mesons started from $D^0$-$D^{\ast0}$ and $B^0$-$B^{\ast0}$, and (e,f) heavy quarkonia started from $\eta_c$-$J/\psi$-$\eta_c^\prime$-$\psi^\prime$ and $\eta_b$-$\Upsilon$-$\eta_b^\prime$-$\Upsilon^\prime$ in imaginary magnetic fields, computed from the effective Lagrangian approach.
The positions of the EPs reflect the mass splittings and M1 transition couplings between pseudoscalar and vector mesons.}
    \label{fig:various-EP}
\end{figure*}

The emergence of EPs arises generally in mesons with different quark flavor contents. In Fig.~\ref{fig:various-EP}, we summarize the locations of EPs for pseudoscalar-vector meson systems across light, heavy-light, and heavy quarkonium sectors. 

A clear systematic trend can be observed. The position of the EP is primarily controlled by the mass splitting between the pseudoscalar and vector states, as well as the corresponding M1 transition strength. 
In the light unflavored meson sector shown in Fig.~\hyperref[fig:charm-strong]{\ref*{fig:various-EP}(a)}, coupled-channel effects are significant because the six states lie relatively close in mass. We find that EPs emerge in the $\eta$--$\rho$ and $\pi$--$\omega$ systems due to their relatively large coupling constants, where the corresponding EPs are located at $eB_I \approx 0.15~\mathrm{GeV}^2$ and $eB_I \approx 0.27~\mathrm{GeV}^2$, respectively.

For the remaining meson systems, coupled-channel effects are less significant because the pseudoscalar and vector states are well separated in mass. Nevertheless, EPs persist and exhibit a systematic flavor dependence. For mesons with unequal quark masses shown in Fig.~\hyperref[fig:charm-strong]{\ref*{fig:various-EP}(b)-(d)}, the EP position decreases with increasing heavier-quark mass, moving from $eB_I \approx 0.31~\mathrm{GeV}^2$ for the kaon system to $0.08~\mathrm{GeV}^2$ and $0.06~\mathrm{GeV}^2$ for the $D$- and $B$-meson systems, respectively. In contrast, the opposite trend is observed in heavy quarkonia shown in Fig.~\hyperref[fig:charm-strong]{\ref*{fig:various-EP}(e)-(f)}, where the EP position increases from $eB_I \approx 0.18~\mathrm{GeV}^2$ in the charmonium sector to $0.48~\mathrm{GeV}^2$ in the bottomonium sector. Similar trend is also found for the excited quarkonium states.
This behavior reflects the competition between the pseudoscalar-vector mass splitting and the M1 transition strength: stronger coupling constants in heavy-light systems favor EP formation at lower fields.
For heavy quarkonia, the smaller coupling constant (despite the smaller mass splitting) in bottomonia requires stronger fields to generate sufficient mixing and EPs. 

Overall, Fig.~\ref{fig:various-EP} highlights a unified picture: the location and nature of EPs are governed by a small set of physical inputs, namely mass splittings and transition couplings, while their existence is insensitive to the specific hadronic system. 
This systematic trend suggests that similar non-Hermitian spectral structures may arise broadly in QCD under external conditions.

\subsection{Strong-field regime in effective Lagrangian approach}

In the main text, using the constituent quark model, we have predicted that mesons are  dissociated, and quarks are deconfined by strong imaginary magnetic fields.
Nevertheless, it is instructive to formally extend the effective Lagrangian approach to the strong-field regime, where hadronic degrees of freedom remain well-defined within the model. 
There are several interesting observations: (i) the significant role of additional channels, (ii) the avoided crossing in the complex eigenvalues, and (iii) the appearance of a second EP. 

\begin{figure}[t!]
    \centering
    \includegraphics[width=0.9\columnwidth]{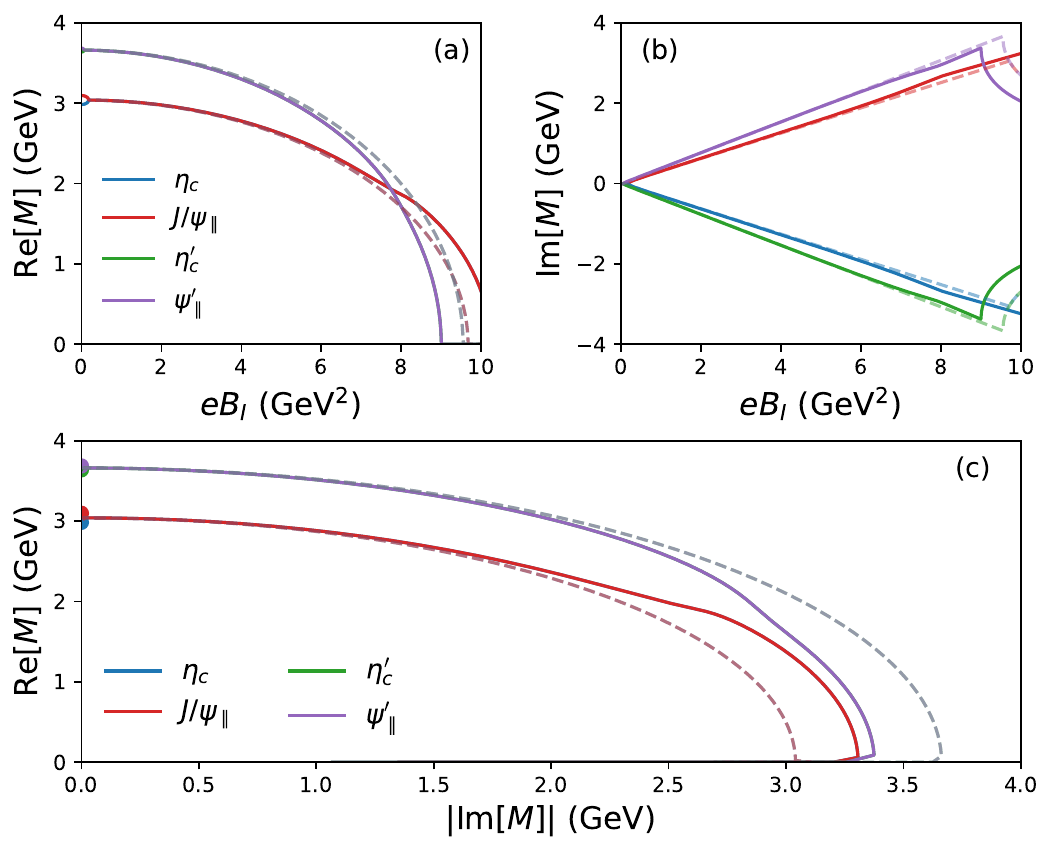}
\caption{Eigenvalues of the longitudinal charmonium states up to $eB_I=10~\mathrm{GeV}^2$, computed from the effective Lagrangian approach (see Fig.~2 for the smaller $eB_I$ regime).
While there is a level crossing in the real parts (a) of the eigenvalues, the trajectories on the complex eigenvalue plane (c) exhibit avoided crossings.
Dashed lines: the results from the two-coupled-channel model.
Solid lines: the four-coupled-channel model.
The impact of the latter model becomes significant in large fields.}
    \label{fig:charm-strong}
\end{figure}

To this end, we first investigate the charmonium spectrum using both a four-coupled-channel model and a two-coupled-channel model in order to examine the impact of including additional channels.
Figure~\hyperref[fig:charm-strong]{\ref*{fig:charm-strong}(a)} shows that noticeable difference between the two models emerge in larger magnetic fields around $eB > 6~\mathrm{GeV}^2$.

Due to the coupling between the ground and excited charmonium states, the four-coupled-channel model exhibits an apparent level crossing in the real part of the mass near $eB \approx 8~\mathrm{GeV}^2$, shifted from $eB \approx 9~\mathrm{GeV}^2$ in the two-coupled-channel model. In addition, the imaginary part of the mass shows an attractive behavior around $eB \approx 8~\mathrm{GeV}^2$ compared to the two-coupled-channel case as depicted in Fig.~\hyperref[fig:charm-strong]{\ref*{fig:charm-strong}(b)}.
In the full complex spectrum beyond EPs shown in Fig.~\hyperref[fig:charm-strong]{\ref*{fig:charm-strong}(c)}, the eigenvalues do not intersect; instead, they undergo an avoided crossing, exhibiting level repulsion in the complex energy plane. Such behavior is a characteristic feature of non-Hermitian dynamics, where the coupling between states prevents the occurrence of a true degeneracy of the complex mass eigenvalues.

As the field is further increased, the real parts of eigenvalues approach zero and develop a second EP, which originates from the coalescence of the positive and negative energy eigenvalues inherent to the Klein-Gordon equation.
The second EP for the two-channel system is located at $eB_I^{\mathrm{EP}} = {(m_{J/\psi} + m_{\eta_c})^2}/{2g_{J/\psi\eta_c\gamma}}\approx 9.7$ GeV$^2$.
Numerically, we are able to trace only one of the two branches in the imaginary part as shown in Fig.~\hyperref[fig:charm-strong]{\ref*{fig:charm-strong}(b)}. Nevertheless, the overall spectral evolution remains consistent with the non-Hermitian nature of the system.
While the physical interpretation of this regime should be treated with caution because it lies beyond the expected domain of confinement, the analysis provides valuable insight into the mathematical structure of non-Hermitian phenomena in hadronic systems.

\bibliography{references}

\begin{thebibliography}{200}%
\makeatletter
\providecommand \@ifxundefined [1]{%
 \@ifx{#1\undefined}
}%
\providecommand \@ifnum [1]{%
 \ifnum #1\expandafter \@firstoftwo
 \else \expandafter \@secondoftwo
 \fi
}%
\providecommand \@ifx [1]{%
 \ifx #1\expandafter \@firstoftwo
 \else \expandafter \@secondoftwo
 \fi
}%
\providecommand \natexlab [1]{#1}%
\providecommand \enquote  [1]{``#1''}%
\providecommand \bibnamefont  [1]{#1}%
\providecommand \bibfnamefont [1]{#1}%
\providecommand \citenamefont [1]{#1}%
\providecommand \href@noop [0]{\@secondoftwo}%
\providecommand \href [0]{\begingroup \@sanitize@url \@href}%
\providecommand \@href[1]{\@@startlink{#1}\@@href}%
\providecommand \@@href[1]{\endgroup#1\@@endlink}%
\providecommand \@sanitize@url [0]{\catcode `\\12\catcode `\$12\catcode `\&12\catcode `\#12\catcode `\^12\catcode `\_12\catcode `\%12\relax}%
\providecommand \@@startlink[1]{}%
\providecommand \@@endlink[0]{}%
\providecommand \url  [0]{\begingroup\@sanitize@url \@url }%
\providecommand \@url [1]{\endgroup\@href {#1}{\urlprefix }}%
\providecommand \urlprefix  [0]{URL }%
\providecommand \Eprint [0]{\href }%
\providecommand \doibase [0]{https://doi.org/}%
\providecommand \selectlanguage [0]{\@gobble}%
\providecommand \bibinfo  [0]{\@secondoftwo}%
\providecommand \bibfield  [0]{\@secondoftwo}%
\providecommand \translation [1]{[#1]}%
\providecommand \BibitemOpen [0]{}%
\providecommand \bibitemStop [0]{}%
\providecommand \bibitemNoStop [0]{.\EOS\space}%
\providecommand \EOS [0]{\spacefactor3000\relax}%
\providecommand \BibitemShut  [1]{\csname bibitem#1\endcsname}%
\let\auto@bib@innerbib\@empty
\bibitem [{\citenamefont {Bender}(2007)}]{Bender:2007nj}%
  \BibitemOpen
  \bibfield  {author} {\bibinfo {author} {\bibfnamefont {C.~M.}\ \bibnamefont {Bender}},\ }\bibfield  {title} {\bibinfo {title} {{Making sense of non-Hermitian Hamiltonians}},\ }\href {https://doi.org/10.1088/0034-4885/70/6/R03} {\bibfield  {journal} {\bibinfo  {journal} {Rept. Prog. Phys.}\ }\textbf {\bibinfo {volume} {70}},\ \bibinfo {pages} {947} (\bibinfo {year} {2007})},\ \Eprint {https://arxiv.org/abs/hep-th/0703096} {arXiv:hep-th/0703096} \BibitemShut {NoStop}%
\bibitem [{\citenamefont {El-Ganainy}\ \emph {et~al.}(2018)\citenamefont {El-Ganainy}, \citenamefont {Makris}, \citenamefont {Khajavikhan}, \citenamefont {Musslimani}, \citenamefont {Rotter},\ and\ \citenamefont {Christodoulides}}]{El-Ganainy:2018ksn}%
  \BibitemOpen
  \bibfield  {author} {\bibinfo {author} {\bibfnamefont {R.}~\bibnamefont {El-Ganainy}}, \bibinfo {author} {\bibfnamefont {K.~G.}\ \bibnamefont {Makris}}, \bibinfo {author} {\bibfnamefont {M.}~\bibnamefont {Khajavikhan}}, \bibinfo {author} {\bibfnamefont {Z.~H.}\ \bibnamefont {Musslimani}}, \bibinfo {author} {\bibfnamefont {S.}~\bibnamefont {Rotter}},\ and\ \bibinfo {author} {\bibfnamefont {D.~N.}\ \bibnamefont {Christodoulides}},\ }\bibfield  {title} {\bibinfo {title} {{Non-Hermitian physics and PT symmetry}},\ }\href {https://doi.org/10.1038/nphys4323} {\bibfield  {journal} {\bibinfo  {journal} {Nature Phys.}\ }\textbf {\bibinfo {volume} {14}},\ \bibinfo {pages} {11} (\bibinfo {year} {2018})}\BibitemShut {NoStop}%
\bibitem [{\citenamefont {Bergholtz}\ \emph {et~al.}(2021)\citenamefont {Bergholtz}, \citenamefont {Budich},\ and\ \citenamefont {Kunst}}]{Bergholtz:2019deh}%
  \BibitemOpen
  \bibfield  {author} {\bibinfo {author} {\bibfnamefont {E.~J.}\ \bibnamefont {Bergholtz}}, \bibinfo {author} {\bibfnamefont {J.~C.}\ \bibnamefont {Budich}},\ and\ \bibinfo {author} {\bibfnamefont {F.~K.}\ \bibnamefont {Kunst}},\ }\bibfield  {title} {\bibinfo {title} {{Exceptional topology of non-Hermitian systems}},\ }\href {https://doi.org/10.1103/revmodphys.93.015005} {\bibfield  {journal} {\bibinfo  {journal} {Rev. Mod. Phys.}\ }\textbf {\bibinfo {volume} {93}},\ \bibinfo {pages} {015005} (\bibinfo {year} {2021})},\ \Eprint {https://arxiv.org/abs/1912.10048} {arXiv:1912.10048 [cond-mat.mes-hall]} \BibitemShut {NoStop}%
\bibitem [{\citenamefont {Ashida}\ \emph {et~al.}(2021)\citenamefont {Ashida}, \citenamefont {Gong},\ and\ \citenamefont {Ueda}}]{Ashida:2020dkc}%
  \BibitemOpen
  \bibfield  {author} {\bibinfo {author} {\bibfnamefont {Y.}~\bibnamefont {Ashida}}, \bibinfo {author} {\bibfnamefont {Z.}~\bibnamefont {Gong}},\ and\ \bibinfo {author} {\bibfnamefont {M.}~\bibnamefont {Ueda}},\ }\bibfield  {title} {\bibinfo {title} {{Non-Hermitian physics}},\ }\href {https://doi.org/10.1080/00018732.2021.1876991} {\bibfield  {journal} {\bibinfo  {journal} {Adv. Phys.}\ }\textbf {\bibinfo {volume} {69}},\ \bibinfo {pages} {249} (\bibinfo {year} {2021})},\ \Eprint {https://arxiv.org/abs/2006.01837} {arXiv:2006.01837 [cond-mat.mes-hall]} \BibitemShut {NoStop}%
\bibitem [{\citenamefont {Ding}\ \emph {et~al.}(2022{\natexlab{a}})\citenamefont {Ding}, \citenamefont {Fang},\ and\ \citenamefont {Ma}}]{Ding:2022juv}%
  \BibitemOpen
  \bibfield  {author} {\bibinfo {author} {\bibfnamefont {K.}~\bibnamefont {Ding}}, \bibinfo {author} {\bibfnamefont {C.}~\bibnamefont {Fang}},\ and\ \bibinfo {author} {\bibfnamefont {G.}~\bibnamefont {Ma}},\ }\bibfield  {title} {\bibinfo {title} {{Non-Hermitian topology and exceptional-point geometries}},\ }\href {https://doi.org/10.1038/s42254-022-00516-5} {\bibfield  {journal} {\bibinfo  {journal} {Nature Rev. Phys.}\ }\textbf {\bibinfo {volume} {4}},\ \bibinfo {pages} {745} (\bibinfo {year} {2022}{\natexlab{a}})},\ \Eprint {https://arxiv.org/abs/2204.11601} {arXiv:2204.11601 [quant-ph]} \BibitemShut {NoStop}%
\bibitem [{\citenamefont {Kato}(1966)}]{Kato1966}%
  \BibitemOpen
  \bibfield  {author} {\bibinfo {author} {\bibfnamefont {T.}~\bibnamefont {Kato}},\ }\href {https://doi.org/10.1007/978-3-662-12678-3} {\emph {\bibinfo {title} {Perturbation Theory for Linear Operators}}}\ (\bibinfo  {publisher} {Springer},\ \bibinfo {year} {1966})\BibitemShut {NoStop}%
\bibitem [{\citenamefont {Dembowski}\ \emph {et~al.}(2001)\citenamefont {Dembowski}, \citenamefont {Graf}, \citenamefont {Harney}, \citenamefont {Heine}, \citenamefont {Heiss}, \citenamefont {Rehfeld},\ and\ \citenamefont {Richter}}]{Dembowski:2001zz}%
  \BibitemOpen
  \bibfield  {author} {\bibinfo {author} {\bibfnamefont {C.}~\bibnamefont {Dembowski}}, \bibinfo {author} {\bibfnamefont {H.~D.}\ \bibnamefont {Graf}}, \bibinfo {author} {\bibfnamefont {H.~L.}\ \bibnamefont {Harney}}, \bibinfo {author} {\bibfnamefont {A.}~\bibnamefont {Heine}}, \bibinfo {author} {\bibfnamefont {W.~D.}\ \bibnamefont {Heiss}}, \bibinfo {author} {\bibfnamefont {H.}~\bibnamefont {Rehfeld}},\ and\ \bibinfo {author} {\bibfnamefont {A.}~\bibnamefont {Richter}},\ }\bibfield  {title} {\bibinfo {title} {{Experimental observation of the topological structure of exceptional points}},\ }\href {https://doi.org/10.1103/PhysRevLett.86.787} {\bibfield  {journal} {\bibinfo  {journal} {Phys. Rev. Lett.}\ }\textbf {\bibinfo {volume} {86}},\ \bibinfo {pages} {787} (\bibinfo {year} {2001})}\BibitemShut {NoStop}%
\bibitem [{\citenamefont {Guo}\ \emph {et~al.}(2009)\citenamefont {Guo}, \citenamefont {Salamo}, \citenamefont {Duchesne}, \citenamefont {Morandotti}, \citenamefont {Volatier-Ravat}, \citenamefont {Aimez}, \citenamefont {Siviloglou},\ and\ \citenamefont {Christodoulides}}]{Guo:2009yqd}%
  \BibitemOpen
  \bibfield  {author} {\bibinfo {author} {\bibfnamefont {A.}~\bibnamefont {Guo}}, \bibinfo {author} {\bibfnamefont {G.~J.}\ \bibnamefont {Salamo}}, \bibinfo {author} {\bibfnamefont {D.}~\bibnamefont {Duchesne}}, \bibinfo {author} {\bibfnamefont {R.}~\bibnamefont {Morandotti}}, \bibinfo {author} {\bibfnamefont {M.}~\bibnamefont {Volatier-Ravat}}, \bibinfo {author} {\bibfnamefont {V.}~\bibnamefont {Aimez}}, \bibinfo {author} {\bibfnamefont {G.~A.}\ \bibnamefont {Siviloglou}},\ and\ \bibinfo {author} {\bibfnamefont {D.~N.}\ \bibnamefont {Christodoulides}},\ }\bibfield  {title} {\bibinfo {title} {{Observation of $\mathcal{P}\mathcal{T}$-symmetry breaking in complex optical potentials}},\ }\href {https://doi.org/10.1103/PhysRevLett.103.093902} {\bibfield  {journal} {\bibinfo  {journal} {Phys. Rev. Lett.}\ }\textbf {\bibinfo {volume} {103}},\ \bibinfo {pages} {093902} (\bibinfo {year} {2009})}\BibitemShut {NoStop}%
\bibitem [{\citenamefont {R{\"u}ter}\ \emph {et~al.}(2010)\citenamefont {R{\"u}ter}, \citenamefont {Makris}, \citenamefont {El-Ganainy}, \citenamefont {Christodoulides}, \citenamefont {Segev},\ and\ \citenamefont {Kip}}]{Ruter:2010qjb}%
  \BibitemOpen
  \bibfield  {author} {\bibinfo {author} {\bibfnamefont {C.~E.}\ \bibnamefont {R{\"u}ter}}, \bibinfo {author} {\bibfnamefont {K.~G.}\ \bibnamefont {Makris}}, \bibinfo {author} {\bibfnamefont {R.}~\bibnamefont {El-Ganainy}}, \bibinfo {author} {\bibfnamefont {D.~N.}\ \bibnamefont {Christodoulides}}, \bibinfo {author} {\bibfnamefont {M.}~\bibnamefont {Segev}},\ and\ \bibinfo {author} {\bibfnamefont {D.}~\bibnamefont {Kip}},\ }\bibfield  {title} {\bibinfo {title} {{Observation of parity{\textendash}time symmetry in optics}},\ }\href {https://doi.org/10.1038/nphys1515} {\bibfield  {journal} {\bibinfo  {journal} {Nature Phys.}\ }\textbf {\bibinfo {volume} {6}},\ \bibinfo {pages} {192} (\bibinfo {year} {2010})}\BibitemShut {NoStop}%
\bibitem [{\citenamefont {Gamow}(1928)}]{Gamow:1928zz}%
  \BibitemOpen
  \bibfield  {author} {\bibinfo {author} {\bibfnamefont {G.}~\bibnamefont {Gamow}},\ }\bibfield  {title} {\bibinfo {title} {{Zur Quantentheorie des Atomkernes}},\ }\href {https://doi.org/10.1007/BF01343196} {\bibfield  {journal} {\bibinfo  {journal} {Z. Phys.}\ }\textbf {\bibinfo {volume} {51}},\ \bibinfo {pages} {204} (\bibinfo {year} {1928})}\BibitemShut {NoStop}%
\bibitem [{\citenamefont {Breit}\ and\ \citenamefont {Wigner}(1936)}]{Breit:1936zzb}%
  \BibitemOpen
  \bibfield  {author} {\bibinfo {author} {\bibfnamefont {G.}~\bibnamefont {Breit}}\ and\ \bibinfo {author} {\bibfnamefont {E.}~\bibnamefont {Wigner}},\ }\bibfield  {title} {\bibinfo {title} {{Capture of Slow Neutrons}},\ }\href {https://doi.org/10.1103/PhysRev.49.519} {\bibfield  {journal} {\bibinfo  {journal} {Phys. Rev.}\ }\textbf {\bibinfo {volume} {49}},\ \bibinfo {pages} {519} (\bibinfo {year} {1936})}\BibitemShut {NoStop}%
\bibitem [{\citenamefont {Siegert}(1939)}]{Siegert:1939zz}%
  \BibitemOpen
  \bibfield  {author} {\bibinfo {author} {\bibfnamefont {A.~J.~F.}\ \bibnamefont {Siegert}},\ }\bibfield  {title} {\bibinfo {title} {{On the Derivation of the Dispersion Formula for Nuclear Reactions}},\ }\href {https://doi.org/10.1103/PhysRev.56.750} {\bibfield  {journal} {\bibinfo  {journal} {Phys. Rev.}\ }\textbf {\bibinfo {volume} {56}},\ \bibinfo {pages} {750} (\bibinfo {year} {1939})}\BibitemShut {NoStop}%
\bibitem [{\citenamefont {Fano}(1961)}]{Fano:1961zz}%
  \BibitemOpen
  \bibfield  {author} {\bibinfo {author} {\bibfnamefont {U.}~\bibnamefont {Fano}},\ }\bibfield  {title} {\bibinfo {title} {{Effects of Configuration Interaction on Intensities and Phase Shifts}},\ }\href {https://doi.org/10.1103/PhysRev.124.1866} {\bibfield  {journal} {\bibinfo  {journal} {Phys. Rev.}\ }\textbf {\bibinfo {volume} {124}},\ \bibinfo {pages} {1866} (\bibinfo {year} {1961})}\BibitemShut {NoStop}%
\bibitem [{\citenamefont {Feshbach}(1958)}]{Feshbach:1958nx}%
  \BibitemOpen
  \bibfield  {author} {\bibinfo {author} {\bibfnamefont {H.}~\bibnamefont {Feshbach}},\ }\bibfield  {title} {\bibinfo {title} {{Unified theory of nuclear reactions}},\ }\href {https://doi.org/10.1016/0003-4916(58)90007-1} {\bibfield  {journal} {\bibinfo  {journal} {Annals Phys.}\ }\textbf {\bibinfo {volume} {5}},\ \bibinfo {pages} {357} (\bibinfo {year} {1958})}\BibitemShut {NoStop}%
\bibitem [{\citenamefont {Aguilar}\ and\ \citenamefont {Combes}(1971)}]{Aguilar:1971ve}%
  \BibitemOpen
  \bibfield  {author} {\bibinfo {author} {\bibfnamefont {J.}~\bibnamefont {Aguilar}}\ and\ \bibinfo {author} {\bibfnamefont {J.~M.}\ \bibnamefont {Combes}},\ }\bibfield  {title} {\bibinfo {title} {{A class of analytic perturbations for one-body schroedinger hamiltonians}},\ }\href {https://doi.org/10.1007/BF01877510} {\bibfield  {journal} {\bibinfo  {journal} {Commun. Math. Phys.}\ }\textbf {\bibinfo {volume} {22}},\ \bibinfo {pages} {269} (\bibinfo {year} {1971})}\BibitemShut {NoStop}%
\bibitem [{\citenamefont {Ho}(1983)}]{Ho:1983lwa}%
  \BibitemOpen
  \bibfield  {author} {\bibinfo {author} {\bibfnamefont {Y.~K.}\ \bibnamefont {Ho}},\ }\bibfield  {title} {\bibinfo {title} {{The method of complex coordinate rotation and its applications to atomic collision processes}},\ }\href {https://doi.org/10.1016/0370-1573(83)90112-6} {\bibfield  {journal} {\bibinfo  {journal} {Phys. Rept.}\ }\textbf {\bibinfo {volume} {99}},\ \bibinfo {pages} {1} (\bibinfo {year} {1983})}\BibitemShut {NoStop}%
\bibitem [{\citenamefont {Moiseyev}(1998)}]{Moiseyev:1998gjp}%
  \BibitemOpen
  \bibfield  {author} {\bibinfo {author} {\bibfnamefont {N.}~\bibnamefont {Moiseyev}},\ }\bibfield  {title} {\bibinfo {title} {{Quantum theory of resonances: calculating energies, widths and cross-sections by complex scaling}},\ }\href {https://doi.org/10.1016/S0370-1573(98)00002-7} {\bibfield  {journal} {\bibinfo  {journal} {Phys. Rept.}\ }\textbf {\bibinfo {volume} {302}},\ \bibinfo {pages} {212} (\bibinfo {year} {1998})}\BibitemShut {NoStop}%
\bibitem [{\citenamefont {Nawa}\ \emph {et~al.}(2013)\citenamefont {Nawa}, \citenamefont {Ozaki}, \citenamefont {Nagahiro}, \citenamefont {Jido},\ and\ \citenamefont {Hosaka}}]{Nawa:2011pz}%
  \BibitemOpen
  \bibfield  {author} {\bibinfo {author} {\bibfnamefont {K.}~\bibnamefont {Nawa}}, \bibinfo {author} {\bibfnamefont {S.}~\bibnamefont {Ozaki}}, \bibinfo {author} {\bibfnamefont {H.}~\bibnamefont {Nagahiro}}, \bibinfo {author} {\bibfnamefont {D.}~\bibnamefont {Jido}},\ and\ \bibinfo {author} {\bibfnamefont {A.}~\bibnamefont {Hosaka}},\ }\bibfield  {title} {\bibinfo {title} {{Complex 2D matrix model and geometrical map on the complex-$N_{c}$ plane}},\ }\href {https://doi.org/10.1093/ptep/ptt051} {\bibfield  {journal} {\bibinfo  {journal} {PTEP}\ }\textbf {\bibinfo {volume} {2013}},\ \bibinfo {pages} {083D01} (\bibinfo {year} {2013})},\ \Eprint {https://arxiv.org/abs/1109.0426} {arXiv:1109.0426 [hep-ph]} \BibitemShut {NoStop}%
\bibitem [{Note1()}]{Note1}%
  \BibitemOpen
  \bibinfo {note} {Other related discussions of exceptional points and hadron physics can be found in Ref.~\cite {Ghodrati:2025fah}.}\BibitemShut {Stop}%
\bibitem [{\citenamefont {Bali}\ \emph {et~al.}(2012)\citenamefont {Bali}, \citenamefont {Bruckmann}, \citenamefont {Endr\ifmmode~\mbox{\H{o}}\else \H{o}\fi{}di}, \citenamefont {Fodor}, \citenamefont {Katz}, \citenamefont {Krieg}, \citenamefont {Sch{\"a}fer},\ and\ \citenamefont {Szab{\'o}}}]{Bali:2011qj}%
  \BibitemOpen
  \bibfield  {author} {\bibinfo {author} {\bibfnamefont {G.~S.}\ \bibnamefont {Bali}}, \bibinfo {author} {\bibfnamefont {F.}~\bibnamefont {Bruckmann}}, \bibinfo {author} {\bibfnamefont {G.}~\bibnamefont {Endr\ifmmode~\mbox{\H{o}}\else \H{o}\fi{}di}}, \bibinfo {author} {\bibfnamefont {Z.}~\bibnamefont {Fodor}}, \bibinfo {author} {\bibfnamefont {S.~D.}\ \bibnamefont {Katz}}, \bibinfo {author} {\bibfnamefont {S.}~\bibnamefont {Krieg}}, \bibinfo {author} {\bibfnamefont {A.}~\bibnamefont {Sch{\"a}fer}},\ and\ \bibinfo {author} {\bibfnamefont {K.~K.}\ \bibnamefont {Szab{\'o}}},\ }\bibfield  {title} {\bibinfo {title} {{The QCD phase diagram for external magnetic fields}},\ }\href {https://doi.org/10.1007/JHEP02(2012)044} {\bibfield  {journal} {\bibinfo  {journal} {JHEP}\ }\textbf {\bibinfo {volume} {02}},\ \bibinfo {pages} {044}},\ \Eprint {https://arxiv.org/abs/1111.4956} {arXiv:1111.4956 [hep-lat]} \BibitemShut {NoStop}%
\bibitem [{\citenamefont {Luschevskaya}\ and\ \citenamefont {Larina}(2014)}]{Luschevskaya:2012xd}%
  \BibitemOpen
  \bibfield  {author} {\bibinfo {author} {\bibfnamefont {E.~V.}\ \bibnamefont {Luschevskaya}}\ and\ \bibinfo {author} {\bibfnamefont {O.~V.}\ \bibnamefont {Larina}},\ }\bibfield  {title} {\bibinfo {title} {{The $\rho$ and $A$ mesons in a strong abelian magnetic field in $SU(2)$ lattice gauge theory}},\ }\href {https://doi.org/10.1016/j.nuclphysb.2014.04.003} {\bibfield  {journal} {\bibinfo  {journal} {Nucl. Phys. B}\ }\textbf {\bibinfo {volume} {884}},\ \bibinfo {pages} {1} (\bibinfo {year} {2014})},\ \Eprint {https://arxiv.org/abs/1203.5699} {arXiv:1203.5699 [hep-lat]} \BibitemShut {NoStop}%
\bibitem [{\citenamefont {Hidaka}\ and\ \citenamefont {Yamamoto}(2013)}]{Hidaka:2012mz}%
  \BibitemOpen
  \bibfield  {author} {\bibinfo {author} {\bibfnamefont {Y.}~\bibnamefont {Hidaka}}\ and\ \bibinfo {author} {\bibfnamefont {A.}~\bibnamefont {Yamamoto}},\ }\bibfield  {title} {\bibinfo {title} {{Charged vector mesons in a strong magnetic field}},\ }\href {https://doi.org/10.1103/PhysRevD.87.094502} {\bibfield  {journal} {\bibinfo  {journal} {Phys. Rev. D}\ }\textbf {\bibinfo {volume} {87}},\ \bibinfo {pages} {094502} (\bibinfo {year} {2013})},\ \Eprint {https://arxiv.org/abs/1209.0007} {arXiv:1209.0007 [hep-ph]} \BibitemShut {NoStop}%
\bibitem [{\citenamefont {Luschevskaya}\ \emph {et~al.}(2015)\citenamefont {Luschevskaya}, \citenamefont {Solovjeva}, \citenamefont {Kochetkov},\ and\ \citenamefont {Teryaev}}]{Luschevskaya:2014lga}%
  \BibitemOpen
  \bibfield  {author} {\bibinfo {author} {\bibfnamefont {E.~V.}\ \bibnamefont {Luschevskaya}}, \bibinfo {author} {\bibfnamefont {O.~E.}\ \bibnamefont {Solovjeva}}, \bibinfo {author} {\bibfnamefont {O.~A.}\ \bibnamefont {Kochetkov}},\ and\ \bibinfo {author} {\bibfnamefont {O.~V.}\ \bibnamefont {Teryaev}},\ }\bibfield  {title} {\bibinfo {title} {{Magnetic polarizabilities of light mesons in $SU(3)$ lattice gauge theory}},\ }\href {https://doi.org/10.1016/j.nuclphysb.2015.07.023} {\bibfield  {journal} {\bibinfo  {journal} {Nucl. Phys. B}\ }\textbf {\bibinfo {volume} {898}},\ \bibinfo {pages} {627} (\bibinfo {year} {2015})},\ \Eprint {https://arxiv.org/abs/1411.4284} {arXiv:1411.4284 [hep-lat]} \BibitemShut {NoStop}%
\bibitem [{\citenamefont {Bonati}\ \emph {et~al.}(2014)\citenamefont {Bonati}, \citenamefont {D'Elia}, \citenamefont {Mariti}, \citenamefont {Mesiti}, \citenamefont {Negro},\ and\ \citenamefont {Sanfilippo}}]{Bonati:2014ksa}%
  \BibitemOpen
  \bibfield  {author} {\bibinfo {author} {\bibfnamefont {C.}~\bibnamefont {Bonati}}, \bibinfo {author} {\bibfnamefont {M.}~\bibnamefont {D'Elia}}, \bibinfo {author} {\bibfnamefont {M.}~\bibnamefont {Mariti}}, \bibinfo {author} {\bibfnamefont {M.}~\bibnamefont {Mesiti}}, \bibinfo {author} {\bibfnamefont {F.}~\bibnamefont {Negro}},\ and\ \bibinfo {author} {\bibfnamefont {F.}~\bibnamefont {Sanfilippo}},\ }\bibfield  {title} {\bibinfo {title} {{Anisotropy of the quark-antiquark potential in a magnetic field}},\ }\href {https://doi.org/10.1103/PhysRevD.89.114502} {\bibfield  {journal} {\bibinfo  {journal} {Phys. Rev. D}\ }\textbf {\bibinfo {volume} {89}},\ \bibinfo {pages} {114502} (\bibinfo {year} {2014})},\ \Eprint {https://arxiv.org/abs/1403.6094} {arXiv:1403.6094 [hep-lat]} \BibitemShut {NoStop}%
\bibitem [{\citenamefont {Luschevskaya}\ \emph {et~al.}(2016)\citenamefont {Luschevskaya}, \citenamefont {Solovjeva},\ and\ \citenamefont {Teryaev}}]{Luschevskaya:2015cko}%
  \BibitemOpen
  \bibfield  {author} {\bibinfo {author} {\bibfnamefont {E.~V.}\ \bibnamefont {Luschevskaya}}, \bibinfo {author} {\bibfnamefont {O.~E.}\ \bibnamefont {Solovjeva}},\ and\ \bibinfo {author} {\bibfnamefont {O.~V.}\ \bibnamefont {Teryaev}},\ }\bibfield  {title} {\bibinfo {title} {{Magnetic polarizability of pion}},\ }\href {https://doi.org/10.1016/j.physletb.2016.08.054} {\bibfield  {journal} {\bibinfo  {journal} {Phys. Lett. B}\ }\textbf {\bibinfo {volume} {761}},\ \bibinfo {pages} {393} (\bibinfo {year} {2016})},\ \Eprint {https://arxiv.org/abs/1511.09316} {arXiv:1511.09316 [hep-lat]} \BibitemShut {NoStop}%
\bibitem [{\citenamefont {Bonati}\ \emph {et~al.}(2016)\citenamefont {Bonati}, \citenamefont {D'Elia}, \citenamefont {Mariti}, \citenamefont {Mesiti}, \citenamefont {Negro}, \citenamefont {Rucci},\ and\ \citenamefont {Sanfilippo}}]{Bonati:2016kxj}%
  \BibitemOpen
  \bibfield  {author} {\bibinfo {author} {\bibfnamefont {C.}~\bibnamefont {Bonati}}, \bibinfo {author} {\bibfnamefont {M.}~\bibnamefont {D'Elia}}, \bibinfo {author} {\bibfnamefont {M.}~\bibnamefont {Mariti}}, \bibinfo {author} {\bibfnamefont {M.}~\bibnamefont {Mesiti}}, \bibinfo {author} {\bibfnamefont {F.}~\bibnamefont {Negro}}, \bibinfo {author} {\bibfnamefont {A.}~\bibnamefont {Rucci}},\ and\ \bibinfo {author} {\bibfnamefont {F.}~\bibnamefont {Sanfilippo}},\ }\bibfield  {title} {\bibinfo {title} {{Magnetic field effects on the static quark potential at zero and finite temperature}},\ }\href {https://doi.org/10.1103/PhysRevD.94.094007} {\bibfield  {journal} {\bibinfo  {journal} {Phys. Rev. D}\ }\textbf {\bibinfo {volume} {94}},\ \bibinfo {pages} {094007} (\bibinfo {year} {2016})},\ \Eprint {https://arxiv.org/abs/1607.08160} {arXiv:1607.08160 [hep-lat]} \BibitemShut {NoStop}%
\bibitem [{\citenamefont {Bali}\ \emph {et~al.}(2018{\natexlab{a}})\citenamefont {Bali}, \citenamefont {Brandt}, \citenamefont {Endr{\H{o}}di},\ and\ \citenamefont {Gl{\"a}{\ss}le}}]{Bali:2017ian}%
  \BibitemOpen
  \bibfield  {author} {\bibinfo {author} {\bibfnamefont {G.~S.}\ \bibnamefont {Bali}}, \bibinfo {author} {\bibfnamefont {B.~B.}\ \bibnamefont {Brandt}}, \bibinfo {author} {\bibfnamefont {G.}~\bibnamefont {Endr{\H{o}}di}},\ and\ \bibinfo {author} {\bibfnamefont {B.}~\bibnamefont {Gl{\"a}{\ss}le}},\ }\bibfield  {title} {\bibinfo {title} {{Meson masses in electromagnetic fields with Wilson fermions}},\ }\href {https://doi.org/10.1103/PhysRevD.97.034505} {\bibfield  {journal} {\bibinfo  {journal} {Phys. Rev. D}\ }\textbf {\bibinfo {volume} {97}},\ \bibinfo {pages} {034505} (\bibinfo {year} {2018}{\natexlab{a}})},\ \Eprint {https://arxiv.org/abs/1707.05600} {arXiv:1707.05600 [hep-lat]} \BibitemShut {NoStop}%
\bibitem [{\citenamefont {Bali}\ \emph {et~al.}(2018{\natexlab{b}})\citenamefont {Bali}, \citenamefont {Brandt}, \citenamefont {Endr{\H{o}}di},\ and\ \citenamefont {Gl{\"a}{\ss}le}}]{Bali:2018sey}%
  \BibitemOpen
  \bibfield  {author} {\bibinfo {author} {\bibfnamefont {G.~S.}\ \bibnamefont {Bali}}, \bibinfo {author} {\bibfnamefont {B.~B.}\ \bibnamefont {Brandt}}, \bibinfo {author} {\bibfnamefont {G.}~\bibnamefont {Endr{\H{o}}di}},\ and\ \bibinfo {author} {\bibfnamefont {B.}~\bibnamefont {Gl{\"a}{\ss}le}},\ }\bibfield  {title} {\bibinfo {title} {{Weak decay of magnetized pions}},\ }\href {https://doi.org/10.1103/PhysRevLett.121.072001} {\bibfield  {journal} {\bibinfo  {journal} {Phys. Rev. Lett.}\ }\textbf {\bibinfo {volume} {121}},\ \bibinfo {pages} {072001} (\bibinfo {year} {2018}{\natexlab{b}})},\ \Eprint {https://arxiv.org/abs/1805.10971} {arXiv:1805.10971 [hep-lat]} \BibitemShut {NoStop}%
\bibitem [{\citenamefont {Luschevskaya}\ \emph {et~al.}(2018)\citenamefont {Luschevskaya}, \citenamefont {Teryaev}, \citenamefont {Golubkov}, \citenamefont {Solovjeva},\ and\ \citenamefont {Ishkuvatov}}]{Luschevskaya:2018chr}%
  \BibitemOpen
  \bibfield  {author} {\bibinfo {author} {\bibfnamefont {E.~V.}\ \bibnamefont {Luschevskaya}}, \bibinfo {author} {\bibfnamefont {O.~V.}\ \bibnamefont {Teryaev}}, \bibinfo {author} {\bibfnamefont {D.~Y.}\ \bibnamefont {Golubkov}}, \bibinfo {author} {\bibfnamefont {O.~V.}\ \bibnamefont {Solovjeva}},\ and\ \bibinfo {author} {\bibfnamefont {R.~A.}\ \bibnamefont {Ishkuvatov}},\ }\bibfield  {title} {\bibinfo {title} {{Tensor polarizability of the vector mesons from $SU(3)$ lattice gauge theory}},\ }\href {https://doi.org/10.1007/JHEP11(2018)186} {\bibfield  {journal} {\bibinfo  {journal} {JHEP}\ }\textbf {\bibinfo {volume} {11}},\ \bibinfo {pages} {186}},\ \Eprint {https://arxiv.org/abs/1811.02344} {arXiv:1811.02344 [hep-lat]} \BibitemShut {NoStop}%
\bibitem [{\citenamefont {Hattori}\ and\ \citenamefont {Yamamoto}(2019)}]{Hattori:2019ijy}%
  \BibitemOpen
  \bibfield  {author} {\bibinfo {author} {\bibfnamefont {K.}~\bibnamefont {Hattori}}\ and\ \bibinfo {author} {\bibfnamefont {A.}~\bibnamefont {Yamamoto}},\ }\bibfield  {title} {\bibinfo {title} {{Meson deformation by magnetic fields in lattice QCD}},\ }\href {https://doi.org/10.1093/ptep/ptz023} {\bibfield  {journal} {\bibinfo  {journal} {PTEP}\ }\textbf {\bibinfo {volume} {2019}},\ \bibinfo {pages} {043B04} (\bibinfo {year} {2019})},\ \Eprint {https://arxiv.org/abs/1901.10182} {arXiv:1901.10182 [hep-lat]} \BibitemShut {NoStop}%
\bibitem [{\citenamefont {Endr{\H{o}}di}\ and\ \citenamefont {Mark{\'o}}(2019)}]{Endrodi:2019whh}%
  \BibitemOpen
  \bibfield  {author} {\bibinfo {author} {\bibfnamefont {G.}~\bibnamefont {Endr{\H{o}}di}}\ and\ \bibinfo {author} {\bibfnamefont {G.}~\bibnamefont {Mark{\'o}}},\ }\bibfield  {title} {\bibinfo {title} {{Magnetized baryons and the QCD phase diagram: NJL model meets the lattice}},\ }\href {https://doi.org/10.1007/JHEP08(2019)036} {\bibfield  {journal} {\bibinfo  {journal} {JHEP}\ }\textbf {\bibinfo {volume} {08}},\ \bibinfo {pages} {036}},\ \Eprint {https://arxiv.org/abs/1905.02103} {arXiv:1905.02103 [hep-lat]} \BibitemShut {NoStop}%
\bibitem [{\citenamefont {Bignell}\ \emph {et~al.}(2019)\citenamefont {Bignell}, \citenamefont {Kamleh},\ and\ \citenamefont {Leinweber}}]{Bignell:2019vpy}%
  \BibitemOpen
  \bibfield  {author} {\bibinfo {author} {\bibfnamefont {R.}~\bibnamefont {Bignell}}, \bibinfo {author} {\bibfnamefont {W.}~\bibnamefont {Kamleh}},\ and\ \bibinfo {author} {\bibfnamefont {D.}~\bibnamefont {Leinweber}},\ }\bibfield  {title} {\bibinfo {title} {{Pion in a uniform background magnetic field with clover fermions}},\ }\href {https://doi.org/10.1103/PhysRevD.100.114518} {\bibfield  {journal} {\bibinfo  {journal} {Phys. Rev. D}\ }\textbf {\bibinfo {volume} {100}},\ \bibinfo {pages} {114518} (\bibinfo {year} {2019})},\ \Eprint {https://arxiv.org/abs/1910.14244} {arXiv:1910.14244 [hep-lat]} \BibitemShut {NoStop}%
\bibitem [{\citenamefont {Bignell}\ \emph {et~al.}(2020)\citenamefont {Bignell}, \citenamefont {Kamleh},\ and\ \citenamefont {Leinweber}}]{Bignell:2020dze}%
  \BibitemOpen
  \bibfield  {author} {\bibinfo {author} {\bibfnamefont {R.}~\bibnamefont {Bignell}}, \bibinfo {author} {\bibfnamefont {W.}~\bibnamefont {Kamleh}},\ and\ \bibinfo {author} {\bibfnamefont {D.}~\bibnamefont {Leinweber}},\ }\bibfield  {title} {\bibinfo {title} {{Pion magnetic polarisability using the background field method}},\ }\href {https://doi.org/10.1016/j.physletb.2020.135853} {\bibfield  {journal} {\bibinfo  {journal} {Phys. Lett. B}\ }\textbf {\bibinfo {volume} {811}},\ \bibinfo {pages} {135853} (\bibinfo {year} {2020})},\ \Eprint {https://arxiv.org/abs/2005.10453} {arXiv:2005.10453 [hep-lat]} \BibitemShut {NoStop}%
\bibitem [{\citenamefont {Ding}\ \emph {et~al.}(2021)\citenamefont {Ding}, \citenamefont {Li}, \citenamefont {Tomiya}, \citenamefont {Wang},\ and\ \citenamefont {Zhang}}]{Ding:2020hxw}%
  \BibitemOpen
  \bibfield  {author} {\bibinfo {author} {\bibfnamefont {H.-T.}\ \bibnamefont {Ding}}, \bibinfo {author} {\bibfnamefont {S.-T.}\ \bibnamefont {Li}}, \bibinfo {author} {\bibfnamefont {A.}~\bibnamefont {Tomiya}}, \bibinfo {author} {\bibfnamefont {X.-D.}\ \bibnamefont {Wang}},\ and\ \bibinfo {author} {\bibfnamefont {Y.}~\bibnamefont {Zhang}},\ }\bibfield  {title} {\bibinfo {title} {{Chiral properties of (2+1)-flavor QCD in strong magnetic fields at zero temperature}},\ }\href {https://doi.org/10.1103/PhysRevD.104.014505} {\bibfield  {journal} {\bibinfo  {journal} {Phys. Rev. D}\ }\textbf {\bibinfo {volume} {104}},\ \bibinfo {pages} {014505} (\bibinfo {year} {2021})},\ \Eprint {https://arxiv.org/abs/2008.00493} {arXiv:2008.00493 [hep-lat]} \BibitemShut {NoStop}%
\bibitem [{\citenamefont {D'Elia}\ \emph {et~al.}(2021)\citenamefont {D'Elia}, \citenamefont {Maio}, \citenamefont {Sanfilippo},\ and\ \citenamefont {Stanzione}}]{DElia:2021tfb}%
  \BibitemOpen
  \bibfield  {author} {\bibinfo {author} {\bibfnamefont {M.}~\bibnamefont {D'Elia}}, \bibinfo {author} {\bibfnamefont {L.}~\bibnamefont {Maio}}, \bibinfo {author} {\bibfnamefont {F.}~\bibnamefont {Sanfilippo}},\ and\ \bibinfo {author} {\bibfnamefont {A.}~\bibnamefont {Stanzione}},\ }\bibfield  {title} {\bibinfo {title} {{Confining and chiral properties of QCD in extremely strong magnetic fields}},\ }\href {https://doi.org/10.1103/PhysRevD.104.114512} {\bibfield  {journal} {\bibinfo  {journal} {Phys. Rev. D}\ }\textbf {\bibinfo {volume} {104}},\ \bibinfo {pages} {114512} (\bibinfo {year} {2021})},\ \Eprint {https://arxiv.org/abs/2109.07456} {arXiv:2109.07456 [hep-lat]} \BibitemShut {NoStop}%
\bibitem [{\citenamefont {Ding}\ \emph {et~al.}(2022{\natexlab{b}})\citenamefont {Ding}, \citenamefont {Li}, \citenamefont {Liu},\ and\ \citenamefont {Wang}}]{Ding:2022tqn}%
  \BibitemOpen
  \bibfield  {author} {\bibinfo {author} {\bibfnamefont {H.-T.}\ \bibnamefont {Ding}}, \bibinfo {author} {\bibfnamefont {S.-T.}\ \bibnamefont {Li}}, \bibinfo {author} {\bibfnamefont {J.-H.}\ \bibnamefont {Liu}},\ and\ \bibinfo {author} {\bibfnamefont {X.-D.}\ \bibnamefont {Wang}},\ }\bibfield  {title} {\bibinfo {title} {{Chiral condensates and screening masses of neutral pseudoscalar mesons in thermomagnetic QCD medium}},\ }\href {https://doi.org/10.1103/PhysRevD.105.034514} {\bibfield  {journal} {\bibinfo  {journal} {Phys. Rev. D}\ }\textbf {\bibinfo {volume} {105}},\ \bibinfo {pages} {034514} (\bibinfo {year} {2022}{\natexlab{b}})},\ \Eprint {https://arxiv.org/abs/2201.02349} {arXiv:2201.02349 [hep-lat]} \BibitemShut {NoStop}%
\bibitem [{\citenamefont {Ding}\ \emph {et~al.}(2025)\citenamefont {Ding}, \citenamefont {Gu}, \citenamefont {Li},\ and\ \citenamefont {Thakkar}}]{Ding:2025pbu}%
  \BibitemOpen
  \bibfield  {author} {\bibinfo {author} {\bibfnamefont {H.-T.}\ \bibnamefont {Ding}}, \bibinfo {author} {\bibfnamefont {J.-B.}\ \bibnamefont {Gu}}, \bibinfo {author} {\bibfnamefont {S.-T.}\ \bibnamefont {Li}},\ and\ \bibinfo {author} {\bibfnamefont {R.}~\bibnamefont {Thakkar}},\ }\bibfield  {title} {\bibinfo {title} {{Chiral condensates and screening masses of neutral pseudoscalar mesons from lattice QCD at physical quark masses}},\ }\href {https://doi.org/10.1103/PhysRevD.111.074513} {\bibfield  {journal} {\bibinfo  {journal} {Phys. Rev. D}\ }\textbf {\bibinfo {volume} {111}},\ \bibinfo {pages} {074513} (\bibinfo {year} {2025})},\ \Eprint {https://arxiv.org/abs/2501.11262} {arXiv:2501.11262 [hep-lat]} \BibitemShut {NoStop}%
\bibitem [{\citenamefont {Ding}\ and\ \citenamefont {Zhang}(2026)}]{Ding:2026qzu}%
  \BibitemOpen
  \bibfield  {author} {\bibinfo {author} {\bibfnamefont {H.-T.}\ \bibnamefont {Ding}}\ and\ \bibinfo {author} {\bibfnamefont {D.}~\bibnamefont {Zhang}},\ }\bibfield  {title} {\bibinfo {title} {{Chiral properties of (2+1)-flavor QCD in magnetic fields at zero temperature}},\ }\href {https://doi.org/10.1103/qpxq-xjqc} {\bibfield  {journal} {\bibinfo  {journal} {Phys. Rev. D}\ }\textbf {\bibinfo {volume} {113}},\ \bibinfo {pages} {094503} (\bibinfo {year} {2026})},\ \Eprint {https://arxiv.org/abs/2601.18354} {arXiv:2601.18354 [hep-lat]} \BibitemShut {NoStop}%
\bibitem [{\citenamefont {Endr\ifmmode~\mbox{\H{o}}\else \H{o}\fi{}di}(2025)}]{Endrodi:2024cqn}%
  \BibitemOpen
  \bibfield  {author} {\bibinfo {author} {\bibfnamefont {G.}~\bibnamefont {Endr\ifmmode~\mbox{\H{o}}\else \H{o}\fi{}di}},\ }\bibfield  {title} {\bibinfo {title} {{QCD with background electromagnetic fields on the lattice: A review}},\ }\href {https://doi.org/10.1016/j.ppnp.2024.104153} {\bibfield  {journal} {\bibinfo  {journal} {Prog. Part. Nucl. Phys.}\ }\textbf {\bibinfo {volume} {141}},\ \bibinfo {pages} {104153} (\bibinfo {year} {2025})},\ \Eprint {https://arxiv.org/abs/2406.19780} {arXiv:2406.19780 [hep-lat]} \BibitemShut {NoStop}%
\bibitem [{\citenamefont {Fiebig}\ \emph {et~al.}(1989)\citenamefont {Fiebig}, \citenamefont {Wilcox},\ and\ \citenamefont {Woloshyn}}]{Fiebig:1988en}%
  \BibitemOpen
  \bibfield  {author} {\bibinfo {author} {\bibfnamefont {H.~R.}\ \bibnamefont {Fiebig}}, \bibinfo {author} {\bibfnamefont {W.}~\bibnamefont {Wilcox}},\ and\ \bibinfo {author} {\bibfnamefont {R.~M.}\ \bibnamefont {Woloshyn}},\ }\bibfield  {title} {\bibinfo {title} {{A study of hadron electric polarizability in quenched lattice {QCD}}},\ }\href {https://doi.org/10.1016/0550-3213(89)90180-6} {\bibfield  {journal} {\bibinfo  {journal} {Nucl. Phys. B}\ }\textbf {\bibinfo {volume} {324}},\ \bibinfo {pages} {47} (\bibinfo {year} {1989})}\BibitemShut {NoStop}%
\bibitem [{\citenamefont {Christensen}\ \emph {et~al.}(2005)\citenamefont {Christensen}, \citenamefont {Wilcox}, \citenamefont {Lee},\ and\ \citenamefont {Zhou}}]{Christensen:2004ca}%
  \BibitemOpen
  \bibfield  {author} {\bibinfo {author} {\bibfnamefont {J.~C.}\ \bibnamefont {Christensen}}, \bibinfo {author} {\bibfnamefont {W.}~\bibnamefont {Wilcox}}, \bibinfo {author} {\bibfnamefont {F.~X.}\ \bibnamefont {Lee}},\ and\ \bibinfo {author} {\bibfnamefont {L.-m.}\ \bibnamefont {Zhou}},\ }\bibfield  {title} {\bibinfo {title} {{Electric polarizability of neutral hadrons from lattice QCD}},\ }\href {https://doi.org/10.1103/PhysRevD.72.034503} {\bibfield  {journal} {\bibinfo  {journal} {Phys. Rev. D}\ }\textbf {\bibinfo {volume} {72}},\ \bibinfo {pages} {034503} (\bibinfo {year} {2005})},\ \Eprint {https://arxiv.org/abs/hep-lat/0408024} {arXiv:hep-lat/0408024} \BibitemShut {NoStop}%
\bibitem [{\citenamefont {Engelhardt}(2007)}]{Engelhardt:2007ub}%
  \BibitemOpen
  \bibfield  {author} {\bibinfo {author} {\bibfnamefont {M.}~\bibnamefont {Engelhardt}} (\bibinfo {collaboration} {LHPC Collaboration}),\ }\bibfield  {title} {\bibinfo {title} {{Neutron electric polarizability from unquenched lattice QCD using the background field approach}},\ }\href {https://doi.org/10.1103/PhysRevD.76.114502} {\bibfield  {journal} {\bibinfo  {journal} {Phys. Rev. D}\ }\textbf {\bibinfo {volume} {76}},\ \bibinfo {pages} {114502} (\bibinfo {year} {2007})},\ \Eprint {https://arxiv.org/abs/0706.3919} {arXiv:0706.3919 [hep-lat]} \BibitemShut {NoStop}%
\bibitem [{\citenamefont {Detmold}\ \emph {et~al.}(2009)\citenamefont {Detmold}, \citenamefont {Tiburzi},\ and\ \citenamefont {Walker-Loud}}]{Detmold:2009dx}%
  \BibitemOpen
  \bibfield  {author} {\bibinfo {author} {\bibfnamefont {W.}~\bibnamefont {Detmold}}, \bibinfo {author} {\bibfnamefont {B.~C.}\ \bibnamefont {Tiburzi}},\ and\ \bibinfo {author} {\bibfnamefont {A.}~\bibnamefont {Walker-Loud}},\ }\bibfield  {title} {\bibinfo {title} {{Extracting Electric Polarizabilities from Lattice QCD}},\ }\href {https://doi.org/10.1103/PhysRevD.79.094505} {\bibfield  {journal} {\bibinfo  {journal} {Phys. Rev. D}\ }\textbf {\bibinfo {volume} {79}},\ \bibinfo {pages} {094505} (\bibinfo {year} {2009})},\ \Eprint {https://arxiv.org/abs/0904.1586} {arXiv:0904.1586 [hep-lat]} \BibitemShut {NoStop}%
\bibitem [{\citenamefont {Detmold}\ \emph {et~al.}(2010)\citenamefont {Detmold}, \citenamefont {Tiburzi},\ and\ \citenamefont {Walker-Loud}}]{Detmold:2010ts}%
  \BibitemOpen
  \bibfield  {author} {\bibinfo {author} {\bibfnamefont {W.}~\bibnamefont {Detmold}}, \bibinfo {author} {\bibfnamefont {B.~C.}\ \bibnamefont {Tiburzi}},\ and\ \bibinfo {author} {\bibfnamefont {A.}~\bibnamefont {Walker-Loud}},\ }\bibfield  {title} {\bibinfo {title} {{Extracting Nucleon Magnetic Moments and Electric Polarizabilities from Lattice QCD in Background Electric Fields}},\ }\href {https://doi.org/10.1103/PhysRevD.81.054502} {\bibfield  {journal} {\bibinfo  {journal} {Phys. Rev. D}\ }\textbf {\bibinfo {volume} {81}},\ \bibinfo {pages} {054502} (\bibinfo {year} {2010})},\ \Eprint {https://arxiv.org/abs/1001.1131} {arXiv:1001.1131 [hep-lat]} \BibitemShut {NoStop}%
\bibitem [{\citenamefont {D'Elia}\ \emph {et~al.}(2013)\citenamefont {D'Elia}, \citenamefont {Mariti},\ and\ \citenamefont {Negro}}]{DElia:2012ifm}%
  \BibitemOpen
  \bibfield  {author} {\bibinfo {author} {\bibfnamefont {M.}~\bibnamefont {D'Elia}}, \bibinfo {author} {\bibfnamefont {M.}~\bibnamefont {Mariti}},\ and\ \bibinfo {author} {\bibfnamefont {F.}~\bibnamefont {Negro}},\ }\bibfield  {title} {\bibinfo {title} {{Susceptibility of the QCD vacuum to CP-odd electromagnetic background fields}},\ }\href {https://doi.org/10.1103/PhysRevLett.110.082002} {\bibfield  {journal} {\bibinfo  {journal} {Phys. Rev. Lett.}\ }\textbf {\bibinfo {volume} {110}},\ \bibinfo {pages} {082002} (\bibinfo {year} {2013})},\ \Eprint {https://arxiv.org/abs/1209.0722} {arXiv:1209.0722 [hep-lat]} \BibitemShut {NoStop}%
\bibitem [{\citenamefont {Lujan}\ \emph {et~al.}(2014)\citenamefont {Lujan}, \citenamefont {Alexandru}, \citenamefont {Freeman},\ and\ \citenamefont {Lee}}]{Lujan:2014kia}%
  \BibitemOpen
  \bibfield  {author} {\bibinfo {author} {\bibfnamefont {M.}~\bibnamefont {Lujan}}, \bibinfo {author} {\bibfnamefont {A.}~\bibnamefont {Alexandru}}, \bibinfo {author} {\bibfnamefont {W.}~\bibnamefont {Freeman}},\ and\ \bibinfo {author} {\bibfnamefont {F.}~\bibnamefont {Lee}},\ }\bibfield  {title} {\bibinfo {title} {{Electric polarizability of neutral hadrons from dynamical lattice QCD ensembles}},\ }\href {https://doi.org/10.1103/PhysRevD.89.074506} {\bibfield  {journal} {\bibinfo  {journal} {Phys. Rev. D}\ }\textbf {\bibinfo {volume} {89}},\ \bibinfo {pages} {074506} (\bibinfo {year} {2014})},\ \Eprint {https://arxiv.org/abs/1402.3025} {arXiv:1402.3025 [hep-lat]} \BibitemShut {NoStop}%
\bibitem [{\citenamefont {Freeman}\ \emph {et~al.}(2014)\citenamefont {Freeman}, \citenamefont {Alexandru}, \citenamefont {Lujan},\ and\ \citenamefont {Lee}}]{Freeman:2014kka}%
  \BibitemOpen
  \bibfield  {author} {\bibinfo {author} {\bibfnamefont {W.}~\bibnamefont {Freeman}}, \bibinfo {author} {\bibfnamefont {A.}~\bibnamefont {Alexandru}}, \bibinfo {author} {\bibfnamefont {M.}~\bibnamefont {Lujan}},\ and\ \bibinfo {author} {\bibfnamefont {F.~X.}\ \bibnamefont {Lee}},\ }\bibfield  {title} {\bibinfo {title} {{Sea quark contributions to the electric polarizability of hadrons}},\ }\href {https://doi.org/10.1103/PhysRevD.90.054507} {\bibfield  {journal} {\bibinfo  {journal} {Phys. Rev. D}\ }\textbf {\bibinfo {volume} {90}},\ \bibinfo {pages} {054507} (\bibinfo {year} {2014})},\ \Eprint {https://arxiv.org/abs/1407.2687} {arXiv:1407.2687 [hep-lat]} \BibitemShut {NoStop}%
\bibitem [{\citenamefont {Lujan}\ \emph {et~al.}(2016)\citenamefont {Lujan}, \citenamefont {Alexandru}, \citenamefont {Freeman},\ and\ \citenamefont {Lee}}]{Lujan:2016ffj}%
  \BibitemOpen
  \bibfield  {author} {\bibinfo {author} {\bibfnamefont {M.}~\bibnamefont {Lujan}}, \bibinfo {author} {\bibfnamefont {A.}~\bibnamefont {Alexandru}}, \bibinfo {author} {\bibfnamefont {W.}~\bibnamefont {Freeman}},\ and\ \bibinfo {author} {\bibfnamefont {F.~X.}\ \bibnamefont {Lee}},\ }\bibfield  {title} {\bibinfo {title} {{Finite volume effects on the electric polarizability of neutral hadrons in lattice QCD}},\ }\href {https://doi.org/10.1103/PhysRevD.94.074506} {\bibfield  {journal} {\bibinfo  {journal} {Phys. Rev. D}\ }\textbf {\bibinfo {volume} {94}},\ \bibinfo {pages} {074506} (\bibinfo {year} {2016})},\ \Eprint {https://arxiv.org/abs/1606.07928} {arXiv:1606.07928 [hep-lat]} \BibitemShut {NoStop}%
\bibitem [{\citenamefont {Niyazi}\ \emph {et~al.}(2021)\citenamefont {Niyazi}, \citenamefont {Alexandru}, \citenamefont {Lee},\ and\ \citenamefont {Lujan}}]{Niyazi:2021jrz}%
  \BibitemOpen
  \bibfield  {author} {\bibinfo {author} {\bibfnamefont {H.}~\bibnamefont {Niyazi}}, \bibinfo {author} {\bibfnamefont {A.}~\bibnamefont {Alexandru}}, \bibinfo {author} {\bibfnamefont {F.~X.}\ \bibnamefont {Lee}},\ and\ \bibinfo {author} {\bibfnamefont {M.}~\bibnamefont {Lujan}},\ }\bibfield  {title} {\bibinfo {title} {{Charged pion electric polarizability from lattice QCD}},\ }\href {https://doi.org/10.1103/PhysRevD.104.014510} {\bibfield  {journal} {\bibinfo  {journal} {Phys. Rev. D}\ }\textbf {\bibinfo {volume} {104}},\ \bibinfo {pages} {014510} (\bibinfo {year} {2021})},\ \Eprint {https://arxiv.org/abs/2105.06906} {arXiv:2105.06906 [hep-lat]} \BibitemShut {NoStop}%
\bibitem [{\citenamefont {Yang}\ \emph {et~al.}(2022)\citenamefont {Yang}, \citenamefont {Chang},\ and\ \citenamefont {Chen}}]{Yang:2022zob}%
  \BibitemOpen
  \bibfield  {author} {\bibinfo {author} {\bibfnamefont {J.-C.}\ \bibnamefont {Yang}}, \bibinfo {author} {\bibfnamefont {X.-T.}\ \bibnamefont {Chang}},\ and\ \bibinfo {author} {\bibfnamefont {J.-X.}\ \bibnamefont {Chen}},\ }\bibfield  {title} {\bibinfo {title} {{Study of the Roberge-Weiss phase caused by external uniform classical electric field using lattice QCD approach}},\ }\href {https://doi.org/10.1007/JHEP10(2022)053} {\bibfield  {journal} {\bibinfo  {journal} {JHEP}\ }\textbf {\bibinfo {volume} {10}},\ \bibinfo {pages} {053}},\ \Eprint {https://arxiv.org/abs/2207.11796} {arXiv:2207.11796 [hep-lat]} \BibitemShut {NoStop}%
\bibitem [{\citenamefont {Endr\ifmmode~\mbox{\H{o}}\else \H{o}\fi{}di}\ and\ \citenamefont {Mark\'o}(2024)}]{Endrodi:2023wwf}%
  \BibitemOpen
  \bibfield  {author} {\bibinfo {author} {\bibfnamefont {G.}~\bibnamefont {Endr\ifmmode~\mbox{\H{o}}\else \H{o}\fi{}di}}\ and\ \bibinfo {author} {\bibfnamefont {G.}~\bibnamefont {Mark\'o}},\ }\bibfield  {title} {\bibinfo {title} {{QCD phase diagram and equation of state in background electric fields}},\ }\href {https://doi.org/10.1103/PhysRevD.109.034506} {\bibfield  {journal} {\bibinfo  {journal} {Phys. Rev. D}\ }\textbf {\bibinfo {volume} {109}},\ \bibinfo {pages} {034506} (\bibinfo {year} {2024})},\ \Eprint {https://arxiv.org/abs/2309.07058} {arXiv:2309.07058 [hep-lat]} \BibitemShut {NoStop}%
\bibitem [{\citenamefont {Yang}\ \emph {et~al.}(2026)\citenamefont {Yang}, \citenamefont {Zhao}, \citenamefont {Li},\ and\ \citenamefont {Li}}]{Yang:2026tzt}%
  \BibitemOpen
  \bibfield  {author} {\bibinfo {author} {\bibfnamefont {J.-C.}\ \bibnamefont {Yang}}, \bibinfo {author} {\bibfnamefont {Z.}~\bibnamefont {Zhao}}, \bibinfo {author} {\bibfnamefont {X.-N.}\ \bibnamefont {Li}},\ and\ \bibinfo {author} {\bibfnamefont {W.-W.}\ \bibnamefont {Li}},\ }\bibfield  {title} {\bibinfo {title} {{Mesonic screening correlators in an external imaginary electric field at finite temperature}},\ }\href@noop {} {\  (\bibinfo {year} {2026})},\ \Eprint {https://arxiv.org/abs/2603.27620} {arXiv:2603.27620 [hep-lat]} \BibitemShut {NoStop}%
\bibitem [{Note2()}]{Note2}%
  \BibitemOpen
  \bibinfo {note} {As other techniques for real electric fields, there are methods with partially quenched real electric fields~\cite {Shintani:2006xr,Shintani:2008nt} and isospin electric fields~\cite {Yamamoto:2012bd}.}\BibitemShut {Stop}%
\bibitem [{Note3()}]{Note3}%
  \BibitemOpen
  \bibinfo {note} {Imaginary magnetic fields have historically been introduced as an artificial but useful parameter, for example, in the context of Lee-Yang zeros~\cite {Yang:1952be,Lee:1952ig}. More recently, a few phenomenological studies~\cite {Ozawa:2023oqc,Montag:2026cqh} have investigated physical consequences of imaginary magnetic fields in non-Hermitian environments.}\BibitemShut {Stop}%
\bibitem [{\citenamefont {Ba{\~n}uls}\ and\ \citenamefont {Cichy}(2020)}]{Banuls:2019rao}%
  \BibitemOpen
  \bibfield  {author} {\bibinfo {author} {\bibfnamefont {M.~C.}\ \bibnamefont {Ba{\~n}uls}}\ and\ \bibinfo {author} {\bibfnamefont {K.}~\bibnamefont {Cichy}},\ }\bibfield  {title} {\bibinfo {title} {{Review on Novel Methods for Lattice Gauge Theories}},\ }\href {https://doi.org/10.1088/1361-6633/ab6311} {\bibfield  {journal} {\bibinfo  {journal} {Rept. Prog. Phys.}\ }\textbf {\bibinfo {volume} {83}},\ \bibinfo {pages} {024401} (\bibinfo {year} {2020})},\ \Eprint {https://arxiv.org/abs/1910.00257} {arXiv:1910.00257 [hep-lat]} \BibitemShut {NoStop}%
\bibitem [{\citenamefont {Marasinghe}\ and\ \citenamefont {Tuchin}(2011)}]{Marasinghe:2011bt}%
  \BibitemOpen
  \bibfield  {author} {\bibinfo {author} {\bibfnamefont {K.}~\bibnamefont {Marasinghe}}\ and\ \bibinfo {author} {\bibfnamefont {K.}~\bibnamefont {Tuchin}},\ }\bibfield  {title} {\bibinfo {title} {{Quarkonium dissociation in quark-gluon plasma via ionization in magnetic field}},\ }\href {https://doi.org/10.1103/PhysRevC.84.044908} {\bibfield  {journal} {\bibinfo  {journal} {Phys. Rev. C}\ }\textbf {\bibinfo {volume} {84}},\ \bibinfo {pages} {044908} (\bibinfo {year} {2011})},\ \Eprint {https://arxiv.org/abs/1103.1329} {arXiv:1103.1329 [hep-ph]} \BibitemShut {NoStop}%
\bibitem [{\citenamefont {Tuchin}(2011)}]{Tuchin:2011cg}%
  \BibitemOpen
  \bibfield  {author} {\bibinfo {author} {\bibfnamefont {K.}~\bibnamefont {Tuchin}},\ }\bibfield  {title} {\bibinfo {title} {{J/{\ensuremath{\psi}} dissociation in parity-odd bubbles}},\ }\href {https://doi.org/10.1016/j.physletb.2011.10.047} {\bibfield  {journal} {\bibinfo  {journal} {Phys. Lett. B}\ }\textbf {\bibinfo {volume} {705}},\ \bibinfo {pages} {482} (\bibinfo {year} {2011})},\ \Eprint {https://arxiv.org/abs/1105.5360} {arXiv:1105.5360 [nucl-th]} \BibitemShut {NoStop}%
\bibitem [{\citenamefont {Yang}\ and\ \citenamefont {M{\"u}ller}(2012)}]{Yang:2011cz}%
  \BibitemOpen
  \bibfield  {author} {\bibinfo {author} {\bibfnamefont {D.-L.}\ \bibnamefont {Yang}}\ and\ \bibinfo {author} {\bibfnamefont {B.}~\bibnamefont {M{\"u}ller}},\ }\bibfield  {title} {\bibinfo {title} {{$J/\psi$ production by magnetic excitation of $\eta_c$}},\ }\href {https://doi.org/10.1088/0954-3899/39/1/015007} {\bibfield  {journal} {\bibinfo  {journal} {J. Phys. G}\ }\textbf {\bibinfo {volume} {39}},\ \bibinfo {pages} {015007} (\bibinfo {year} {2012})},\ \Eprint {https://arxiv.org/abs/1108.2525} {arXiv:1108.2525 [hep-ph]} \BibitemShut {NoStop}%
\bibitem [{\citenamefont {Tuchin}(2013)}]{Tuchin:2013ie}%
  \BibitemOpen
  \bibfield  {author} {\bibinfo {author} {\bibfnamefont {K.}~\bibnamefont {Tuchin}},\ }\bibfield  {title} {\bibinfo {title} {{Particle production in strong electromagnetic fields in relativistic heavy-ion collisions}},\ }\href {https://doi.org/10.1155/2013/490495} {\bibfield  {journal} {\bibinfo  {journal} {Adv. High Energy Phys.}\ }\textbf {\bibinfo {volume} {2013}},\ \bibinfo {pages} {490495} (\bibinfo {year} {2013})},\ \Eprint {https://arxiv.org/abs/1301.0099} {arXiv:1301.0099 [hep-ph]} \BibitemShut {NoStop}%
\bibitem [{\citenamefont {Machado}\ \emph {et~al.}(2013)\citenamefont {Machado}, \citenamefont {Navarra}, \citenamefont {de~Oliveira}, \citenamefont {Noronha},\ and\ \citenamefont {Strickland}}]{Machado:2013rta}%
  \BibitemOpen
  \bibfield  {author} {\bibinfo {author} {\bibfnamefont {C.~S.}\ \bibnamefont {Machado}}, \bibinfo {author} {\bibfnamefont {F.~S.}\ \bibnamefont {Navarra}}, \bibinfo {author} {\bibfnamefont {E.~G.}\ \bibnamefont {de~Oliveira}}, \bibinfo {author} {\bibfnamefont {J.}~\bibnamefont {Noronha}},\ and\ \bibinfo {author} {\bibfnamefont {M.}~\bibnamefont {Strickland}},\ }\bibfield  {title} {\bibinfo {title} {{Heavy quarkonium production in a strong magnetic field}},\ }\href {https://doi.org/10.1103/PhysRevD.88.034009} {\bibfield  {journal} {\bibinfo  {journal} {Phys. Rev. D}\ }\textbf {\bibinfo {volume} {88}},\ \bibinfo {pages} {034009} (\bibinfo {year} {2013})},\ \Eprint {https://arxiv.org/abs/1305.3308} {arXiv:1305.3308 [hep-ph]} \BibitemShut {NoStop}%
\bibitem [{\citenamefont {Alford}\ and\ \citenamefont {Strickland}(2013)}]{Alford:2013jva}%
  \BibitemOpen
  \bibfield  {author} {\bibinfo {author} {\bibfnamefont {J.}~\bibnamefont {Alford}}\ and\ \bibinfo {author} {\bibfnamefont {M.}~\bibnamefont {Strickland}},\ }\bibfield  {title} {\bibinfo {title} {{Charmonia and bottomonia in a magnetic field}},\ }\href {https://doi.org/10.1103/PhysRevD.88.105017} {\bibfield  {journal} {\bibinfo  {journal} {Phys. Rev. D}\ }\textbf {\bibinfo {volume} {88}},\ \bibinfo {pages} {105017} (\bibinfo {year} {2013})},\ \Eprint {https://arxiv.org/abs/1309.3003} {arXiv:1309.3003 [hep-ph]} \BibitemShut {NoStop}%
\bibitem [{\citenamefont {Cho}\ \emph {et~al.}(2014)\citenamefont {Cho}, \citenamefont {Hattori}, \citenamefont {Lee}, \citenamefont {Morita},\ and\ \citenamefont {Ozaki}}]{Cho:2014exa}%
  \BibitemOpen
  \bibfield  {author} {\bibinfo {author} {\bibfnamefont {S.}~\bibnamefont {Cho}}, \bibinfo {author} {\bibfnamefont {K.}~\bibnamefont {Hattori}}, \bibinfo {author} {\bibfnamefont {S.~H.}\ \bibnamefont {Lee}}, \bibinfo {author} {\bibfnamefont {K.}~\bibnamefont {Morita}},\ and\ \bibinfo {author} {\bibfnamefont {S.}~\bibnamefont {Ozaki}},\ }\bibfield  {title} {\bibinfo {title} {{QCD sum rules for magnetically induced mixing between $\eta_c$ and $J/\psi$}},\ }\href {https://doi.org/10.1103/PhysRevLett.113.172301} {\bibfield  {journal} {\bibinfo  {journal} {Phys. Rev. Lett.}\ }\textbf {\bibinfo {volume} {113}},\ \bibinfo {pages} {172301} (\bibinfo {year} {2014})},\ \Eprint {https://arxiv.org/abs/1406.4586} {arXiv:1406.4586 [hep-ph]} \BibitemShut {NoStop}%
\bibitem [{\citenamefont {Cho}\ \emph {et~al.}(2015)\citenamefont {Cho}, \citenamefont {Hattori}, \citenamefont {Lee}, \citenamefont {Morita},\ and\ \citenamefont {Ozaki}}]{Cho:2014loa}%
  \BibitemOpen
  \bibfield  {author} {\bibinfo {author} {\bibfnamefont {S.}~\bibnamefont {Cho}}, \bibinfo {author} {\bibfnamefont {K.}~\bibnamefont {Hattori}}, \bibinfo {author} {\bibfnamefont {S.~H.}\ \bibnamefont {Lee}}, \bibinfo {author} {\bibfnamefont {K.}~\bibnamefont {Morita}},\ and\ \bibinfo {author} {\bibfnamefont {S.}~\bibnamefont {Ozaki}},\ }\bibfield  {title} {\bibinfo {title} {{Charmonium Spectroscopy in Strong Magnetic Fields by QCD Sum Rules: S-Wave Ground States}},\ }\href {https://doi.org/10.1103/PhysRevD.91.045025} {\bibfield  {journal} {\bibinfo  {journal} {Phys. Rev. D}\ }\textbf {\bibinfo {volume} {91}},\ \bibinfo {pages} {045025} (\bibinfo {year} {2015})},\ \Eprint {https://arxiv.org/abs/1411.7675} {arXiv:1411.7675 [hep-ph]} \BibitemShut {NoStop}%
\bibitem [{\citenamefont {Dudal}\ and\ \citenamefont {Mertens}(2015)}]{Dudal:2014jfa}%
  \BibitemOpen
  \bibfield  {author} {\bibinfo {author} {\bibfnamefont {D.}~\bibnamefont {Dudal}}\ and\ \bibinfo {author} {\bibfnamefont {T.~G.}\ \bibnamefont {Mertens}},\ }\bibfield  {title} {\bibinfo {title} {{Melting of charmonium in a magnetic field from an effective AdS/QCD model}},\ }\href {https://doi.org/10.1103/PhysRevD.91.086002} {\bibfield  {journal} {\bibinfo  {journal} {Phys. Rev. D}\ }\textbf {\bibinfo {volume} {91}},\ \bibinfo {pages} {086002} (\bibinfo {year} {2015})},\ \Eprint {https://arxiv.org/abs/1410.3297} {arXiv:1410.3297 [hep-th]} \BibitemShut {NoStop}%
\bibitem [{\citenamefont {Bonati}\ \emph {et~al.}(2015)\citenamefont {Bonati}, \citenamefont {D'Elia},\ and\ \citenamefont {Rucci}}]{Bonati:2015dka}%
  \BibitemOpen
  \bibfield  {author} {\bibinfo {author} {\bibfnamefont {C.}~\bibnamefont {Bonati}}, \bibinfo {author} {\bibfnamefont {M.}~\bibnamefont {D'Elia}},\ and\ \bibinfo {author} {\bibfnamefont {A.}~\bibnamefont {Rucci}},\ }\bibfield  {title} {\bibinfo {title} {{Heavy quarkonia in strong magnetic fields}},\ }\href {https://doi.org/10.1103/PhysRevD.92.054014} {\bibfield  {journal} {\bibinfo  {journal} {Phys. Rev. D}\ }\textbf {\bibinfo {volume} {92}},\ \bibinfo {pages} {054014} (\bibinfo {year} {2015})},\ \Eprint {https://arxiv.org/abs/1506.07890} {arXiv:1506.07890 [hep-ph]} \BibitemShut {NoStop}%
\bibitem [{\citenamefont {Guo}\ \emph {et~al.}(2015)\citenamefont {Guo}, \citenamefont {Shi}, \citenamefont {Xu}, \citenamefont {Xu},\ and\ \citenamefont {Zhuang}}]{Guo:2015nsa}%
  \BibitemOpen
  \bibfield  {author} {\bibinfo {author} {\bibfnamefont {X.}~\bibnamefont {Guo}}, \bibinfo {author} {\bibfnamefont {S.}~\bibnamefont {Shi}}, \bibinfo {author} {\bibfnamefont {N.}~\bibnamefont {Xu}}, \bibinfo {author} {\bibfnamefont {Z.}~\bibnamefont {Xu}},\ and\ \bibinfo {author} {\bibfnamefont {P.}~\bibnamefont {Zhuang}},\ }\bibfield  {title} {\bibinfo {title} {{Magnetic field effect on charmonium production in high energy nuclear collisions}},\ }\href {https://doi.org/10.1016/j.physletb.2015.10.038} {\bibfield  {journal} {\bibinfo  {journal} {Phys. Lett. B}\ }\textbf {\bibinfo {volume} {751}},\ \bibinfo {pages} {215} (\bibinfo {year} {2015})},\ \Eprint {https://arxiv.org/abs/1502.04407} {arXiv:1502.04407 [hep-ph]} \BibitemShut {NoStop}%
\bibitem [{\citenamefont {Sadofyev}\ and\ \citenamefont {Yin}(2016)}]{Sadofyev:2015hxa}%
  \BibitemOpen
  \bibfield  {author} {\bibinfo {author} {\bibfnamefont {A.~V.}\ \bibnamefont {Sadofyev}}\ and\ \bibinfo {author} {\bibfnamefont {Y.}~\bibnamefont {Yin}},\ }\bibfield  {title} {\bibinfo {title} {{The charmonium dissociation in an {\textquotedblleft}anomalous wind{\textquotedblright}}},\ }\href {https://doi.org/10.1007/JHEP01(2016)052} {\bibfield  {journal} {\bibinfo  {journal} {JHEP}\ }\textbf {\bibinfo {volume} {01}},\ \bibinfo {pages} {052}},\ \Eprint {https://arxiv.org/abs/1510.06760} {arXiv:1510.06760 [hep-th]} \BibitemShut {NoStop}%
\bibitem [{\citenamefont {Suzuki}\ and\ \citenamefont {Yoshida}(2016)}]{Suzuki:2016kcs}%
  \BibitemOpen
  \bibfield  {author} {\bibinfo {author} {\bibfnamefont {K.}~\bibnamefont {Suzuki}}\ and\ \bibinfo {author} {\bibfnamefont {T.}~\bibnamefont {Yoshida}},\ }\bibfield  {title} {\bibinfo {title} {{Cigar-shaped quarkonia under strong magnetic field}},\ }\href {https://doi.org/10.1103/PhysRevD.93.051502} {\bibfield  {journal} {\bibinfo  {journal} {Phys. Rev. D}\ }\textbf {\bibinfo {volume} {93}},\ \bibinfo {pages} {051502} (\bibinfo {year} {2016})},\ \Eprint {https://arxiv.org/abs/1601.02178} {arXiv:1601.02178 [hep-ph]} \BibitemShut {NoStop}%
\bibitem [{\citenamefont {Yoshida}\ and\ \citenamefont {Suzuki}(2016)}]{Yoshida:2016xgm}%
  \BibitemOpen
  \bibfield  {author} {\bibinfo {author} {\bibfnamefont {T.}~\bibnamefont {Yoshida}}\ and\ \bibinfo {author} {\bibfnamefont {K.}~\bibnamefont {Suzuki}},\ }\bibfield  {title} {\bibinfo {title} {{Heavy meson spectroscopy under strong magnetic field}},\ }\href {https://doi.org/10.1103/PhysRevD.94.074043} {\bibfield  {journal} {\bibinfo  {journal} {Phys. Rev. D}\ }\textbf {\bibinfo {volume} {94}},\ \bibinfo {pages} {074043} (\bibinfo {year} {2016})},\ \Eprint {https://arxiv.org/abs/1607.04935} {arXiv:1607.04935 [hep-ph]} \BibitemShut {NoStop}%
\bibitem [{\citenamefont {Suzuki}\ and\ \citenamefont {Lee}(2017)}]{Suzuki:2016fof}%
  \BibitemOpen
  \bibfield  {author} {\bibinfo {author} {\bibfnamefont {K.}~\bibnamefont {Suzuki}}\ and\ \bibinfo {author} {\bibfnamefont {S.~H.}\ \bibnamefont {Lee}},\ }\bibfield  {title} {\bibinfo {title} {{Delayed versus accelerated quarkonium formation in a magnetic field}},\ }\href {https://doi.org/10.1103/PhysRevC.96.035203} {\bibfield  {journal} {\bibinfo  {journal} {Phys. Rev. C}\ }\textbf {\bibinfo {volume} {96}},\ \bibinfo {pages} {035203} (\bibinfo {year} {2017})},\ \Eprint {https://arxiv.org/abs/1610.09853} {arXiv:1610.09853 [hep-ph]} \BibitemShut {NoStop}%
\bibitem [{\citenamefont {Hasan}\ \emph {et~al.}(2017)\citenamefont {Hasan}, \citenamefont {Chatterjee},\ and\ \citenamefont {Patra}}]{Hasan:2017fmf}%
  \BibitemOpen
  \bibfield  {author} {\bibinfo {author} {\bibfnamefont {M.}~\bibnamefont {Hasan}}, \bibinfo {author} {\bibfnamefont {B.}~\bibnamefont {Chatterjee}},\ and\ \bibinfo {author} {\bibfnamefont {B.~K.}\ \bibnamefont {Patra}},\ }\bibfield  {title} {\bibinfo {title} {{Heavy quark potential in a static and strong homogeneous magnetic field}},\ }\href {https://doi.org/10.1140/epjc/s10052-017-5346-z} {\bibfield  {journal} {\bibinfo  {journal} {Eur. Phys. J. C}\ }\textbf {\bibinfo {volume} {77}},\ \bibinfo {pages} {767} (\bibinfo {year} {2017})},\ \Eprint {https://arxiv.org/abs/1703.10508} {arXiv:1703.10508 [hep-ph]} \BibitemShut {NoStop}%
\bibitem [{\citenamefont {Dutta}\ and\ \citenamefont {Mazumder}(2018)}]{Dutta:2017pya}%
  \BibitemOpen
  \bibfield  {author} {\bibinfo {author} {\bibfnamefont {N.}~\bibnamefont {Dutta}}\ and\ \bibinfo {author} {\bibfnamefont {S.}~\bibnamefont {Mazumder}},\ }\bibfield  {title} {\bibinfo {title} {{Majorana flipping of quarkonium spin states in transient magnetic field}},\ }\href {https://doi.org/10.1140/epjc/s10052-018-6000-0} {\bibfield  {journal} {\bibinfo  {journal} {Eur. Phys. J. C}\ }\textbf {\bibinfo {volume} {78}},\ \bibinfo {pages} {525} (\bibinfo {year} {2018})},\ \Eprint {https://arxiv.org/abs/1704.04094} {arXiv:1704.04094 [nucl-th]} \BibitemShut {NoStop}%
\bibitem [{\citenamefont {Singh}\ \emph {et~al.}(2018)\citenamefont {Singh}, \citenamefont {Thakur},\ and\ \citenamefont {Mishra}}]{Singh:2017nfa}%
  \BibitemOpen
  \bibfield  {author} {\bibinfo {author} {\bibfnamefont {B.}~\bibnamefont {Singh}}, \bibinfo {author} {\bibfnamefont {L.}~\bibnamefont {Thakur}},\ and\ \bibinfo {author} {\bibfnamefont {H.}~\bibnamefont {Mishra}},\ }\bibfield  {title} {\bibinfo {title} {{Heavy quark complex potential in a strongly magnetized hot QGP medium}},\ }\href {https://doi.org/10.1103/PhysRevD.97.096011} {\bibfield  {journal} {\bibinfo  {journal} {Phys. Rev. D}\ }\textbf {\bibinfo {volume} {97}},\ \bibinfo {pages} {096011} (\bibinfo {year} {2018})},\ \Eprint {https://arxiv.org/abs/1711.03071} {arXiv:1711.03071 [hep-ph]} \BibitemShut {NoStop}%
\bibitem [{\citenamefont {Hoelck}\ and\ \citenamefont {Wolschin}(2017)}]{Hoelck:2017dby}%
  \BibitemOpen
  \bibfield  {author} {\bibinfo {author} {\bibfnamefont {J.}~\bibnamefont {Hoelck}}\ and\ \bibinfo {author} {\bibfnamefont {G.}~\bibnamefont {Wolschin}},\ }\bibfield  {title} {\bibinfo {title} {{Electromagnetic field effects on $\Upsilon$-meson dissociation in PbPb collisions at LHC energies}},\ }\href {https://doi.org/10.1140/epja/i2017-12441-0} {\bibfield  {journal} {\bibinfo  {journal} {Eur. Phys. J. A}\ }\textbf {\bibinfo {volume} {53}},\ \bibinfo {pages} {241} (\bibinfo {year} {2017})},\ \Eprint {https://arxiv.org/abs/1712.06871} {arXiv:1712.06871 [hep-ph]} \BibitemShut {NoStop}%
\bibitem [{\citenamefont {Braga}\ and\ \citenamefont {Ferreira}(2018)}]{Braga:2018zlu}%
  \BibitemOpen
  \bibfield  {author} {\bibinfo {author} {\bibfnamefont {N.~R.~F.}\ \bibnamefont {Braga}}\ and\ \bibinfo {author} {\bibfnamefont {L.~F.}\ \bibnamefont {Ferreira}},\ }\bibfield  {title} {\bibinfo {title} {{Heavy meson dissociation in a plasma with magnetic fields}},\ }\href {https://doi.org/10.1016/j.physletb.2018.06.053} {\bibfield  {journal} {\bibinfo  {journal} {Phys. Lett. B}\ }\textbf {\bibinfo {volume} {783}},\ \bibinfo {pages} {186} (\bibinfo {year} {2018})},\ \Eprint {https://arxiv.org/abs/1802.02084} {arXiv:1802.02084 [hep-ph]} \BibitemShut {NoStop}%
\bibitem [{\citenamefont {Iwasaki}\ \emph {et~al.}(2019)\citenamefont {Iwasaki}, \citenamefont {Oka}, \citenamefont {Suzuki},\ and\ \citenamefont {Yoshida}}]{Iwasaki:2018pby}%
  \BibitemOpen
  \bibfield  {author} {\bibinfo {author} {\bibfnamefont {S.}~\bibnamefont {Iwasaki}}, \bibinfo {author} {\bibfnamefont {M.}~\bibnamefont {Oka}}, \bibinfo {author} {\bibfnamefont {K.}~\bibnamefont {Suzuki}},\ and\ \bibinfo {author} {\bibfnamefont {T.}~\bibnamefont {Yoshida}},\ }\bibfield  {title} {\bibinfo {title} {{Hadronic Paschen{\textendash}Back effect}},\ }\href {https://doi.org/10.1016/j.physletb.2018.10.024} {\bibfield  {journal} {\bibinfo  {journal} {Phys. Lett. B}\ }\textbf {\bibinfo {volume} {790}},\ \bibinfo {pages} {71} (\bibinfo {year} {2019})},\ \Eprint {https://arxiv.org/abs/1802.04971} {arXiv:1802.04971 [hep-ph]} \BibitemShut {NoStop}%
\bibitem [{\citenamefont {Hasan}\ \emph {et~al.}(2020)\citenamefont {Hasan}, \citenamefont {Patra}, \citenamefont {Chatterjee},\ and\ \citenamefont {Bagchi}}]{Hasan:2018kvx}%
  \BibitemOpen
  \bibfield  {author} {\bibinfo {author} {\bibfnamefont {M.}~\bibnamefont {Hasan}}, \bibinfo {author} {\bibfnamefont {B.~K.}\ \bibnamefont {Patra}}, \bibinfo {author} {\bibfnamefont {B.}~\bibnamefont {Chatterjee}},\ and\ \bibinfo {author} {\bibfnamefont {P.}~\bibnamefont {Bagchi}},\ }\bibfield  {title} {\bibinfo {title} {{Landau Damping in a strong magnetic field: Dissociation of Quarkonia}},\ }\href {https://doi.org/10.1016/j.nuclphysa.2019.121688} {\bibfield  {journal} {\bibinfo  {journal} {Nucl. Phys. A}\ }\textbf {\bibinfo {volume} {995}},\ \bibinfo {pages} {121688} (\bibinfo {year} {2020})},\ \Eprint {https://arxiv.org/abs/1802.06874} {arXiv:1802.06874 [hep-ph]} \BibitemShut {NoStop}%
\bibitem [{\citenamefont {Jahan C.~S.}\ \emph {et~al.}(2018)\citenamefont {Jahan C.~S.}, \citenamefont {Dhale}, \citenamefont {Reddy~P.}, \citenamefont {Kesarwani},\ and\ \citenamefont {Mishra}}]{Amal:2018qln}%
  \BibitemOpen
  \bibfield  {author} {\bibinfo {author} {\bibfnamefont {A.}~\bibnamefont {Jahan C.~S.}}, \bibinfo {author} {\bibfnamefont {N.}~\bibnamefont {Dhale}}, \bibinfo {author} {\bibfnamefont {S.}~\bibnamefont {Reddy~P.}}, \bibinfo {author} {\bibfnamefont {S.}~\bibnamefont {Kesarwani}},\ and\ \bibinfo {author} {\bibfnamefont {A.}~\bibnamefont {Mishra}},\ }\bibfield  {title} {\bibinfo {title} {{Charmonium states in strong magnetic fields}},\ }\href {https://doi.org/10.1103/PhysRevC.98.065202} {\bibfield  {journal} {\bibinfo  {journal} {Phys. Rev. C}\ }\textbf {\bibinfo {volume} {98}},\ \bibinfo {pages} {065202} (\bibinfo {year} {2018})},\ \Eprint {https://arxiv.org/abs/1803.04322} {arXiv:1803.04322 [nucl-th]} \BibitemShut {NoStop}%
\bibitem [{\citenamefont {Bagchi}\ \emph {et~al.}(2023)\citenamefont {Bagchi}, \citenamefont {Dutta}, \citenamefont {Chatterjee},\ and\ \citenamefont {Adhya}}]{Bagchi:2018olp}%
  \BibitemOpen
  \bibfield  {author} {\bibinfo {author} {\bibfnamefont {P.}~\bibnamefont {Bagchi}}, \bibinfo {author} {\bibfnamefont {N.}~\bibnamefont {Dutta}}, \bibinfo {author} {\bibfnamefont {B.}~\bibnamefont {Chatterjee}},\ and\ \bibinfo {author} {\bibfnamefont {S.~P.}\ \bibnamefont {Adhya}},\ }\bibfield  {title} {\bibinfo {title} {{Dissociation of heavy quarkonium states in a rapidly varying strong magnetic field}},\ }\href {https://doi.org/10.1142/S0217732323500359} {\bibfield  {journal} {\bibinfo  {journal} {Mod. Phys. Lett. A}\ }\textbf {\bibinfo {volume} {38}},\ \bibinfo {pages} {2350035} (\bibinfo {year} {2023})},\ \Eprint {https://arxiv.org/abs/1805.04082} {arXiv:1805.04082 [nucl-th]} \BibitemShut {NoStop}%
\bibitem [{\citenamefont {Iwasaki}\ and\ \citenamefont {Suzuki}(2018)}]{Iwasaki:2018czv}%
  \BibitemOpen
  \bibfield  {author} {\bibinfo {author} {\bibfnamefont {S.}~\bibnamefont {Iwasaki}}\ and\ \bibinfo {author} {\bibfnamefont {K.}~\bibnamefont {Suzuki}},\ }\bibfield  {title} {\bibinfo {title} {{Quarkonium radiative decays from the hadronic Paschen-Back effect}},\ }\href {https://doi.org/10.1103/PhysRevD.98.054017} {\bibfield  {journal} {\bibinfo  {journal} {Phys. Rev. D}\ }\textbf {\bibinfo {volume} {98}},\ \bibinfo {pages} {054017} (\bibinfo {year} {2018})},\ \Eprint {https://arxiv.org/abs/1805.09787} {arXiv:1805.09787 [hep-ph]} \BibitemShut {NoStop}%
\bibitem [{\citenamefont {Braga}\ and\ \citenamefont {Ferreira}(2019)}]{Braga:2019yeh}%
  \BibitemOpen
  \bibfield  {author} {\bibinfo {author} {\bibfnamefont {N.~R.~F.}\ \bibnamefont {Braga}}\ and\ \bibinfo {author} {\bibfnamefont {L.~F.}\ \bibnamefont {Ferreira}},\ }\bibfield  {title} {\bibinfo {title} {{Quasinormal modes for quarkonium in a plasma with magnetic fields}},\ }\href {https://doi.org/10.1016/j.physletb.2019.06.050} {\bibfield  {journal} {\bibinfo  {journal} {Phys. Lett. B}\ }\textbf {\bibinfo {volume} {795}},\ \bibinfo {pages} {462} (\bibinfo {year} {2019})},\ \Eprint {https://arxiv.org/abs/1905.11309} {arXiv:1905.11309 [hep-ph]} \BibitemShut {NoStop}%
\bibitem [{\citenamefont {Hasan}\ and\ \citenamefont {Patra}(2020)}]{Hasan:2020iwa}%
  \BibitemOpen
  \bibfield  {author} {\bibinfo {author} {\bibfnamefont {M.}~\bibnamefont {Hasan}}\ and\ \bibinfo {author} {\bibfnamefont {B.~K.}\ \bibnamefont {Patra}},\ }\bibfield  {title} {\bibinfo {title} {{Dissociation of heavy quarkonia in a weak magnetic field}},\ }\href {https://doi.org/10.1103/PhysRevD.102.036020} {\bibfield  {journal} {\bibinfo  {journal} {Phys. Rev. D}\ }\textbf {\bibinfo {volume} {102}},\ \bibinfo {pages} {036020} (\bibinfo {year} {2020})},\ \Eprint {https://arxiv.org/abs/2004.12857} {arXiv:2004.12857 [hep-ph]} \BibitemShut {NoStop}%
\bibitem [{\citenamefont {Chen}\ \emph {et~al.}(2021)\citenamefont {Chen}, \citenamefont {Zhao},\ and\ \citenamefont {Zhuang}}]{Chen:2020xsr}%
  \BibitemOpen
  \bibfield  {author} {\bibinfo {author} {\bibfnamefont {S.}~\bibnamefont {Chen}}, \bibinfo {author} {\bibfnamefont {J.}~\bibnamefont {Zhao}},\ and\ \bibinfo {author} {\bibfnamefont {P.}~\bibnamefont {Zhuang}},\ }\bibfield  {title} {\bibinfo {title} {{Charmonium transition in electromagnetic and rotational fields}},\ }\href {https://doi.org/10.1103/PhysRevC.103.L031902} {\bibfield  {journal} {\bibinfo  {journal} {Phys. Rev. C}\ }\textbf {\bibinfo {volume} {103}},\ \bibinfo {pages} {L031902} (\bibinfo {year} {2021})},\ \Eprint {https://arxiv.org/abs/2005.08473} {arXiv:2005.08473 [nucl-th]} \BibitemShut {NoStop}%
\bibitem [{\citenamefont {Zhou}\ \emph {et~al.}(2020)\citenamefont {Zhou}, \citenamefont {Chen}, \citenamefont {Zhao},\ and\ \citenamefont {Ping}}]{Zhou:2020ssi}%
  \BibitemOpen
  \bibfield  {author} {\bibinfo {author} {\bibfnamefont {J.}~\bibnamefont {Zhou}}, \bibinfo {author} {\bibfnamefont {X.}~\bibnamefont {Chen}}, \bibinfo {author} {\bibfnamefont {Y.-Q.}\ \bibnamefont {Zhao}},\ and\ \bibinfo {author} {\bibfnamefont {J.}~\bibnamefont {Ping}},\ }\bibfield  {title} {\bibinfo {title} {{Thermodynamics of heavy quarkonium in a magnetic field background}},\ }\href {https://doi.org/10.1103/PhysRevD.102.086020} {\bibfield  {journal} {\bibinfo  {journal} {Phys. Rev. D}\ }\textbf {\bibinfo {volume} {102}},\ \bibinfo {pages} {086020} (\bibinfo {year} {2020})},\ \Eprint {https://arxiv.org/abs/2006.09062} {arXiv:2006.09062 [hep-ph]} \BibitemShut {NoStop}%
\bibitem [{\citenamefont {Braga}\ and\ \citenamefont {da~Mata}(2020)}]{Braga:2020hhs}%
  \BibitemOpen
  \bibfield  {author} {\bibinfo {author} {\bibfnamefont {N.~R.~F.}\ \bibnamefont {Braga}}\ and\ \bibinfo {author} {\bibfnamefont {R.}~\bibnamefont {da~Mata}},\ }\bibfield  {title} {\bibinfo {title} {{Configuration entropy description of charmonium dissociation under the influence of magnetic fields}},\ }\href {https://doi.org/10.1016/j.physletb.2020.135918} {\bibfield  {journal} {\bibinfo  {journal} {Phys. Lett. B}\ }\textbf {\bibinfo {volume} {811}},\ \bibinfo {pages} {135918} (\bibinfo {year} {2020})},\ \Eprint {https://arxiv.org/abs/2008.10457} {arXiv:2008.10457 [hep-th]} \BibitemShut {NoStop}%
\bibitem [{\citenamefont {Iwasaki}\ \emph {et~al.}(2021{\natexlab{a}})\citenamefont {Iwasaki}, \citenamefont {Jido}, \citenamefont {Oka},\ and\ \citenamefont {Suzuki}}]{Iwasaki:2021kms}%
  \BibitemOpen
  \bibfield  {author} {\bibinfo {author} {\bibfnamefont {S.}~\bibnamefont {Iwasaki}}, \bibinfo {author} {\bibfnamefont {D.}~\bibnamefont {Jido}}, \bibinfo {author} {\bibfnamefont {M.}~\bibnamefont {Oka}},\ and\ \bibinfo {author} {\bibfnamefont {K.}~\bibnamefont {Suzuki}},\ }\bibfield  {title} {\bibinfo {title} {{Survival probabilities of charmonia as a clue to measure transient magnetic fields}},\ }\href {https://doi.org/10.1016/j.physletb.2021.136498} {\bibfield  {journal} {\bibinfo  {journal} {Phys. Lett. B}\ }\textbf {\bibinfo {volume} {820}},\ \bibinfo {pages} {136498} (\bibinfo {year} {2021}{\natexlab{a}})},\ \Eprint {https://arxiv.org/abs/2104.13989} {arXiv:2104.13989 [hep-ph]} \BibitemShut {NoStop}%
\bibitem [{\citenamefont {Braga}\ \emph {et~al.}(2022)\citenamefont {Braga}, \citenamefont {Ferreira},\ and\ \citenamefont {Ferreira}}]{Braga:2021fey}%
  \BibitemOpen
  \bibfield  {author} {\bibinfo {author} {\bibfnamefont {N.~R.~F.}\ \bibnamefont {Braga}}, \bibinfo {author} {\bibfnamefont {Y.~F.}\ \bibnamefont {Ferreira}},\ and\ \bibinfo {author} {\bibfnamefont {L.~F.}\ \bibnamefont {Ferreira}},\ }\bibfield  {title} {\bibinfo {title} {{Configuration entropy and stability of bottomonium radial excitations in a plasma with magnetic fields}},\ }\href {https://doi.org/10.1103/PhysRevD.105.114044} {\bibfield  {journal} {\bibinfo  {journal} {Phys. Rev. D}\ }\textbf {\bibinfo {volume} {105}},\ \bibinfo {pages} {114044} (\bibinfo {year} {2022})},\ \Eprint {https://arxiv.org/abs/2110.04560} {arXiv:2110.04560 [hep-th]} \BibitemShut {NoStop}%
\bibitem [{\citenamefont {Jena}\ \emph {et~al.}(2022)\citenamefont {Jena}, \citenamefont {Shukla}, \citenamefont {Dudal},\ and\ \citenamefont {Mahapatra}}]{Jena:2022nzw}%
  \BibitemOpen
  \bibfield  {author} {\bibinfo {author} {\bibfnamefont {S.~S.}\ \bibnamefont {Jena}}, \bibinfo {author} {\bibfnamefont {B.}~\bibnamefont {Shukla}}, \bibinfo {author} {\bibfnamefont {D.}~\bibnamefont {Dudal}},\ and\ \bibinfo {author} {\bibfnamefont {S.}~\bibnamefont {Mahapatra}},\ }\bibfield  {title} {\bibinfo {title} {{Entropic force and real-time dynamics of holographic quarkonium in a magnetic field}},\ }\href {https://doi.org/10.1103/PhysRevD.105.086011} {\bibfield  {journal} {\bibinfo  {journal} {Phys. Rev. D}\ }\textbf {\bibinfo {volume} {105}},\ \bibinfo {pages} {086011} (\bibinfo {year} {2022})},\ \Eprint {https://arxiv.org/abs/2202.01486} {arXiv:2202.01486 [hep-th]} \BibitemShut {NoStop}%
\bibitem [{\citenamefont {Hu}\ \emph {et~al.}(2022)\citenamefont {Hu}, \citenamefont {Shi}, \citenamefont {Xu}, \citenamefont {Zhao},\ and\ \citenamefont {Zhuang}}]{Hu:2022ofv}%
  \BibitemOpen
  \bibfield  {author} {\bibinfo {author} {\bibfnamefont {J.}~\bibnamefont {Hu}}, \bibinfo {author} {\bibfnamefont {S.}~\bibnamefont {Shi}}, \bibinfo {author} {\bibfnamefont {Z.}~\bibnamefont {Xu}}, \bibinfo {author} {\bibfnamefont {J.}~\bibnamefont {Zhao}},\ and\ \bibinfo {author} {\bibfnamefont {P.}~\bibnamefont {Zhuang}},\ }\bibfield  {title} {\bibinfo {title} {{Magnetic field induced hair structure in the charmonium gluon dissociation}},\ }\href {https://doi.org/10.1103/PhysRevD.105.094013} {\bibfield  {journal} {\bibinfo  {journal} {Phys. Rev. D}\ }\textbf {\bibinfo {volume} {105}},\ \bibinfo {pages} {094013} (\bibinfo {year} {2022})},\ \Eprint {https://arxiv.org/abs/2202.07938} {arXiv:2202.07938 [nucl-th]} \BibitemShut {NoStop}%
\bibitem [{\citenamefont {Ghosh}\ \emph {et~al.}(2022)\citenamefont {Ghosh}, \citenamefont {Bandyopadhyay}, \citenamefont {Nilima},\ and\ \citenamefont {Ghosh}}]{Ghosh:2022sxi}%
  \BibitemOpen
  \bibfield  {author} {\bibinfo {author} {\bibfnamefont {R.}~\bibnamefont {Ghosh}}, \bibinfo {author} {\bibfnamefont {A.}~\bibnamefont {Bandyopadhyay}}, \bibinfo {author} {\bibfnamefont {I.}~\bibnamefont {Nilima}},\ and\ \bibinfo {author} {\bibfnamefont {S.}~\bibnamefont {Ghosh}},\ }\bibfield  {title} {\bibinfo {title} {{Anisotropic tomography of heavy quark dissociation by using the general propagator structure in a finite magnetic field}},\ }\href {https://doi.org/10.1103/PhysRevD.106.054010} {\bibfield  {journal} {\bibinfo  {journal} {Phys. Rev. D}\ }\textbf {\bibinfo {volume} {106}},\ \bibinfo {pages} {054010} (\bibinfo {year} {2022})},\ \Eprint {https://arxiv.org/abs/2204.02312} {arXiv:2204.02312 [hep-ph]} \BibitemShut {NoStop}%
\bibitem [{\citenamefont {Parui}\ \emph {et~al.}(2022)\citenamefont {Parui}, \citenamefont {De}, \citenamefont {Kumar},\ and\ \citenamefont {Mishra}}]{Parui:2022msu}%
  \BibitemOpen
  \bibfield  {author} {\bibinfo {author} {\bibfnamefont {P.}~\bibnamefont {Parui}}, \bibinfo {author} {\bibfnamefont {S.}~\bibnamefont {De}}, \bibinfo {author} {\bibfnamefont {A.}~\bibnamefont {Kumar}},\ and\ \bibinfo {author} {\bibfnamefont {A.}~\bibnamefont {Mishra}},\ }\bibfield  {title} {\bibinfo {title} {{QCD sum rule analysis of heavy quarkonium states in magnetized matter: Effects of magnetic and inverse magnetic catalysis}},\ }\href {https://doi.org/10.1103/PhysRevD.106.114033} {\bibfield  {journal} {\bibinfo  {journal} {Phys. Rev. D}\ }\textbf {\bibinfo {volume} {106}},\ \bibinfo {pages} {114033} (\bibinfo {year} {2022})},\ \Eprint {https://arxiv.org/abs/2208.05856} {arXiv:2208.05856 [hep-ph]} \BibitemShut {NoStop}%
\bibitem [{\citenamefont {Sebastian}\ \emph {et~al.}(2023)\citenamefont {Sebastian}, \citenamefont {Thakur}, \citenamefont {Mishra},\ and\ \citenamefont {Haque}}]{Sebastian:2023tlw}%
  \BibitemOpen
  \bibfield  {author} {\bibinfo {author} {\bibfnamefont {J.}~\bibnamefont {Sebastian}}, \bibinfo {author} {\bibfnamefont {L.}~\bibnamefont {Thakur}}, \bibinfo {author} {\bibfnamefont {H.}~\bibnamefont {Mishra}},\ and\ \bibinfo {author} {\bibfnamefont {N.}~\bibnamefont {Haque}},\ }\bibfield  {title} {\bibinfo {title} {{Heavy quarkonia in QGP medium in an arbitrary magnetic field}},\ }\href {https://doi.org/10.1103/PhysRevD.108.094001} {\bibfield  {journal} {\bibinfo  {journal} {Phys. Rev. D}\ }\textbf {\bibinfo {volume} {108}},\ \bibinfo {pages} {094001} (\bibinfo {year} {2023})},\ \Eprint {https://arxiv.org/abs/2308.04410} {arXiv:2308.04410 [hep-ph]} \BibitemShut {NoStop}%
\bibitem [{\citenamefont {Nilima}\ \emph {et~al.}(2024)\citenamefont {Nilima}, \citenamefont {Hasan}, \citenamefont {Singh},\ and\ \citenamefont {Jamal}}]{Nilima:2024nvd}%
  \BibitemOpen
  \bibfield  {author} {\bibinfo {author} {\bibfnamefont {I.}~\bibnamefont {Nilima}}, \bibinfo {author} {\bibfnamefont {M.}~\bibnamefont {Hasan}}, \bibinfo {author} {\bibfnamefont {B.~K.}\ \bibnamefont {Singh}},\ and\ \bibinfo {author} {\bibfnamefont {M.~Y.}\ \bibnamefont {Jamal}},\ }\bibfield  {title} {\bibinfo {title} {{Quarkonia dissociation at finite magnetic field in the presence of momentum anisotropy}},\ }\href {https://doi.org/10.1140/epjc/s10052-024-12525-y} {\bibfield  {journal} {\bibinfo  {journal} {Eur. Phys. J. C}\ }\textbf {\bibinfo {volume} {84}},\ \bibinfo {pages} {160} (\bibinfo {year} {2024})},\ \Eprint {https://arxiv.org/abs/2402.07848} {arXiv:2402.07848 [hep-ph]} \BibitemShut {NoStop}%
\bibitem [{\citenamefont {Jena}\ \emph {et~al.}(2024)\citenamefont {Jena}, \citenamefont {Barman}, \citenamefont {Toniato}, \citenamefont {Dudal},\ and\ \citenamefont {Mahapatra}}]{Jena:2024cqs}%
  \BibitemOpen
  \bibfield  {author} {\bibinfo {author} {\bibfnamefont {S.~S.}\ \bibnamefont {Jena}}, \bibinfo {author} {\bibfnamefont {J.}~\bibnamefont {Barman}}, \bibinfo {author} {\bibfnamefont {B.}~\bibnamefont {Toniato}}, \bibinfo {author} {\bibfnamefont {D.}~\bibnamefont {Dudal}},\ and\ \bibinfo {author} {\bibfnamefont {S.}~\bibnamefont {Mahapatra}},\ }\bibfield  {title} {\bibinfo {title} {{A dynamical Einstein-Born-Infeld-dilaton model and holographic quarkonium melting in a magnetic field}},\ }\href {https://doi.org/10.1007/JHEP12(2024)096} {\bibfield  {journal} {\bibinfo  {journal} {JHEP}\ }\textbf {\bibinfo {volume} {12}},\ \bibinfo {pages} {096}},\ \Eprint {https://arxiv.org/abs/2408.14813} {arXiv:2408.14813 [hep-th]} \BibitemShut {NoStop}%
\bibitem [{\citenamefont {Shukla}\ \emph {et~al.}(2025)\citenamefont {Shukla}, \citenamefont {Nongmaithem}, \citenamefont {Dudal},\ and\ \citenamefont {Mahapatra}}]{Shukla:2024qlf}%
  \BibitemOpen
  \bibfield  {author} {\bibinfo {author} {\bibfnamefont {B.}~\bibnamefont {Shukla}}, \bibinfo {author} {\bibfnamefont {J.}~\bibnamefont {Nongmaithem}}, \bibinfo {author} {\bibfnamefont {D.}~\bibnamefont {Dudal}},\ and\ \bibinfo {author} {\bibfnamefont {S.}~\bibnamefont {Mahapatra}},\ }\bibfield  {title} {\bibinfo {title} {{Interplay of magnetic field and chemical potential induced anisotropy and frame dependent chaos of a $Q\bar{Q}$ pair in holographic QCD}},\ }\href {https://doi.org/10.1103/PhysRevD.111.106002} {\bibfield  {journal} {\bibinfo  {journal} {Phys. Rev. D}\ }\textbf {\bibinfo {volume} {111}},\ \bibinfo {pages} {106002} (\bibinfo {year} {2025})},\ \Eprint {https://arxiv.org/abs/2411.17279} {arXiv:2411.17279 [hep-th]} \BibitemShut {NoStop}%
\bibitem [{\citenamefont {Wen}\ \emph {et~al.}(2025)\citenamefont {Wen}, \citenamefont {Li}, \citenamefont {Zhou}, \citenamefont {Li},\ and\ \citenamefont {Vary}}]{Wen:2025dwy}%
  \BibitemOpen
  \bibfield  {author} {\bibinfo {author} {\bibfnamefont {L.}~\bibnamefont {Wen}}, \bibinfo {author} {\bibfnamefont {M.}~\bibnamefont {Li}}, \bibinfo {author} {\bibfnamefont {Y.}~\bibnamefont {Zhou}}, \bibinfo {author} {\bibfnamefont {Y.}~\bibnamefont {Li}},\ and\ \bibinfo {author} {\bibfnamefont {J.~P.}\ \bibnamefont {Vary}},\ }\bibfield  {title} {\bibinfo {title} {{Relativistic dynamics of charmonia in strong magnetic fields}},\ }\href {https://doi.org/10.1103/dnq8-ncd7} {\bibfield  {journal} {\bibinfo  {journal} {Phys. Rev. D}\ }\textbf {\bibinfo {volume} {112}},\ \bibinfo {pages} {014019} (\bibinfo {year} {2025})},\ \Eprint {https://arxiv.org/abs/2504.03294} {arXiv:2504.03294 [hep-ph]} \BibitemShut {NoStop}%
\bibitem [{\citenamefont {Jena}\ \emph {et~al.}(2025)\citenamefont {Jena}, \citenamefont {Bhattacharjee}, \citenamefont {Dudal},\ and\ \citenamefont {Mahapatra}}]{Jena:2025xcf}%
  \BibitemOpen
  \bibfield  {author} {\bibinfo {author} {\bibfnamefont {S.~S.}\ \bibnamefont {Jena}}, \bibinfo {author} {\bibfnamefont {A.}~\bibnamefont {Bhattacharjee}}, \bibinfo {author} {\bibfnamefont {D.}~\bibnamefont {Dudal}},\ and\ \bibinfo {author} {\bibfnamefont {S.}~\bibnamefont {Mahapatra}},\ }\bibfield  {title} {\bibinfo {title} {{Probing quarkonium diffusion in a magnetized quark-gluon plasma}},\ }\href {https://doi.org/10.1103/15z1-bxlm} {\bibfield  {journal} {\bibinfo  {journal} {Phys. Rev. D}\ }\textbf {\bibinfo {volume} {112}},\ \bibinfo {pages} {086010} (\bibinfo {year} {2025})},\ \Eprint {https://arxiv.org/abs/2507.00746} {arXiv:2507.00746 [hep-th]} \BibitemShut {NoStop}%
\bibitem [{\citenamefont {Arifi}\ and\ \citenamefont {Suzuki}(2025)}]{Arifi:2025ivt}%
  \BibitemOpen
  \bibfield  {author} {\bibinfo {author} {\bibfnamefont {A.~J.}\ \bibnamefont {Arifi}}\ and\ \bibinfo {author} {\bibfnamefont {K.}~\bibnamefont {Suzuki}},\ }\bibfield  {title} {\bibinfo {title} {{Structure of heavy quarkonia in a strong magnetic field}},\ }\href {https://doi.org/10.1103/pwsl-xrq5} {\bibfield  {journal} {\bibinfo  {journal} {Phys. Rev. D}\ }\textbf {\bibinfo {volume} {112}},\ \bibinfo {pages} {094013} (\bibinfo {year} {2025})},\ \Eprint {https://arxiv.org/abs/2507.18894} {arXiv:2507.18894 [hep-ph]} \BibitemShut {NoStop}%
\bibitem [{\citenamefont {Yan}\ and\ \citenamefont {Lin}(2026)}]{Yan:2025tlx}%
  \BibitemOpen
  \bibfield  {author} {\bibinfo {author} {\bibfnamefont {G.}~\bibnamefont {Yan}}\ and\ \bibinfo {author} {\bibfnamefont {S.}~\bibnamefont {Lin}},\ }\bibfield  {title} {\bibinfo {title} {{Distorted quarkonia and spin alignment}},\ }\href {https://doi.org/10.1103/jg98-9vx9} {\bibfield  {journal} {\bibinfo  {journal} {Phys. Rev. D}\ }\textbf {\bibinfo {volume} {113}},\ \bibinfo {pages} {054045} (\bibinfo {year} {2026})},\ \Eprint {https://arxiv.org/abs/2507.22684} {arXiv:2507.22684 [hep-ph]} \BibitemShut {NoStop}%
\bibitem [{\citenamefont {Dominguez}\ \emph {et~al.}(2025)\citenamefont {Dominguez}, \citenamefont {Koning},\ and\ \citenamefont {Hern{\'a}ndez}}]{Dominguez:2025nar}%
  \BibitemOpen
  \bibfield  {author} {\bibinfo {author} {\bibfnamefont {C.~A.}\ \bibnamefont {Dominguez}}, \bibinfo {author} {\bibfnamefont {M.}~\bibnamefont {Koning}},\ and\ \bibinfo {author} {\bibfnamefont {L.~A.}\ \bibnamefont {Hern{\'a}ndez}},\ }\bibfield  {title} {\bibinfo {title} {{Magnetic catalysis of charmonium in the vector channel}},\ }\href {https://doi.org/10.1103/6nz8-lwc1} {\bibfield  {journal} {\bibinfo  {journal} {Phys. Rev. D}\ }\textbf {\bibinfo {volume} {112}},\ \bibinfo {pages} {094049} (\bibinfo {year} {2025})},\ \Eprint {https://arxiv.org/abs/2510.09927} {arXiv:2510.09927 [hep-ph]} \BibitemShut {NoStop}%
\bibitem [{\citenamefont {Arifi}\ and\ \citenamefont {Suzuki}(2026{\natexlab{a}})}]{Arifi:2025atv}%
  \BibitemOpen
  \bibfield  {author} {\bibinfo {author} {\bibfnamefont {A.~J.}\ \bibnamefont {Arifi}}\ and\ \bibinfo {author} {\bibfnamefont {K.}~\bibnamefont {Suzuki}},\ }\bibfield  {title} {\bibinfo {title} {{Landau-Zener-St{\"u}ckelberg-Majorana dynamics of magnetized quarkonia}},\ }\href {https://doi.org/10.1103/xc28-3v8r} {\bibfield  {journal} {\bibinfo  {journal} {Phys. Rev. D}\ }\textbf {\bibinfo {volume} {113}},\ \bibinfo {pages} {054047} (\bibinfo {year} {2026}{\natexlab{a}})},\ \Eprint {https://arxiv.org/abs/2512.24072} {arXiv:2512.24072 [hep-ph]} \BibitemShut {NoStop}%
\bibitem [{\citenamefont {Arifi}\ and\ \citenamefont {Suzuki}(2026{\natexlab{b}})}]{Arifi:2026asz}%
  \BibitemOpen
  \bibfield  {author} {\bibinfo {author} {\bibfnamefont {A.~J.}\ \bibnamefont {Arifi}}\ and\ \bibinfo {author} {\bibfnamefont {K.}~\bibnamefont {Suzuki}},\ }\bibfield  {title} {\bibinfo {title} {{Quarkonium spectra with magnetically-induced anisotropic confinement}},\ }\href@noop {} {\  (\bibinfo {year} {2026}{\natexlab{b}})},\ \Eprint {https://arxiv.org/abs/2603.12589} {arXiv:2603.12589 [hep-ph]} \BibitemShut {NoStop}%
\bibitem [{\citenamefont {Hattori}\ and\ \citenamefont {Huang}(2017)}]{Hattori:2016emy}%
  \BibitemOpen
  \bibfield  {author} {\bibinfo {author} {\bibfnamefont {K.}~\bibnamefont {Hattori}}\ and\ \bibinfo {author} {\bibfnamefont {X.-G.}\ \bibnamefont {Huang}},\ }\bibfield  {title} {\bibinfo {title} {{Novel quantum phenomena induced by strong magnetic fields in heavy-ion collisions}},\ }\href {https://doi.org/10.1007/s41365-016-0178-3} {\bibfield  {journal} {\bibinfo  {journal} {Nucl. Sci. Tech.}\ }\textbf {\bibinfo {volume} {28}},\ \bibinfo {pages} {26} (\bibinfo {year} {2017})},\ \Eprint {https://arxiv.org/abs/1609.00747} {arXiv:1609.00747 [nucl-th]} \BibitemShut {NoStop}%
\bibitem [{\citenamefont {Zhao}\ \emph {et~al.}(2020)\citenamefont {Zhao}, \citenamefont {Zhou}, \citenamefont {Chen},\ and\ \citenamefont {Zhuang}}]{Zhao:2020jqu}%
  \BibitemOpen
  \bibfield  {author} {\bibinfo {author} {\bibfnamefont {J.}~\bibnamefont {Zhao}}, \bibinfo {author} {\bibfnamefont {K.}~\bibnamefont {Zhou}}, \bibinfo {author} {\bibfnamefont {S.}~\bibnamefont {Chen}},\ and\ \bibinfo {author} {\bibfnamefont {P.}~\bibnamefont {Zhuang}},\ }\bibfield  {title} {\bibinfo {title} {{Heavy flavors under extreme conditions in high energy nuclear collisions}},\ }\href {https://doi.org/10.1016/j.ppnp.2020.103801} {\bibfield  {journal} {\bibinfo  {journal} {Prog. Part. Nucl. Phys.}\ }\textbf {\bibinfo {volume} {114}},\ \bibinfo {pages} {103801} (\bibinfo {year} {2020})},\ \Eprint {https://arxiv.org/abs/2005.08277} {arXiv:2005.08277 [nucl-th]} \BibitemShut {NoStop}%
\bibitem [{\citenamefont {Iwasaki}\ \emph {et~al.}(2021{\natexlab{b}})\citenamefont {Iwasaki}, \citenamefont {Oka},\ and\ \citenamefont {Suzuki}}]{Iwasaki:2021nrz}%
  \BibitemOpen
  \bibfield  {author} {\bibinfo {author} {\bibfnamefont {S.}~\bibnamefont {Iwasaki}}, \bibinfo {author} {\bibfnamefont {M.}~\bibnamefont {Oka}},\ and\ \bibinfo {author} {\bibfnamefont {K.}~\bibnamefont {Suzuki}},\ }\bibfield  {title} {\bibinfo {title} {{A review of quarkonia under strong magnetic fields}},\ }\href {https://doi.org/10.1140/epja/s10050-021-00533-5} {\bibfield  {journal} {\bibinfo  {journal} {Eur. Phys. J. A}\ }\textbf {\bibinfo {volume} {57}},\ \bibinfo {pages} {222} (\bibinfo {year} {2021}{\natexlab{b}})},\ \Eprint {https://arxiv.org/abs/2104.13990} {arXiv:2104.13990 [hep-ph]} \BibitemShut {NoStop}%
\bibitem [{\citenamefont {Gusynin}\ \emph {et~al.}(1995)\citenamefont {Gusynin}, \citenamefont {Miransky},\ and\ \citenamefont {Shovkovy}}]{Gusynin:1995gt}%
  \BibitemOpen
  \bibfield  {author} {\bibinfo {author} {\bibfnamefont {V.~P.}\ \bibnamefont {Gusynin}}, \bibinfo {author} {\bibfnamefont {V.~A.}\ \bibnamefont {Miransky}},\ and\ \bibinfo {author} {\bibfnamefont {I.~A.}\ \bibnamefont {Shovkovy}},\ }\bibfield  {title} {\bibinfo {title} {{Dynamical chiral symmetry breaking by a magnetic field in QED}},\ }\href {https://doi.org/10.1103/PhysRevD.52.4747} {\bibfield  {journal} {\bibinfo  {journal} {Phys. Rev. D}\ }\textbf {\bibinfo {volume} {52}},\ \bibinfo {pages} {4747} (\bibinfo {year} {1995})},\ \Eprint {https://arxiv.org/abs/hep-ph/9501304} {arXiv:hep-ph/9501304} \BibitemShut {NoStop}%
\bibitem [{\citenamefont {Gusynin}\ \emph {et~al.}(1996)\citenamefont {Gusynin}, \citenamefont {Miransky},\ and\ \citenamefont {Shovkovy}}]{Gusynin:1995nb}%
  \BibitemOpen
  \bibfield  {author} {\bibinfo {author} {\bibfnamefont {V.~P.}\ \bibnamefont {Gusynin}}, \bibinfo {author} {\bibfnamefont {V.~A.}\ \bibnamefont {Miransky}},\ and\ \bibinfo {author} {\bibfnamefont {I.~A.}\ \bibnamefont {Shovkovy}},\ }\bibfield  {title} {\bibinfo {title} {{Dimensional reduction and catalysis of dynamical symmetry breaking by a magnetic field}},\ }\href {https://doi.org/10.1016/0550-3213(96)00021-1} {\bibfield  {journal} {\bibinfo  {journal} {Nucl. Phys. B}\ }\textbf {\bibinfo {volume} {462}},\ \bibinfo {pages} {249} (\bibinfo {year} {1996})},\ \Eprint {https://arxiv.org/abs/hep-ph/9509320} {arXiv:hep-ph/9509320} \BibitemShut {NoStop}%
\bibitem [{\citenamefont {Shushpanov}\ and\ \citenamefont {Smilga}(1997)}]{Shushpanov:1997sf}%
  \BibitemOpen
  \bibfield  {author} {\bibinfo {author} {\bibfnamefont {I.~A.}\ \bibnamefont {Shushpanov}}\ and\ \bibinfo {author} {\bibfnamefont {A.~V.}\ \bibnamefont {Smilga}},\ }\bibfield  {title} {\bibinfo {title} {{Quark condensate in a magnetic field}},\ }\href {https://doi.org/10.1016/S0370-2693(97)00441-3} {\bibfield  {journal} {\bibinfo  {journal} {Phys. Lett. B}\ }\textbf {\bibinfo {volume} {402}},\ \bibinfo {pages} {351} (\bibinfo {year} {1997})},\ \Eprint {https://arxiv.org/abs/hep-ph/9703201} {arXiv:hep-ph/9703201} \BibitemShut {NoStop}%
\bibitem [{\citenamefont {Agasian}\ and\ \citenamefont {Shushpanov}(2001)}]{Agasian:2001ym}%
  \BibitemOpen
  \bibfield  {author} {\bibinfo {author} {\bibfnamefont {N.~O.}\ \bibnamefont {Agasian}}\ and\ \bibinfo {author} {\bibfnamefont {I.~A.}\ \bibnamefont {Shushpanov}},\ }\bibfield  {title} {\bibinfo {title} {{Gell-Mann-Oakes-Renner relation in a magnetic field at finite temperature}},\ }\href {https://doi.org/10.1088/1126-6708/2001/10/006} {\bibfield  {journal} {\bibinfo  {journal} {JHEP}\ }\textbf {\bibinfo {volume} {10}},\ \bibinfo {pages} {006}},\ \Eprint {https://arxiv.org/abs/hep-ph/0107128} {arXiv:hep-ph/0107128} \BibitemShut {NoStop}%
\bibitem [{\citenamefont {Chernodub}(2010)}]{Chernodub:2010qx}%
  \BibitemOpen
  \bibfield  {author} {\bibinfo {author} {\bibfnamefont {M.~N.}\ \bibnamefont {Chernodub}},\ }\bibfield  {title} {\bibinfo {title} {{Superconductivity of QCD vacuum in strong magnetic field}},\ }\href {https://doi.org/10.1103/PhysRevD.82.085011} {\bibfield  {journal} {\bibinfo  {journal} {Phys. Rev. D}\ }\textbf {\bibinfo {volume} {82}},\ \bibinfo {pages} {085011} (\bibinfo {year} {2010})},\ \Eprint {https://arxiv.org/abs/1008.1055} {arXiv:1008.1055 [hep-ph]} \BibitemShut {NoStop}%
\bibitem [{\citenamefont {Chernodub}(2011)}]{Chernodub:2011mc}%
  \BibitemOpen
  \bibfield  {author} {\bibinfo {author} {\bibfnamefont {M.~N.}\ \bibnamefont {Chernodub}},\ }\bibfield  {title} {\bibinfo {title} {{Spontaneous electromagnetic superconductivity of vacuum in strong magnetic field: evidence from the Nambu--Jona-Lasinio model}},\ }\href {https://doi.org/10.1103/PhysRevLett.106.142003} {\bibfield  {journal} {\bibinfo  {journal} {Phys. Rev. Lett.}\ }\textbf {\bibinfo {volume} {106}},\ \bibinfo {pages} {142003} (\bibinfo {year} {2011})},\ \Eprint {https://arxiv.org/abs/1101.0117} {arXiv:1101.0117 [hep-ph]} \BibitemShut {NoStop}%
\bibitem [{\citenamefont {Callebaut}\ \emph {et~al.}(2013)\citenamefont {Callebaut}, \citenamefont {Dudal},\ and\ \citenamefont {Verschelde}}]{Callebaut:2011ab}%
  \BibitemOpen
  \bibfield  {author} {\bibinfo {author} {\bibfnamefont {N.}~\bibnamefont {Callebaut}}, \bibinfo {author} {\bibfnamefont {D.}~\bibnamefont {Dudal}},\ and\ \bibinfo {author} {\bibfnamefont {H.}~\bibnamefont {Verschelde}},\ }\bibfield  {title} {\bibinfo {title} {{Holographic rho mesons in an external magnetic field}},\ }\href {https://doi.org/10.1007/JHEP03(2013)033} {\bibfield  {journal} {\bibinfo  {journal} {JHEP}\ }\textbf {\bibinfo {volume} {03}},\ \bibinfo {pages} {033}},\ \Eprint {https://arxiv.org/abs/1105.2217} {arXiv:1105.2217 [hep-th]} \BibitemShut {NoStop}%
\bibitem [{\citenamefont {Ammon}\ \emph {et~al.}(2011)\citenamefont {Ammon}, \citenamefont {Erdmenger}, \citenamefont {Kerner},\ and\ \citenamefont {Strydom}}]{Ammon:2011je}%
  \BibitemOpen
  \bibfield  {author} {\bibinfo {author} {\bibfnamefont {M.}~\bibnamefont {Ammon}}, \bibinfo {author} {\bibfnamefont {J.}~\bibnamefont {Erdmenger}}, \bibinfo {author} {\bibfnamefont {P.}~\bibnamefont {Kerner}},\ and\ \bibinfo {author} {\bibfnamefont {M.}~\bibnamefont {Strydom}},\ }\bibfield  {title} {\bibinfo {title} {{Black Hole Instability Induced by a Magnetic Field}},\ }\href {https://doi.org/10.1016/j.physletb.2011.10.067} {\bibfield  {journal} {\bibinfo  {journal} {Phys. Lett. B}\ }\textbf {\bibinfo {volume} {706}},\ \bibinfo {pages} {94} (\bibinfo {year} {2011})},\ \Eprint {https://arxiv.org/abs/1106.4551} {arXiv:1106.4551 [hep-th]} \BibitemShut {NoStop}%
\bibitem [{\citenamefont {Chernodub}\ \emph {et~al.}(2012)\citenamefont {Chernodub}, \citenamefont {Van~Doorsselaere},\ and\ \citenamefont {Verschelde}}]{Chernodub:2011gs}%
  \BibitemOpen
  \bibfield  {author} {\bibinfo {author} {\bibfnamefont {M.~N.}\ \bibnamefont {Chernodub}}, \bibinfo {author} {\bibfnamefont {J.}~\bibnamefont {Van~Doorsselaere}},\ and\ \bibinfo {author} {\bibfnamefont {H.}~\bibnamefont {Verschelde}},\ }\bibfield  {title} {\bibinfo {title} {{Electromagnetically superconducting phase of vacuum in strong magnetic field: structure of superconductor and superfluid vortex lattices in the ground state}},\ }\href {https://doi.org/10.1103/PhysRevD.85.045002} {\bibfield  {journal} {\bibinfo  {journal} {Phys. Rev. D}\ }\textbf {\bibinfo {volume} {85}},\ \bibinfo {pages} {045002} (\bibinfo {year} {2012})},\ \Eprint {https://arxiv.org/abs/1111.4401} {arXiv:1111.4401 [hep-ph]} \BibitemShut {NoStop}%
\bibitem [{\citenamefont {Andersen}(2012{\natexlab{a}})}]{Andersen:2012dz}%
  \BibitemOpen
  \bibfield  {author} {\bibinfo {author} {\bibfnamefont {J.~O.}\ \bibnamefont {Andersen}},\ }\bibfield  {title} {\bibinfo {title} {{Thermal pions in a magnetic background}},\ }\href {https://doi.org/10.1103/PhysRevD.86.025020} {\bibfield  {journal} {\bibinfo  {journal} {Phys. Rev. D}\ }\textbf {\bibinfo {volume} {86}},\ \bibinfo {pages} {025020} (\bibinfo {year} {2012}{\natexlab{a}})},\ \Eprint {https://arxiv.org/abs/1202.2051} {arXiv:1202.2051 [hep-ph]} \BibitemShut {NoStop}%
\bibitem [{\citenamefont {Andersen}(2012{\natexlab{b}})}]{Andersen:2012zc}%
  \BibitemOpen
  \bibfield  {author} {\bibinfo {author} {\bibfnamefont {J.~O.}\ \bibnamefont {Andersen}},\ }\bibfield  {title} {\bibinfo {title} {{Chiral perturbation theory in a magnetic background - finite-temperature effects}},\ }\href {https://doi.org/10.1007/JHEP10(2012)005} {\bibfield  {journal} {\bibinfo  {journal} {JHEP}\ }\textbf {\bibinfo {volume} {10}},\ \bibinfo {pages} {005}},\ \Eprint {https://arxiv.org/abs/1205.6978} {arXiv:1205.6978 [hep-ph]} \BibitemShut {NoStop}%
\bibitem [{\citenamefont {Kojo}\ and\ \citenamefont {Su}(2013)}]{Kojo:2012js}%
  \BibitemOpen
  \bibfield  {author} {\bibinfo {author} {\bibfnamefont {T.}~\bibnamefont {Kojo}}\ and\ \bibinfo {author} {\bibfnamefont {N.}~\bibnamefont {Su}},\ }\bibfield  {title} {\bibinfo {title} {{The quark mass gap in a magnetic field}},\ }\href {https://doi.org/10.1016/j.physletb.2013.02.024} {\bibfield  {journal} {\bibinfo  {journal} {Phys. Lett. B}\ }\textbf {\bibinfo {volume} {720}},\ \bibinfo {pages} {192} (\bibinfo {year} {2013})},\ \Eprint {https://arxiv.org/abs/1211.7318} {arXiv:1211.7318 [hep-ph]} \BibitemShut {NoStop}%
\bibitem [{\citenamefont {Andreichikov}\ \emph {et~al.}(2013)\citenamefont {Andreichikov}, \citenamefont {Kerbikov}, \citenamefont {Orlovsky},\ and\ \citenamefont {Simonov}}]{Andreichikov:2013zba}%
  \BibitemOpen
  \bibfield  {author} {\bibinfo {author} {\bibfnamefont {M.~A.}\ \bibnamefont {Andreichikov}}, \bibinfo {author} {\bibfnamefont {B.~O.}\ \bibnamefont {Kerbikov}}, \bibinfo {author} {\bibfnamefont {V.~D.}\ \bibnamefont {Orlovsky}},\ and\ \bibinfo {author} {\bibfnamefont {Y.~A.}\ \bibnamefont {Simonov}},\ }\bibfield  {title} {\bibinfo {title} {{Meson spectrum in strong magnetic fields}},\ }\href {https://doi.org/10.1103/PhysRevD.87.094029} {\bibfield  {journal} {\bibinfo  {journal} {Phys. Rev. D}\ }\textbf {\bibinfo {volume} {87}},\ \bibinfo {pages} {094029} (\bibinfo {year} {2013})},\ \Eprint {https://arxiv.org/abs/1304.2533} {arXiv:1304.2533 [hep-ph]} \BibitemShut {NoStop}%
\bibitem [{\citenamefont {Orlovsky}\ and\ \citenamefont {Simonov}(2013)}]{Orlovsky:2013gha}%
  \BibitemOpen
  \bibfield  {author} {\bibinfo {author} {\bibfnamefont {V.~D.}\ \bibnamefont {Orlovsky}}\ and\ \bibinfo {author} {\bibfnamefont {Y.~A.}\ \bibnamefont {Simonov}},\ }\bibfield  {title} {\bibinfo {title} {{Nambu-Goldstone mesons in strong magnetic field}},\ }\href {https://doi.org/10.1007/JHEP09(2013)136} {\bibfield  {journal} {\bibinfo  {journal} {JHEP}\ }\textbf {\bibinfo {volume} {09}},\ \bibinfo {pages} {136}},\ \Eprint {https://arxiv.org/abs/1306.2232} {arXiv:1306.2232 [hep-ph]} \BibitemShut {NoStop}%
\bibitem [{\citenamefont {Fayazbakhsh}\ and\ \citenamefont {Sadooghi}(2013)}]{Fayazbakhsh:2013cha}%
  \BibitemOpen
  \bibfield  {author} {\bibinfo {author} {\bibfnamefont {S.}~\bibnamefont {Fayazbakhsh}}\ and\ \bibinfo {author} {\bibfnamefont {N.}~\bibnamefont {Sadooghi}},\ }\bibfield  {title} {\bibinfo {title} {{Weak decay constant of neutral pions in a hot and magnetized quark matter}},\ }\href {https://doi.org/10.1103/PhysRevD.88.065030} {\bibfield  {journal} {\bibinfo  {journal} {Phys. Rev. D}\ }\textbf {\bibinfo {volume} {88}},\ \bibinfo {pages} {065030} (\bibinfo {year} {2013})},\ \Eprint {https://arxiv.org/abs/1306.2098} {arXiv:1306.2098 [hep-ph]} \BibitemShut {NoStop}%
\bibitem [{\citenamefont {Frasca}(2013)}]{Frasca:2013kka}%
  \BibitemOpen
  \bibfield  {author} {\bibinfo {author} {\bibfnamefont {M.}~\bibnamefont {Frasca}},\ }\bibfield  {title} {\bibinfo {title} {{$\rho$ condensation and physical parameters}},\ }\href {https://doi.org/10.1007/JHEP11(2013)099} {\bibfield  {journal} {\bibinfo  {journal} {JHEP}\ }\textbf {\bibinfo {volume} {11}},\ \bibinfo {pages} {099}},\ \Eprint {https://arxiv.org/abs/1309.3966} {arXiv:1309.3966 [hep-ph]} \BibitemShut {NoStop}%
\bibitem [{\citenamefont {Andreichikov}\ \emph {et~al.}(2014)\citenamefont {Andreichikov}, \citenamefont {Kerbikov}, \citenamefont {Orlovsky},\ and\ \citenamefont {Simonov}}]{Andreichikov:2013pga}%
  \BibitemOpen
  \bibfield  {author} {\bibinfo {author} {\bibfnamefont {M.~A.}\ \bibnamefont {Andreichikov}}, \bibinfo {author} {\bibfnamefont {B.~O.}\ \bibnamefont {Kerbikov}}, \bibinfo {author} {\bibfnamefont {V.~D.}\ \bibnamefont {Orlovsky}},\ and\ \bibinfo {author} {\bibfnamefont {Y.~A.}\ \bibnamefont {Simonov}},\ }\bibfield  {title} {\bibinfo {title} {{Neutron in Strong Magnetic Fields}},\ }\href {https://doi.org/10.1103/PhysRevD.89.074033} {\bibfield  {journal} {\bibinfo  {journal} {Phys. Rev. D}\ }\textbf {\bibinfo {volume} {89}},\ \bibinfo {pages} {074033} (\bibinfo {year} {2014})},\ \Eprint {https://arxiv.org/abs/1312.2212} {arXiv:1312.2212 [hep-ph]} \BibitemShut {NoStop}%
\bibitem [{\citenamefont {Kamikado}\ and\ \citenamefont {Kanazawa}(2014)}]{Kamikado:2013pya}%
  \BibitemOpen
  \bibfield  {author} {\bibinfo {author} {\bibfnamefont {K.}~\bibnamefont {Kamikado}}\ and\ \bibinfo {author} {\bibfnamefont {T.}~\bibnamefont {Kanazawa}},\ }\bibfield  {title} {\bibinfo {title} {{Chiral dynamics in a magnetic field from the functional renormalization group}},\ }\href {https://doi.org/10.1007/JHEP03(2014)009} {\bibfield  {journal} {\bibinfo  {journal} {JHEP}\ }\textbf {\bibinfo {volume} {03}},\ \bibinfo {pages} {009}},\ \Eprint {https://arxiv.org/abs/1312.3124} {arXiv:1312.3124 [hep-ph]} \BibitemShut {NoStop}%
\bibitem [{\citenamefont {Liu}\ \emph {et~al.}(2015)\citenamefont {Liu}, \citenamefont {Yu},\ and\ \citenamefont {Huang}}]{Liu:2014uwa}%
  \BibitemOpen
  \bibfield  {author} {\bibinfo {author} {\bibfnamefont {H.}~\bibnamefont {Liu}}, \bibinfo {author} {\bibfnamefont {L.}~\bibnamefont {Yu}},\ and\ \bibinfo {author} {\bibfnamefont {M.}~\bibnamefont {Huang}},\ }\bibfield  {title} {\bibinfo {title} {{Charged and neutral vector $\rho$ mesons in a magnetic field}},\ }\href {https://doi.org/10.1103/PhysRevD.91.014017} {\bibfield  {journal} {\bibinfo  {journal} {Phys. Rev. D}\ }\textbf {\bibinfo {volume} {91}},\ \bibinfo {pages} {014017} (\bibinfo {year} {2015})},\ \Eprint {https://arxiv.org/abs/1408.1318} {arXiv:1408.1318 [hep-ph]} \BibitemShut {NoStop}%
\bibitem [{\citenamefont {Haber}\ \emph {et~al.}(2014)\citenamefont {Haber}, \citenamefont {Preis},\ and\ \citenamefont {Schmitt}}]{Haber:2014ula}%
  \BibitemOpen
  \bibfield  {author} {\bibinfo {author} {\bibfnamefont {A.}~\bibnamefont {Haber}}, \bibinfo {author} {\bibfnamefont {F.}~\bibnamefont {Preis}},\ and\ \bibinfo {author} {\bibfnamefont {A.}~\bibnamefont {Schmitt}},\ }\bibfield  {title} {\bibinfo {title} {{Magnetic catalysis in nuclear matter}},\ }\href {https://doi.org/10.1103/PhysRevD.90.125036} {\bibfield  {journal} {\bibinfo  {journal} {Phys. Rev. D}\ }\textbf {\bibinfo {volume} {90}},\ \bibinfo {pages} {125036} (\bibinfo {year} {2014})},\ \Eprint {https://arxiv.org/abs/1409.0425} {arXiv:1409.0425 [nucl-th]} \BibitemShut {NoStop}%
\bibitem [{\citenamefont {Kamikado}\ and\ \citenamefont {Kanazawa}(2015)}]{Kamikado:2014bua}%
  \BibitemOpen
  \bibfield  {author} {\bibinfo {author} {\bibfnamefont {K.}~\bibnamefont {Kamikado}}\ and\ \bibinfo {author} {\bibfnamefont {T.}~\bibnamefont {Kanazawa}},\ }\bibfield  {title} {\bibinfo {title} {{Magnetic susceptibility of a strongly interacting thermal medium with 2$+$1 quark flavors}},\ }\href {https://doi.org/10.1007/JHEP01(2015)129} {\bibfield  {journal} {\bibinfo  {journal} {JHEP}\ }\textbf {\bibinfo {volume} {01}},\ \bibinfo {pages} {129}},\ \Eprint {https://arxiv.org/abs/1410.6253} {arXiv:1410.6253 [hep-ph]} \BibitemShut {NoStop}%
\bibitem [{\citenamefont {Taya}(2015)}]{Taya:2014nha}%
  \BibitemOpen
  \bibfield  {author} {\bibinfo {author} {\bibfnamefont {H.}~\bibnamefont {Taya}},\ }\bibfield  {title} {\bibinfo {title} {{Hadron Masses in Strong Magnetic Fields}},\ }\href {https://doi.org/10.1103/PhysRevD.92.014038} {\bibfield  {journal} {\bibinfo  {journal} {Phys. Rev. D}\ }\textbf {\bibinfo {volume} {92}},\ \bibinfo {pages} {014038} (\bibinfo {year} {2015})},\ \Eprint {https://arxiv.org/abs/1412.6877} {arXiv:1412.6877 [hep-ph]} \BibitemShut {NoStop}%
\bibitem [{\citenamefont {He}(2015)}]{He:2015zca}%
  \BibitemOpen
  \bibfield  {author} {\bibinfo {author} {\bibfnamefont {B.-R.}\ \bibnamefont {He}},\ }\bibfield  {title} {\bibinfo {title} {{Magnetic response of baryon properties in a Skyrme model}},\ }\href {https://doi.org/10.1103/PhysRevD.92.111503} {\bibfield  {journal} {\bibinfo  {journal} {Phys. Rev. D}\ }\textbf {\bibinfo {volume} {92}},\ \bibinfo {pages} {111503} (\bibinfo {year} {2015})},\ \Eprint {https://arxiv.org/abs/1510.04683} {arXiv:1510.04683 [hep-ph]} \BibitemShut {NoStop}%
\bibitem [{\citenamefont {Avancini}\ \emph {et~al.}(2016)\citenamefont {Avancini}, \citenamefont {Tavares},\ and\ \citenamefont {Pinto}}]{Avancini:2015ady}%
  \BibitemOpen
  \bibfield  {author} {\bibinfo {author} {\bibfnamefont {S.~S.}\ \bibnamefont {Avancini}}, \bibinfo {author} {\bibfnamefont {W.~R.}\ \bibnamefont {Tavares}},\ and\ \bibinfo {author} {\bibfnamefont {M.~B.}\ \bibnamefont {Pinto}},\ }\bibfield  {title} {\bibinfo {title} {{Properties of magnetized neutral mesons within a full RPA evaluation}},\ }\href {https://doi.org/10.1103/PhysRevD.93.014010} {\bibfield  {journal} {\bibinfo  {journal} {Phys. Rev. D}\ }\textbf {\bibinfo {volume} {93}},\ \bibinfo {pages} {014010} (\bibinfo {year} {2016})},\ \Eprint {https://arxiv.org/abs/1511.06261} {arXiv:1511.06261 [hep-ph]} \BibitemShut {NoStop}%
\bibitem [{\citenamefont {Kawaguchi}\ and\ \citenamefont {Matsuzaki}(2016)}]{Kawaguchi:2015gpt}%
  \BibitemOpen
  \bibfield  {author} {\bibinfo {author} {\bibfnamefont {M.}~\bibnamefont {Kawaguchi}}\ and\ \bibinfo {author} {\bibfnamefont {S.}~\bibnamefont {Matsuzaki}},\ }\bibfield  {title} {\bibinfo {title} {{Vector meson masses from a hidden local symmetry in a constant magnetic field}},\ }\href {https://doi.org/10.1103/PhysRevD.93.125027} {\bibfield  {journal} {\bibinfo  {journal} {Phys. Rev. D}\ }\textbf {\bibinfo {volume} {93}},\ \bibinfo {pages} {125027} (\bibinfo {year} {2016})},\ \Eprint {https://arxiv.org/abs/1511.06990} {arXiv:1511.06990 [hep-ph]} \BibitemShut {NoStop}%
\bibitem [{\citenamefont {Hattori}\ \emph {et~al.}(2016)\citenamefont {Hattori}, \citenamefont {Kojo},\ and\ \citenamefont {Su}}]{Hattori:2015aki}%
  \BibitemOpen
  \bibfield  {author} {\bibinfo {author} {\bibfnamefont {K.}~\bibnamefont {Hattori}}, \bibinfo {author} {\bibfnamefont {T.}~\bibnamefont {Kojo}},\ and\ \bibinfo {author} {\bibfnamefont {N.}~\bibnamefont {Su}},\ }\bibfield  {title} {\bibinfo {title} {{Mesons in strong magnetic fields: (I) General analyses}},\ }\href {https://doi.org/10.1016/j.nuclphysa.2016.03.016} {\bibfield  {journal} {\bibinfo  {journal} {Nucl. Phys. A}\ }\textbf {\bibinfo {volume} {951}},\ \bibinfo {pages} {1} (\bibinfo {year} {2016})},\ \Eprint {https://arxiv.org/abs/1512.07361} {arXiv:1512.07361 [hep-ph]} \BibitemShut {NoStop}%
\bibitem [{\citenamefont {Zhang}\ \emph {et~al.}(2016)\citenamefont {Zhang}, \citenamefont {Fu},\ and\ \citenamefont {Liu}}]{Zhang:2016qrl}%
  \BibitemOpen
  \bibfield  {author} {\bibinfo {author} {\bibfnamefont {R.}~\bibnamefont {Zhang}}, \bibinfo {author} {\bibfnamefont {W.-j.}\ \bibnamefont {Fu}},\ and\ \bibinfo {author} {\bibfnamefont {Y.-x.}\ \bibnamefont {Liu}},\ }\bibfield  {title} {\bibinfo {title} {{Properties of Mesons in a Strong Magnetic Field}},\ }\href {https://doi.org/10.1140/epjc/s10052-016-4123-8} {\bibfield  {journal} {\bibinfo  {journal} {Eur. Phys. J. C}\ }\textbf {\bibinfo {volume} {76}},\ \bibinfo {pages} {307} (\bibinfo {year} {2016})},\ \Eprint {https://arxiv.org/abs/1604.08888} {arXiv:1604.08888 [hep-ph]} \BibitemShut {NoStop}%
\bibitem [{\citenamefont {Avancini}\ \emph {et~al.}(2017)\citenamefont {Avancini}, \citenamefont {Farias}, \citenamefont {Benghi~Pinto}, \citenamefont {Tavares},\ and\ \citenamefont {Tim{\'o}teo}}]{Avancini:2016fgq}%
  \BibitemOpen
  \bibfield  {author} {\bibinfo {author} {\bibfnamefont {S.~S.}\ \bibnamefont {Avancini}}, \bibinfo {author} {\bibfnamefont {R.~L.~S.}\ \bibnamefont {Farias}}, \bibinfo {author} {\bibfnamefont {M.}~\bibnamefont {Benghi~Pinto}}, \bibinfo {author} {\bibfnamefont {W.~R.}\ \bibnamefont {Tavares}},\ and\ \bibinfo {author} {\bibfnamefont {V.~S.}\ \bibnamefont {Tim{\'o}teo}},\ }\bibfield  {title} {\bibinfo {title} {{$\pi_0$ pole mass calculation in a strong magnetic field and lattice constraints}},\ }\href {https://doi.org/10.1016/j.physletb.2017.02.002} {\bibfield  {journal} {\bibinfo  {journal} {Phys. Lett. B}\ }\textbf {\bibinfo {volume} {767}},\ \bibinfo {pages} {247} (\bibinfo {year} {2017})},\ \Eprint {https://arxiv.org/abs/1606.05754} {arXiv:1606.05754 [hep-ph]} \BibitemShut {NoStop}%
\bibitem [{\citenamefont {He}(2017)}]{He:2016oqk}%
  \BibitemOpen
  \bibfield  {author} {\bibinfo {author} {\bibfnamefont {B.-R.}\ \bibnamefont {He}},\ }\bibfield  {title} {\bibinfo {title} {{Skyrme model study of proton and neutron properties in a strong magnetic field}},\ }\href {https://doi.org/10.1016/j.physletb.2016.12.019} {\bibfield  {journal} {\bibinfo  {journal} {Phys. Lett. B}\ }\textbf {\bibinfo {volume} {765}},\ \bibinfo {pages} {109} (\bibinfo {year} {2017})},\ \Eprint {https://arxiv.org/abs/1609.09055} {arXiv:1609.09055 [hep-ph]} \BibitemShut {NoStop}%
\bibitem [{\citenamefont {Li}\ \emph {et~al.}(2017)\citenamefont {Li}, \citenamefont {Huang}, \citenamefont {Yang},\ and\ \citenamefont {Yuan}}]{Li:2016gfn}%
  \BibitemOpen
  \bibfield  {author} {\bibinfo {author} {\bibfnamefont {D.}~\bibnamefont {Li}}, \bibinfo {author} {\bibfnamefont {M.}~\bibnamefont {Huang}}, \bibinfo {author} {\bibfnamefont {Y.}~\bibnamefont {Yang}},\ and\ \bibinfo {author} {\bibfnamefont {P.-H.}\ \bibnamefont {Yuan}},\ }\bibfield  {title} {\bibinfo {title} {{Inverse Magnetic Catalysis in the Soft-Wall Model of AdS/QCD}},\ }\href {https://doi.org/10.1007/JHEP02(2017)030} {\bibfield  {journal} {\bibinfo  {journal} {JHEP}\ }\textbf {\bibinfo {volume} {02}},\ \bibinfo {pages} {030}},\ \Eprint {https://arxiv.org/abs/1610.04618} {arXiv:1610.04618 [hep-th]} \BibitemShut {NoStop}%
\bibitem [{\citenamefont {Andreichikov}\ \emph {et~al.}(2017)\citenamefont {Andreichikov}, \citenamefont {Kerbikov}, \citenamefont {Luschevskaya}, \citenamefont {Simonov},\ and\ \citenamefont {Solovjeva}}]{Andreichikov:2016ayj}%
  \BibitemOpen
  \bibfield  {author} {\bibinfo {author} {\bibfnamefont {M.~A.}\ \bibnamefont {Andreichikov}}, \bibinfo {author} {\bibfnamefont {B.~O.}\ \bibnamefont {Kerbikov}}, \bibinfo {author} {\bibfnamefont {E.~V.}\ \bibnamefont {Luschevskaya}}, \bibinfo {author} {\bibfnamefont {{\relax Yu}.~A.}\ \bibnamefont {Simonov}},\ and\ \bibinfo {author} {\bibfnamefont {O.~E.}\ \bibnamefont {Solovjeva}},\ }\bibfield  {title} {\bibinfo {title} {{The Evolution of Meson Masses in a Strong Magnetic Field}},\ }\href {https://doi.org/10.1007/JHEP05(2017)007} {\bibfield  {journal} {\bibinfo  {journal} {JHEP}\ }\textbf {\bibinfo {volume} {05}},\ \bibinfo {pages} {007}},\ \Eprint {https://arxiv.org/abs/1610.06887} {arXiv:1610.06887 [hep-ph]} \BibitemShut {NoStop}%
\bibitem [{\citenamefont {Kawaguchi}\ and\ \citenamefont {Matsuzaki}(2017)}]{Kawaguchi:2016gbf}%
  \BibitemOpen
  \bibfield  {author} {\bibinfo {author} {\bibfnamefont {M.}~\bibnamefont {Kawaguchi}}\ and\ \bibinfo {author} {\bibfnamefont {S.}~\bibnamefont {Matsuzaki}},\ }\bibfield  {title} {\bibinfo {title} {{Lifetime of rho meson in correlation with magnetic-dimensional reduction}},\ }\href {https://doi.org/10.1140/epja/i2017-12254-1} {\bibfield  {journal} {\bibinfo  {journal} {Eur. Phys. J. A}\ }\textbf {\bibinfo {volume} {53}},\ \bibinfo {pages} {68} (\bibinfo {year} {2017})},\ \Eprint {https://arxiv.org/abs/1610.08942} {arXiv:1610.08942 [hep-ph]} \BibitemShut {NoStop}%
\bibitem [{\citenamefont {Ghosh}\ \emph {et~al.}(2016)\citenamefont {Ghosh}, \citenamefont {Mukherjee}, \citenamefont {Mandal}, \citenamefont {Sarkar},\ and\ \citenamefont {Roy}}]{Ghosh:2016evc}%
  \BibitemOpen
  \bibfield  {author} {\bibinfo {author} {\bibfnamefont {S.}~\bibnamefont {Ghosh}}, \bibinfo {author} {\bibfnamefont {A.}~\bibnamefont {Mukherjee}}, \bibinfo {author} {\bibfnamefont {M.}~\bibnamefont {Mandal}}, \bibinfo {author} {\bibfnamefont {S.}~\bibnamefont {Sarkar}},\ and\ \bibinfo {author} {\bibfnamefont {P.}~\bibnamefont {Roy}},\ }\bibfield  {title} {\bibinfo {title} {{Spectral properties of $\rho$ meson in a magnetic field}},\ }\href {https://doi.org/10.1103/PhysRevD.94.094043} {\bibfield  {journal} {\bibinfo  {journal} {Phys. Rev. D}\ }\textbf {\bibinfo {volume} {94}},\ \bibinfo {pages} {094043} (\bibinfo {year} {2016})},\ \Eprint {https://arxiv.org/abs/1612.02966} {arXiv:1612.02966 [nucl-th]} \BibitemShut {NoStop}%
\bibitem [{\citenamefont {Mao}\ and\ \citenamefont {Wang}(2017)}]{Mao:2017wmq}%
  \BibitemOpen
  \bibfield  {author} {\bibinfo {author} {\bibfnamefont {S.}~\bibnamefont {Mao}}\ and\ \bibinfo {author} {\bibfnamefont {Y.}~\bibnamefont {Wang}},\ }\bibfield  {title} {\bibinfo {title} {{Effect of discrete quark momenta on the Goldstone mode in a magnetic field}},\ }\href {https://doi.org/10.1103/PhysRevD.96.034004} {\bibfield  {journal} {\bibinfo  {journal} {Phys. Rev. D}\ }\textbf {\bibinfo {volume} {96}},\ \bibinfo {pages} {034004} (\bibinfo {year} {2017})},\ \Eprint {https://arxiv.org/abs/1702.04868} {arXiv:1702.04868 [hep-ph]} \BibitemShut {NoStop}%
\bibitem [{\citenamefont {Ghosh}\ \emph {et~al.}(2017)\citenamefont {Ghosh}, \citenamefont {Mukherjee}, \citenamefont {Mandal}, \citenamefont {Sarkar},\ and\ \citenamefont {Roy}}]{Ghosh:2017rjo}%
  \BibitemOpen
  \bibfield  {author} {\bibinfo {author} {\bibfnamefont {S.}~\bibnamefont {Ghosh}}, \bibinfo {author} {\bibfnamefont {A.}~\bibnamefont {Mukherjee}}, \bibinfo {author} {\bibfnamefont {M.}~\bibnamefont {Mandal}}, \bibinfo {author} {\bibfnamefont {S.}~\bibnamefont {Sarkar}},\ and\ \bibinfo {author} {\bibfnamefont {P.}~\bibnamefont {Roy}},\ }\bibfield  {title} {\bibinfo {title} {{Thermal effects on $\rho$ meson properties in an external magnetic field}},\ }\href {https://doi.org/10.1103/PhysRevD.96.116020} {\bibfield  {journal} {\bibinfo  {journal} {Phys. Rev. D}\ }\textbf {\bibinfo {volume} {96}},\ \bibinfo {pages} {116020} (\bibinfo {year} {2017})},\ \Eprint {https://arxiv.org/abs/1704.05319} {arXiv:1704.05319 [hep-ph]} \BibitemShut {NoStop}%
\bibitem [{\citenamefont {G{\'o}mez~Dumm}\ \emph {et~al.}(2018)\citenamefont {G{\'o}mez~Dumm}, \citenamefont {Izzo~Villafa{\~n}e},\ and\ \citenamefont {Scoccola}}]{GomezDumm:2017jij}%
  \BibitemOpen
  \bibfield  {author} {\bibinfo {author} {\bibfnamefont {D.}~\bibnamefont {G{\'o}mez~Dumm}}, \bibinfo {author} {\bibfnamefont {M.~F.}\ \bibnamefont {Izzo~Villafa{\~n}e}},\ and\ \bibinfo {author} {\bibfnamefont {N.~N.}\ \bibnamefont {Scoccola}},\ }\bibfield  {title} {\bibinfo {title} {{Neutral meson properties under an external magnetic field in nonlocal chiral quark models}},\ }\href {https://doi.org/10.1103/PhysRevD.97.034025} {\bibfield  {journal} {\bibinfo  {journal} {Phys. Rev. D}\ }\textbf {\bibinfo {volume} {97}},\ \bibinfo {pages} {034025} (\bibinfo {year} {2018})},\ \Eprint {https://arxiv.org/abs/1710.08950} {arXiv:1710.08950 [hep-ph]} \BibitemShut {NoStop}%
\bibitem [{\citenamefont {Aguirre}(2017)}]{Aguirre:2017dht}%
  \BibitemOpen
  \bibfield  {author} {\bibinfo {author} {\bibfnamefont {R.~M.}\ \bibnamefont {Aguirre}},\ }\bibfield  {title} {\bibinfo {title} {{Modification of the masses of the lightest neutral mesons in a hadronic medium under an external magnetic field}},\ }\href {https://doi.org/10.1103/PhysRevD.96.096013} {\bibfield  {journal} {\bibinfo  {journal} {Phys. Rev. D}\ }\textbf {\bibinfo {volume} {96}},\ \bibinfo {pages} {096013} (\bibinfo {year} {2017})},\ \Eprint {https://arxiv.org/abs/1710.10707} {arXiv:1710.10707 [nucl-th]} \BibitemShut {NoStop}%
\bibitem [{\citenamefont {Wang}\ and\ \citenamefont {Zhuang}(2018)}]{Wang:2017vtn}%
  \BibitemOpen
  \bibfield  {author} {\bibinfo {author} {\bibfnamefont {Z.}~\bibnamefont {Wang}}\ and\ \bibinfo {author} {\bibfnamefont {P.}~\bibnamefont {Zhuang}},\ }\bibfield  {title} {\bibinfo {title} {{Meson properties in magnetized quark matter}},\ }\href {https://doi.org/10.1103/PhysRevD.97.034026} {\bibfield  {journal} {\bibinfo  {journal} {Phys. Rev. D}\ }\textbf {\bibinfo {volume} {97}},\ \bibinfo {pages} {034026} (\bibinfo {year} {2018})},\ \Eprint {https://arxiv.org/abs/1712.00554} {arXiv:1712.00554 [hep-ph]} \BibitemShut {NoStop}%
\bibitem [{\citenamefont {Liu}\ \emph {et~al.}(2018)\citenamefont {Liu}, \citenamefont {Wang}, \citenamefont {Yu},\ and\ \citenamefont {Huang}}]{Liu:2018zag}%
  \BibitemOpen
  \bibfield  {author} {\bibinfo {author} {\bibfnamefont {H.}~\bibnamefont {Liu}}, \bibinfo {author} {\bibfnamefont {X.}~\bibnamefont {Wang}}, \bibinfo {author} {\bibfnamefont {L.}~\bibnamefont {Yu}},\ and\ \bibinfo {author} {\bibfnamefont {M.}~\bibnamefont {Huang}},\ }\bibfield  {title} {\bibinfo {title} {{Neutral and charged scalar mesons, pseudoscalar mesons, and diquarks in magnetic fields}},\ }\href {https://doi.org/10.1103/PhysRevD.97.076008} {\bibfield  {journal} {\bibinfo  {journal} {Phys. Rev. D}\ }\textbf {\bibinfo {volume} {97}},\ \bibinfo {pages} {076008} (\bibinfo {year} {2018})},\ \Eprint {https://arxiv.org/abs/1801.02174} {arXiv:1801.02174 [hep-ph]} \BibitemShut {NoStop}%
\bibitem [{\citenamefont {Coppola}\ \emph {et~al.}(2018)\citenamefont {Coppola}, \citenamefont {G{\'o}mez~Dumm},\ and\ \citenamefont {Scoccola}}]{Coppola:2018vkw}%
  \BibitemOpen
  \bibfield  {author} {\bibinfo {author} {\bibfnamefont {M.}~\bibnamefont {Coppola}}, \bibinfo {author} {\bibfnamefont {D.}~\bibnamefont {G{\'o}mez~Dumm}},\ and\ \bibinfo {author} {\bibfnamefont {N.~N.}\ \bibnamefont {Scoccola}},\ }\bibfield  {title} {\bibinfo {title} {{Charged pion masses under strong magnetic fields in the NJL model}},\ }\href {https://doi.org/10.1016/j.physletb.2018.04.043} {\bibfield  {journal} {\bibinfo  {journal} {Phys. Lett. B}\ }\textbf {\bibinfo {volume} {782}},\ \bibinfo {pages} {155} (\bibinfo {year} {2018})},\ \Eprint {https://arxiv.org/abs/1802.08041} {arXiv:1802.08041 [hep-ph]} \BibitemShut {NoStop}%
\bibitem [{\citenamefont {Andreichikov}\ and\ \citenamefont {Simonov}(2018)}]{Andreichikov:2018wrc}%
  \BibitemOpen
  \bibfield  {author} {\bibinfo {author} {\bibfnamefont {M.~A.}\ \bibnamefont {Andreichikov}}\ and\ \bibinfo {author} {\bibfnamefont {Y.~A.}\ \bibnamefont {Simonov}},\ }\bibfield  {title} {\bibinfo {title} {{Chiral physics in the magnetic field with quark confinement contribution}},\ }\href {https://doi.org/10.1140/epjc/s10052-018-6384-x} {\bibfield  {journal} {\bibinfo  {journal} {Eur. Phys. J. C}\ }\textbf {\bibinfo {volume} {78}},\ \bibinfo {pages} {902} (\bibinfo {year} {2018})},\ \Eprint {https://arxiv.org/abs/1805.11896} {arXiv:1805.11896 [hep-ph]} \BibitemShut {NoStop}%
\bibitem [{\citenamefont {Mao}(2019)}]{Mao:2018dqe}%
  \BibitemOpen
  \bibfield  {author} {\bibinfo {author} {\bibfnamefont {S.}~\bibnamefont {Mao}},\ }\bibfield  {title} {\bibinfo {title} {{Pions in magnetic field at finite temperature}},\ }\href {https://doi.org/10.1103/PhysRevD.99.056005} {\bibfield  {journal} {\bibinfo  {journal} {Phys. Rev. D}\ }\textbf {\bibinfo {volume} {99}},\ \bibinfo {pages} {056005} (\bibinfo {year} {2019})},\ \Eprint {https://arxiv.org/abs/1808.10242} {arXiv:1808.10242 [nucl-th]} \BibitemShut {NoStop}%
\bibitem [{\citenamefont {Ayala}\ \emph {et~al.}(2018)\citenamefont {Ayala}, \citenamefont {Farias}, \citenamefont {Hern{\'a}ndez-Ortiz}, \citenamefont {Hern{\'a}ndez}, \citenamefont {Paret},\ and\ \citenamefont {Zamora}}]{Ayala:2018zat}%
  \BibitemOpen
  \bibfield  {author} {\bibinfo {author} {\bibfnamefont {A.}~\bibnamefont {Ayala}}, \bibinfo {author} {\bibfnamefont {R.~L.~S.}\ \bibnamefont {Farias}}, \bibinfo {author} {\bibfnamefont {S.}~\bibnamefont {Hern{\'a}ndez-Ortiz}}, \bibinfo {author} {\bibfnamefont {L.~A.}\ \bibnamefont {Hern{\'a}ndez}}, \bibinfo {author} {\bibfnamefont {D.~M.}\ \bibnamefont {Paret}},\ and\ \bibinfo {author} {\bibfnamefont {R.}~\bibnamefont {Zamora}},\ }\bibfield  {title} {\bibinfo {title} {{Magnetic field-dependence of the neutral pion mass in the linear sigma model coupled to quarks: The weak field case}},\ }\href {https://doi.org/10.1103/PhysRevD.98.114008} {\bibfield  {journal} {\bibinfo  {journal} {Phys. Rev. D}\ }\textbf {\bibinfo {volume} {98}},\ \bibinfo {pages} {114008} (\bibinfo {year} {2018})},\ \Eprint {https://arxiv.org/abs/1809.08312} {arXiv:1809.08312 [hep-ph]} \BibitemShut {NoStop}%
\bibitem [{\citenamefont {He}(2019)}]{He:2018vfc}%
  \BibitemOpen
  \bibfield  {author} {\bibinfo {author} {\bibfnamefont {B.-R.}\ \bibnamefont {He}},\ }\bibfield  {title} {\bibinfo {title} {{Magnetic field dependence of Delta isobars properties in a Skyrme model}},\ }\href {https://doi.org/10.1103/PhysRevD.99.034019} {\bibfield  {journal} {\bibinfo  {journal} {Phys. Rev. D}\ }\textbf {\bibinfo {volume} {99}},\ \bibinfo {pages} {034019} (\bibinfo {year} {2019})},\ \Eprint {https://arxiv.org/abs/1810.01289} {arXiv:1810.01289 [hep-ph]} \BibitemShut {NoStop}%
\bibitem [{\citenamefont {Coppola}\ \emph {et~al.}(2019)\citenamefont {Coppola}, \citenamefont {Gomez~Dumm}, \citenamefont {Noguera},\ and\ \citenamefont {Scoccola}}]{Coppola:2018ygv}%
  \BibitemOpen
  \bibfield  {author} {\bibinfo {author} {\bibfnamefont {M.}~\bibnamefont {Coppola}}, \bibinfo {author} {\bibfnamefont {D.}~\bibnamefont {Gomez~Dumm}}, \bibinfo {author} {\bibfnamefont {S.}~\bibnamefont {Noguera}},\ and\ \bibinfo {author} {\bibfnamefont {N.~N.}\ \bibnamefont {Scoccola}},\ }\bibfield  {title} {\bibinfo {title} {{Pion-to-vacuum vector and axial vector amplitudes and weak decays of pions in a magnetic field}},\ }\href {https://doi.org/10.1103/PhysRevD.99.054031} {\bibfield  {journal} {\bibinfo  {journal} {Phys. Rev. D}\ }\textbf {\bibinfo {volume} {99}},\ \bibinfo {pages} {054031} (\bibinfo {year} {2019})},\ \Eprint {https://arxiv.org/abs/1810.08110} {arXiv:1810.08110 [hep-ph]} \BibitemShut {NoStop}%
\bibitem [{\citenamefont {Avancini}\ \emph {et~al.}(2019)\citenamefont {Avancini}, \citenamefont {Farias},\ and\ \citenamefont {Tavares}}]{Avancini:2018svs}%
  \BibitemOpen
  \bibfield  {author} {\bibinfo {author} {\bibfnamefont {S.~S.}\ \bibnamefont {Avancini}}, \bibinfo {author} {\bibfnamefont {R.~L.~S.}\ \bibnamefont {Farias}},\ and\ \bibinfo {author} {\bibfnamefont {W.~R.}\ \bibnamefont {Tavares}},\ }\bibfield  {title} {\bibinfo {title} {{Neutral meson properties in hot and magnetized quark matter: a new magnetic field independent regularization scheme applied to NJL-type model}},\ }\href {https://doi.org/10.1103/PhysRevD.99.056009} {\bibfield  {journal} {\bibinfo  {journal} {Phys. Rev. D}\ }\textbf {\bibinfo {volume} {99}},\ \bibinfo {pages} {056009} (\bibinfo {year} {2019})},\ \Eprint {https://arxiv.org/abs/1812.00945} {arXiv:1812.00945 [hep-ph]} \BibitemShut {NoStop}%
\bibitem [{\citenamefont {Ghosh}\ \emph {et~al.}(2019)\citenamefont {Ghosh}, \citenamefont {Mukherjee}, \citenamefont {Roy},\ and\ \citenamefont {Sarkar}}]{Ghosh:2019fet}%
  \BibitemOpen
  \bibfield  {author} {\bibinfo {author} {\bibfnamefont {S.}~\bibnamefont {Ghosh}}, \bibinfo {author} {\bibfnamefont {A.}~\bibnamefont {Mukherjee}}, \bibinfo {author} {\bibfnamefont {P.}~\bibnamefont {Roy}},\ and\ \bibinfo {author} {\bibfnamefont {S.}~\bibnamefont {Sarkar}},\ }\bibfield  {title} {\bibinfo {title} {{General structure of the neutral {\ensuremath{\rho}} meson self-energy and its spectral properties in a hot and dense magnetized medium}},\ }\href {https://doi.org/10.1103/PhysRevD.99.096004} {\bibfield  {journal} {\bibinfo  {journal} {Phys. Rev. D}\ }\textbf {\bibinfo {volume} {99}},\ \bibinfo {pages} {096004} (\bibinfo {year} {2019})},\ \Eprint {https://arxiv.org/abs/1901.02290} {arXiv:1901.02290 [hep-ph]} \BibitemShut {NoStop}%
\bibitem [{\citenamefont {Yakhshiev}\ \emph {et~al.}(2019)\citenamefont {Yakhshiev}, \citenamefont {Kim},\ and\ \citenamefont {Oka}}]{Yakhshiev:2019gvb}%
  \BibitemOpen
  \bibfield  {author} {\bibinfo {author} {\bibfnamefont {U.}~\bibnamefont {Yakhshiev}}, \bibinfo {author} {\bibfnamefont {H.-C.}\ \bibnamefont {Kim}},\ and\ \bibinfo {author} {\bibfnamefont {M.}~\bibnamefont {Oka}},\ }\bibfield  {title} {\bibinfo {title} {{Nucleon and $\Delta$ isobar in a strong magnetic field}},\ }\href {https://doi.org/10.1103/PhysRevD.99.054027} {\bibfield  {journal} {\bibinfo  {journal} {Phys. Rev. D}\ }\textbf {\bibinfo {volume} {99}},\ \bibinfo {pages} {054027} (\bibinfo {year} {2019})},\ \Eprint {https://arxiv.org/abs/1902.00212} {arXiv:1902.00212 [hep-ph]} \BibitemShut {NoStop}%
\bibitem [{\citenamefont {Das}\ and\ \citenamefont {Haque}(2020)}]{Das:2019ehv}%
  \BibitemOpen
  \bibfield  {author} {\bibinfo {author} {\bibfnamefont {A.}~\bibnamefont {Das}}\ and\ \bibinfo {author} {\bibfnamefont {N.}~\bibnamefont {Haque}},\ }\bibfield  {title} {\bibinfo {title} {{Neutral pion mass in the linear sigma model coupled to quarks at arbitrary magnetic field}},\ }\href {https://doi.org/10.1103/PhysRevD.101.074033} {\bibfield  {journal} {\bibinfo  {journal} {Phys. Rev. D}\ }\textbf {\bibinfo {volume} {101}},\ \bibinfo {pages} {074033} (\bibinfo {year} {2020})},\ \Eprint {https://arxiv.org/abs/1908.10323} {arXiv:1908.10323 [hep-ph]} \BibitemShut {NoStop}%
\bibitem [{\citenamefont {G{\'o}mez~Dumm}\ \emph {et~al.}(2020)\citenamefont {G{\'o}mez~Dumm}, \citenamefont {Izzo~Villafa{\~n}e},\ and\ \citenamefont {Scoccola}}]{GomezDumm:2020bxj}%
  \BibitemOpen
  \bibfield  {author} {\bibinfo {author} {\bibfnamefont {D.}~\bibnamefont {G{\'o}mez~Dumm}}, \bibinfo {author} {\bibfnamefont {M.~F.}\ \bibnamefont {Izzo~Villafa{\~n}e}},\ and\ \bibinfo {author} {\bibfnamefont {N.~N.}\ \bibnamefont {Scoccola}},\ }\bibfield  {title} {\bibinfo {title} {{Properties of magnetized neutral pions at zero and finite temperature in nonlocal chiral quark models}},\ }\href {https://doi.org/10.1103/PhysRevD.101.116018} {\bibfield  {journal} {\bibinfo  {journal} {Phys. Rev. D}\ }\textbf {\bibinfo {volume} {101}},\ \bibinfo {pages} {116018} (\bibinfo {year} {2020})},\ \Eprint {https://arxiv.org/abs/2004.10052} {arXiv:2004.10052 [hep-ph]} \BibitemShut {NoStop}%
\bibitem [{\citenamefont {Li}\ \emph {et~al.}(2021)\citenamefont {Li}, \citenamefont {Cao},\ and\ \citenamefont {He}}]{Li:2020hlp}%
  \BibitemOpen
  \bibfield  {author} {\bibinfo {author} {\bibfnamefont {J.}~\bibnamefont {Li}}, \bibinfo {author} {\bibfnamefont {G.}~\bibnamefont {Cao}},\ and\ \bibinfo {author} {\bibfnamefont {L.}~\bibnamefont {He}},\ }\bibfield  {title} {\bibinfo {title} {{Gauge independence of pion masses in a magnetic field within the Nambu{\textendash}Jona-Lasinio model}},\ }\href {https://doi.org/10.1103/PhysRevD.104.074026} {\bibfield  {journal} {\bibinfo  {journal} {Phys. Rev. D}\ }\textbf {\bibinfo {volume} {104}},\ \bibinfo {pages} {074026} (\bibinfo {year} {2021})},\ \Eprint {https://arxiv.org/abs/2009.04697} {arXiv:2009.04697 [nucl-th]} \BibitemShut {NoStop}%
\bibitem [{\citenamefont {Coppola}\ \emph {et~al.}(2020)\citenamefont {Coppola}, \citenamefont {Gomez~Dumm},\ and\ \citenamefont {Scoccola}}]{Coppola:2020mon}%
  \BibitemOpen
  \bibfield  {author} {\bibinfo {author} {\bibfnamefont {M.}~\bibnamefont {Coppola}}, \bibinfo {author} {\bibfnamefont {D.}~\bibnamefont {Gomez~Dumm}},\ and\ \bibinfo {author} {\bibfnamefont {N.~N.}\ \bibnamefont {Scoccola}},\ }\bibfield  {title} {\bibinfo {title} {{Diquarks and nucleons under strong magnetic fields in the NJL model}},\ }\href {https://doi.org/10.1103/PhysRevD.102.094020} {\bibfield  {journal} {\bibinfo  {journal} {Phys. Rev. D}\ }\textbf {\bibinfo {volume} {102}},\ \bibinfo {pages} {094020} (\bibinfo {year} {2020})},\ \Eprint {https://arxiv.org/abs/2009.14105} {arXiv:2009.14105 [hep-ph]} \BibitemShut {NoStop}%
\bibitem [{\citenamefont {Sheng}\ \emph {et~al.}(2021)\citenamefont {Sheng}, \citenamefont {Wang}, \citenamefont {Wang},\ and\ \citenamefont {Yu}}]{Sheng:2020hge}%
  \BibitemOpen
  \bibfield  {author} {\bibinfo {author} {\bibfnamefont {B.}~\bibnamefont {Sheng}}, \bibinfo {author} {\bibfnamefont {Y.}~\bibnamefont {Wang}}, \bibinfo {author} {\bibfnamefont {X.}~\bibnamefont {Wang}},\ and\ \bibinfo {author} {\bibfnamefont {L.}~\bibnamefont {Yu}},\ }\bibfield  {title} {\bibinfo {title} {{Pole and screening masses of neutral pions in a hot and magnetized medium: A comprehensive study in the Nambu{\textendash}Jona-Lasinio model}},\ }\href {https://doi.org/10.1103/PhysRevD.103.094001} {\bibfield  {journal} {\bibinfo  {journal} {Phys. Rev. D}\ }\textbf {\bibinfo {volume} {103}},\ \bibinfo {pages} {094001} (\bibinfo {year} {2021})},\ \Eprint {https://arxiv.org/abs/2010.05716} {arXiv:2010.05716 [hep-ph]} \BibitemShut {NoStop}%
\bibitem [{\citenamefont {Ayala}\ \emph {et~al.}(2021)\citenamefont {Ayala}, \citenamefont {Hern{\'a}ndez}, \citenamefont {Hern{\'a}ndez}, \citenamefont {Farias},\ and\ \citenamefont {Zamora}}]{Ayala:2020dxs}%
  \BibitemOpen
  \bibfield  {author} {\bibinfo {author} {\bibfnamefont {A.}~\bibnamefont {Ayala}}, \bibinfo {author} {\bibfnamefont {J.~L.}\ \bibnamefont {Hern{\'a}ndez}}, \bibinfo {author} {\bibfnamefont {L.~A.}\ \bibnamefont {Hern{\'a}ndez}}, \bibinfo {author} {\bibfnamefont {R.~L.~S.}\ \bibnamefont {Farias}},\ and\ \bibinfo {author} {\bibfnamefont {R.}~\bibnamefont {Zamora}},\ }\bibfield  {title} {\bibinfo {title} {{Magnetic field dependence of the neutral pion mass in the linear sigma model with quarks: The strong field case}},\ }\href {https://doi.org/10.1103/PhysRevD.103.054038} {\bibfield  {journal} {\bibinfo  {journal} {Phys. Rev. D}\ }\textbf {\bibinfo {volume} {103}},\ \bibinfo {pages} {054038} (\bibinfo {year} {2021})},\ \Eprint {https://arxiv.org/abs/2011.03673} {arXiv:2011.03673 [hep-ph]} \BibitemShut {NoStop}%
\bibitem [{\citenamefont {Kojo}(2021)}]{Kojo:2021gvm}%
  \BibitemOpen
  \bibfield  {author} {\bibinfo {author} {\bibfnamefont {T.}~\bibnamefont {Kojo}},\ }\bibfield  {title} {\bibinfo {title} {{Neutral and charged mesons in magnetic fields: A resonance gas in a non-relativistic quark model}},\ }\href {https://doi.org/10.1140/epja/s10050-021-00629-y} {\bibfield  {journal} {\bibinfo  {journal} {Eur. Phys. J. A}\ }\textbf {\bibinfo {volume} {57}},\ \bibinfo {pages} {317} (\bibinfo {year} {2021})},\ \Eprint {https://arxiv.org/abs/2104.00376} {arXiv:2104.00376 [hep-ph]} \BibitemShut {NoStop}%
\bibitem [{\citenamefont {Moreira}\ and\ \citenamefont {Braghin}(2022)}]{Moreira:2022dwo}%
  \BibitemOpen
  \bibfield  {author} {\bibinfo {author} {\bibfnamefont {T.~H.}\ \bibnamefont {Moreira}}\ and\ \bibinfo {author} {\bibfnamefont {F.~L.}\ \bibnamefont {Braghin}},\ }\bibfield  {title} {\bibinfo {title} {{Magnetic field induced corrections to the NJL model coupling constant from vacuum polarization}},\ }\href {https://doi.org/10.1103/PhysRevD.105.114009} {\bibfield  {journal} {\bibinfo  {journal} {Phys. Rev. D}\ }\textbf {\bibinfo {volume} {105}},\ \bibinfo {pages} {114009} (\bibinfo {year} {2022})},\ \Eprint {https://arxiv.org/abs/2202.10559} {arXiv:2202.10559 [hep-ph]} \BibitemShut {NoStop}%
\bibitem [{\citenamefont {Carlomagno}\ \emph {et~al.}(2022{\natexlab{a}})\citenamefont {Carlomagno}, \citenamefont {Gomez~Dumm}, \citenamefont {Noguera},\ and\ \citenamefont {Scoccola}}]{Carlomagno:2022inu}%
  \BibitemOpen
  \bibfield  {author} {\bibinfo {author} {\bibfnamefont {J.~P.}\ \bibnamefont {Carlomagno}}, \bibinfo {author} {\bibfnamefont {D.}~\bibnamefont {Gomez~Dumm}}, \bibinfo {author} {\bibfnamefont {S.}~\bibnamefont {Noguera}},\ and\ \bibinfo {author} {\bibfnamefont {N.~N.}\ \bibnamefont {Scoccola}},\ }\bibfield  {title} {\bibinfo {title} {{Neutral pseudoscalar and vector meson masses under strong magnetic fields in an extended NJL model: Mixing effects}},\ }\href {https://doi.org/10.1103/PhysRevD.106.074002} {\bibfield  {journal} {\bibinfo  {journal} {Phys. Rev. D}\ }\textbf {\bibinfo {volume} {106}},\ \bibinfo {pages} {074002} (\bibinfo {year} {2022}{\natexlab{a}})},\ \Eprint {https://arxiv.org/abs/2205.15928} {arXiv:2205.15928 [hep-ph]} \BibitemShut {NoStop}%
\bibitem [{\citenamefont {Sheng}\ \emph {et~al.}(2024)\citenamefont {Sheng}, \citenamefont {Yang}, \citenamefont {Zou},\ and\ \citenamefont {Hou}}]{Sheng:2022ssp}%
  \BibitemOpen
  \bibfield  {author} {\bibinfo {author} {\bibfnamefont {X.-L.}\ \bibnamefont {Sheng}}, \bibinfo {author} {\bibfnamefont {S.-Y.}\ \bibnamefont {Yang}}, \bibinfo {author} {\bibfnamefont {Y.-L.}\ \bibnamefont {Zou}},\ and\ \bibinfo {author} {\bibfnamefont {D.}~\bibnamefont {Hou}},\ }\bibfield  {title} {\bibinfo {title} {{Mass splitting and spin alignment for $\phi $ mesons in a magnetic field in NJL model}},\ }\href {https://doi.org/10.1140/epjc/s10052-024-12643-7} {\bibfield  {journal} {\bibinfo  {journal} {Eur. Phys. J. C}\ }\textbf {\bibinfo {volume} {84}},\ \bibinfo {pages} {299} (\bibinfo {year} {2024})},\ \Eprint {https://arxiv.org/abs/2209.01872} {arXiv:2209.01872 [nucl-th]} \BibitemShut {NoStop}%
\bibitem [{\citenamefont {Carlomagno}\ \emph {et~al.}(2022{\natexlab{b}})\citenamefont {Carlomagno}, \citenamefont {Gomez~Dumm}, \citenamefont {Villafa{\~n}e}, \citenamefont {Noguera},\ and\ \citenamefont {Scoccola}}]{Carlomagno:2022arc}%
  \BibitemOpen
  \bibfield  {author} {\bibinfo {author} {\bibfnamefont {J.~P.}\ \bibnamefont {Carlomagno}}, \bibinfo {author} {\bibfnamefont {D.}~\bibnamefont {Gomez~Dumm}}, \bibinfo {author} {\bibfnamefont {M.~F.~I.}\ \bibnamefont {Villafa{\~n}e}}, \bibinfo {author} {\bibfnamefont {S.}~\bibnamefont {Noguera}},\ and\ \bibinfo {author} {\bibfnamefont {N.~N.}\ \bibnamefont {Scoccola}},\ }\bibfield  {title} {\bibinfo {title} {{Charged pseudoscalar and vector meson masses in strong magnetic fields in an extended NJL model}},\ }\href {https://doi.org/10.1103/PhysRevD.106.094035} {\bibfield  {journal} {\bibinfo  {journal} {Phys. Rev. D}\ }\textbf {\bibinfo {volume} {106}},\ \bibinfo {pages} {094035} (\bibinfo {year} {2022}{\natexlab{b}})},\ \Eprint {https://arxiv.org/abs/2209.10679} {arXiv:2209.10679 [hep-ph]} \BibitemShut {NoStop}%
\bibitem [{\citenamefont {Mei}\ \emph {et~al.}(2023)\citenamefont {Mei}, \citenamefont {Xia},\ and\ \citenamefont {Mao}}]{Mei:2022dkd}%
  \BibitemOpen
  \bibfield  {author} {\bibinfo {author} {\bibfnamefont {J.}~\bibnamefont {Mei}}, \bibinfo {author} {\bibfnamefont {T.}~\bibnamefont {Xia}},\ and\ \bibinfo {author} {\bibfnamefont {S.}~\bibnamefont {Mao}},\ }\bibfield  {title} {\bibinfo {title} {Mass spectra of neutral mesons ${K}_{0},{\ensuremath{\pi}}_{0},\ensuremath{\eta},{\ensuremath{\eta}}^{\ensuremath{'}}$ at finite magnetic field, temperature and quark chemical potential},\ }\href {https://doi.org/10.1103/PhysRevD.107.074018} {\bibfield  {journal} {\bibinfo  {journal} {Phys. Rev. D}\ }\textbf {\bibinfo {volume} {107}},\ \bibinfo {pages} {074018} (\bibinfo {year} {2023})},\ \bibinfo {note} {[Erratum: Phys.Rev.D 110, 119901 (2024)]},\ \Eprint {https://arxiv.org/abs/2212.04778} {arXiv:2212.04778 [hep-ph]} \BibitemShut {NoStop}%
\bibitem [{\citenamefont {Braghin}(2023)}]{Braghin:2022uih}%
  \BibitemOpen
  \bibfield  {author} {\bibinfo {author} {\bibfnamefont {F.~L.}\ \bibnamefont {Braghin}},\ }\bibfield  {title} {\bibinfo {title} {{Quark-antiquark states of the lightest scalar mesons within the Nambu-Jona-Lasinio model with flavor-dependent coupling constants}},\ }\href {https://doi.org/10.1088/1361-6471/acdaea} {\bibfield  {journal} {\bibinfo  {journal} {J. Phys. G}\ }\textbf {\bibinfo {volume} {50}},\ \bibinfo {pages} {095101} (\bibinfo {year} {2023})},\ \Eprint {https://arxiv.org/abs/2212.06616} {arXiv:2212.06616 [hep-ph]} \BibitemShut {NoStop}%
\bibitem [{\citenamefont {Chen}\ and\ \citenamefont {Jiang}(2023)}]{Chen:2023gws}%
  \BibitemOpen
  \bibfield  {author} {\bibinfo {author} {\bibfnamefont {Y.}~\bibnamefont {Chen}}\ and\ \bibinfo {author} {\bibfnamefont {W.}~\bibnamefont {Jiang}},\ }\bibfield  {title} {\bibinfo {title} {{Nucleons and {\ensuremath{\Delta}} isobars under strong magnetic fields in Skyrme models}},\ }\href {https://doi.org/10.1103/PhysRevC.107.025201} {\bibfield  {journal} {\bibinfo  {journal} {Phys. Rev. C}\ }\textbf {\bibinfo {volume} {107}},\ \bibinfo {pages} {025201} (\bibinfo {year} {2023})}\BibitemShut {NoStop}%
\bibitem [{\citenamefont {Chen}\ \emph {et~al.}(2023)\citenamefont {Chen}, \citenamefont {Fukushima},\ and\ \citenamefont {Qiu}}]{Chen:2023jbq}%
  \BibitemOpen
  \bibfield  {author} {\bibinfo {author} {\bibfnamefont {S.}~\bibnamefont {Chen}}, \bibinfo {author} {\bibfnamefont {K.}~\bibnamefont {Fukushima}},\ and\ \bibinfo {author} {\bibfnamefont {Z.}~\bibnamefont {Qiu}},\ }\bibfield  {title} {\bibinfo {title} {{Magnetic enhancement of baryon confinement modeled via a deformed Skyrmion}},\ }\href {https://doi.org/10.1016/j.physletb.2023.137992} {\bibfield  {journal} {\bibinfo  {journal} {Phys. Lett. B}\ }\textbf {\bibinfo {volume} {843}},\ \bibinfo {pages} {137992} (\bibinfo {year} {2023})},\ \Eprint {https://arxiv.org/abs/2303.04692} {arXiv:2303.04692 [hep-th]} \BibitemShut {NoStop}%
\bibitem [{\citenamefont {Li}\ and\ \citenamefont {Mao}(2023)}]{Li:2023rsy}%
  \BibitemOpen
  \bibfield  {author} {\bibinfo {author} {\bibfnamefont {L.}~\bibnamefont {Li}}\ and\ \bibinfo {author} {\bibfnamefont {S.}~\bibnamefont {Mao}},\ }\bibfield  {title} {\bibinfo {title} {{Inverse magnetic catalysis effect and current quark mass effect on mass spectra and Mott transitions of pions under external magnetic field}},\ }\href {https://doi.org/10.1103/PhysRevD.108.054001} {\bibfield  {journal} {\bibinfo  {journal} {Phys. Rev. D}\ }\textbf {\bibinfo {volume} {108}},\ \bibinfo {pages} {054001} (\bibinfo {year} {2023})},\ \Eprint {https://arxiv.org/abs/2308.12491} {arXiv:2308.12491 [hep-ph]} \BibitemShut {NoStop}%
\bibitem [{\citenamefont {Ayala}\ \emph {et~al.}(2024)\citenamefont {Ayala}, \citenamefont {Farias}, \citenamefont {Hern{\'a}ndez}, \citenamefont {Mizher}, \citenamefont {Rend{\'o}n}, \citenamefont {Villavicencio},\ and\ \citenamefont {Zamora}}]{Ayala:2023llp}%
  \BibitemOpen
  \bibfield  {author} {\bibinfo {author} {\bibfnamefont {A.}~\bibnamefont {Ayala}}, \bibinfo {author} {\bibfnamefont {R.~L.~S.}\ \bibnamefont {Farias}}, \bibinfo {author} {\bibfnamefont {L.~A.}\ \bibnamefont {Hern{\'a}ndez}}, \bibinfo {author} {\bibfnamefont {A.~J.}\ \bibnamefont {Mizher}}, \bibinfo {author} {\bibfnamefont {J.}~\bibnamefont {Rend{\'o}n}}, \bibinfo {author} {\bibfnamefont {C.}~\bibnamefont {Villavicencio}},\ and\ \bibinfo {author} {\bibfnamefont {R.}~\bibnamefont {Zamora}},\ }\bibfield  {title} {\bibinfo {title} {{Magnetic field dependence of the neutral pion longitudinal screening mass in the linear sigma model with quarks}},\ }\href {https://doi.org/10.1103/PhysRevD.109.074019} {\bibfield  {journal} {\bibinfo  {journal} {Phys. Rev. D}\ }\textbf {\bibinfo {volume} {109}},\ \bibinfo {pages} {074019} (\bibinfo {year} {2024})},\ \Eprint {https://arxiv.org/abs/2311.13068} {arXiv:2311.13068 [hep-ph]} \BibitemShut {NoStop}%
\bibitem [{\citenamefont {Coppola}\ \emph {et~al.}(2024{\natexlab{a}})\citenamefont {Coppola}, \citenamefont {Gomez~Dumm}, \citenamefont {Noguera},\ and\ \citenamefont {Scoccola}}]{Coppola:2023mmq}%
  \BibitemOpen
  \bibfield  {author} {\bibinfo {author} {\bibfnamefont {M.}~\bibnamefont {Coppola}}, \bibinfo {author} {\bibfnamefont {D.}~\bibnamefont {Gomez~Dumm}}, \bibinfo {author} {\bibfnamefont {S.}~\bibnamefont {Noguera}},\ and\ \bibinfo {author} {\bibfnamefont {N.~N.}\ \bibnamefont {Scoccola}},\ }\bibfield  {title} {\bibinfo {title} {{Masses of magnetized pseudoscalar and vector mesons in an extended NJL model: The role of axial vector mesons}},\ }\href {https://doi.org/10.1103/PhysRevD.109.054014} {\bibfield  {journal} {\bibinfo  {journal} {Phys. Rev. D}\ }\textbf {\bibinfo {volume} {109}},\ \bibinfo {pages} {054014} (\bibinfo {year} {2024}{\natexlab{a}})},\ \Eprint {https://arxiv.org/abs/2312.16675} {arXiv:2312.16675 [hep-ph]} \BibitemShut {NoStop}%
\bibitem [{\citenamefont {Mei}\ \emph {et~al.}(2024)\citenamefont {Mei}, \citenamefont {Wen}, \citenamefont {Mao}, \citenamefont {Huang},\ and\ \citenamefont {Xu}}]{Mei:2024rjg}%
  \BibitemOpen
  \bibfield  {author} {\bibinfo {author} {\bibfnamefont {J.}~\bibnamefont {Mei}}, \bibinfo {author} {\bibfnamefont {R.}~\bibnamefont {Wen}}, \bibinfo {author} {\bibfnamefont {S.}~\bibnamefont {Mao}}, \bibinfo {author} {\bibfnamefont {M.}~\bibnamefont {Huang}},\ and\ \bibinfo {author} {\bibfnamefont {K.}~\bibnamefont {Xu}},\ }\bibfield  {title} {\bibinfo {title} {{Magnetic catalysis and diamagnetism from pion fluctuations}},\ }\href {https://doi.org/10.1103/PhysRevD.110.034024} {\bibfield  {journal} {\bibinfo  {journal} {Phys. Rev. D}\ }\textbf {\bibinfo {volume} {110}},\ \bibinfo {pages} {034024} (\bibinfo {year} {2024})},\ \Eprint {https://arxiv.org/abs/2402.19193} {arXiv:2402.19193 [hep-ph]} \BibitemShut {NoStop}%
\bibitem [{\citenamefont {Coppola}\ \emph {et~al.}(2024{\natexlab{b}})\citenamefont {Coppola}, \citenamefont {Tavares}, \citenamefont {Avancini}, \citenamefont {Sodr{\'e}},\ and\ \citenamefont {Scoccola}}]{Coppola:2024uvz}%
  \BibitemOpen
  \bibfield  {author} {\bibinfo {author} {\bibfnamefont {M.}~\bibnamefont {Coppola}}, \bibinfo {author} {\bibfnamefont {W.~R.}\ \bibnamefont {Tavares}}, \bibinfo {author} {\bibfnamefont {S.~S.}\ \bibnamefont {Avancini}}, \bibinfo {author} {\bibfnamefont {J.~C.}\ \bibnamefont {Sodr{\'e}}},\ and\ \bibinfo {author} {\bibfnamefont {N.~N.}\ \bibnamefont {Scoccola}},\ }\bibfield  {title} {\bibinfo {title} {{Thermomagnetic effects on light pseudoscalar meson masses within the SU(3) Nambu{\textendash}Jona-Lasinio model}},\ }\href {https://doi.org/10.1103/PhysRevD.110.114036} {\bibfield  {journal} {\bibinfo  {journal} {Phys. Rev. D}\ }\textbf {\bibinfo {volume} {110}},\ \bibinfo {pages} {114036} (\bibinfo {year} {2024}{\natexlab{b}})},\ \Eprint {https://arxiv.org/abs/2410.05568} {arXiv:2410.05568 [hep-ph]} \BibitemShut {NoStop}%
\bibitem [{\citenamefont {Hern{\'a}ndez}\ \emph {et~al.}(2025)\citenamefont {Hern{\'a}ndez}, \citenamefont {Mart{\'\i}nez-S{\'a}nchez},\ and\ \citenamefont {Zamora}}]{Hernandez:2025inu}%
  \BibitemOpen
  \bibfield  {author} {\bibinfo {author} {\bibfnamefont {L.~A.}\ \bibnamefont {Hern{\'a}ndez}}, \bibinfo {author} {\bibfnamefont {J.~D.}\ \bibnamefont {Mart{\'\i}nez-S{\'a}nchez}},\ and\ \bibinfo {author} {\bibfnamefont {R.}~\bibnamefont {Zamora}},\ }\bibfield  {title} {\bibinfo {title} {{Rho-meson screening mass in the presence of strong magnetic fields}},\ }\href {https://doi.org/10.1103/PhysRevD.111.096019} {\bibfield  {journal} {\bibinfo  {journal} {Phys. Rev. D}\ }\textbf {\bibinfo {volume} {111}},\ \bibinfo {pages} {096019} (\bibinfo {year} {2025})},\ \Eprint {https://arxiv.org/abs/2502.08051} {arXiv:2502.08051 [hep-ph]} \BibitemShut {NoStop}%
\bibitem [{\citenamefont {Coppola}\ \emph {et~al.}(2025)\citenamefont {Coppola}, \citenamefont {Gomez~Dumm},\ and\ \citenamefont {Scoccola}}]{Coppola:2025nus}%
  \BibitemOpen
  \bibfield  {author} {\bibinfo {author} {\bibfnamefont {M.}~\bibnamefont {Coppola}}, \bibinfo {author} {\bibfnamefont {D.}~\bibnamefont {Gomez~Dumm}},\ and\ \bibinfo {author} {\bibfnamefont {N.~N.}\ \bibnamefont {Scoccola}},\ }\bibfield  {title} {\bibinfo {title} {${\ensuremath{\pi}}^{0}\ensuremath{\rightarrow}2\ensuremath{\gamma}$ decay under strong magnetic fields in the njl model},\ }\href {https://doi.org/10.1103/3hqb-28zb} {\bibfield  {journal} {\bibinfo  {journal} {Phys. Rev. D}\ }\textbf {\bibinfo {volume} {112}},\ \bibinfo {pages} {054043} (\bibinfo {year} {2025})},\ \Eprint {https://arxiv.org/abs/2507.13560} {arXiv:2507.13560 [hep-ph]} \BibitemShut {NoStop}%
\bibitem [{\citenamefont {Mei}\ \emph {et~al.}(2026)\citenamefont {Mei}, \citenamefont {Wen}, \citenamefont {Zhou}, \citenamefont {Mao},\ and\ \citenamefont {Huang}}]{Mei:2026xlj}%
  \BibitemOpen
  \bibfield  {author} {\bibinfo {author} {\bibfnamefont {J.}~\bibnamefont {Mei}}, \bibinfo {author} {\bibfnamefont {R.}~\bibnamefont {Wen}}, \bibinfo {author} {\bibfnamefont {M.}~\bibnamefont {Zhou}}, \bibinfo {author} {\bibfnamefont {S.}~\bibnamefont {Mao}},\ and\ \bibinfo {author} {\bibfnamefont {M.}~\bibnamefont {Huang}},\ }\bibfield  {title} {\bibinfo {title} {{Spectral function for pions in a magnetic field}},\ }\href {https://doi.org/10.1103/skmg-ql8c} {\bibfield  {journal} {\bibinfo  {journal} {Phys. Rev. D}\ }\textbf {\bibinfo {volume} {113}},\ \bibinfo {pages} {074031} (\bibinfo {year} {2026})},\ \Eprint {https://arxiv.org/abs/2601.22422} {arXiv:2601.22422 [hep-ph]} \BibitemShut {NoStop}%
\bibitem [{\citenamefont {Navas}\ \emph {et~al.}(2024)\citenamefont {Navas} \emph {et~al.}}]{ParticleDataGroup:2024cfk}%
  \BibitemOpen
  \bibfield  {author} {\bibinfo {author} {\bibfnamefont {S.}~\bibnamefont {Navas}} \emph {et~al.} (\bibinfo {collaboration} {Particle Data Group}),\ }\bibfield  {title} {\bibinfo {title} {{Review of particle physics}},\ }\href {https://doi.org/10.1103/PhysRevD.110.030001} {\bibfield  {journal} {\bibinfo  {journal} {Phys. Rev. D}\ }\textbf {\bibinfo {volume} {110}},\ \bibinfo {pages} {030001} (\bibinfo {year} {2024})}\BibitemShut {NoStop}%
\bibitem [{\citenamefont {Mostafazadeh}(2002{\natexlab{a}})}]{Mostafazadeh:2001jk}%
  \BibitemOpen
  \bibfield  {author} {\bibinfo {author} {\bibfnamefont {A.}~\bibnamefont {Mostafazadeh}},\ }\bibfield  {title} {\bibinfo {title} {{Pseudo-Hermiticity versus $PT$ symmetry: The necessary condition for the reality of the spectrum of a non-Hermitian Hamiltonian}},\ }\href {https://doi.org/10.1063/1.1418246} {\bibfield  {journal} {\bibinfo  {journal} {J. Math. Phys.}\ }\textbf {\bibinfo {volume} {43}},\ \bibinfo {pages} {205} (\bibinfo {year} {2002}{\natexlab{a}})},\ \Eprint {https://arxiv.org/abs/math-ph/0107001} {arXiv:math-ph/0107001} \BibitemShut {NoStop}%
\bibitem [{\citenamefont {Mostafazadeh}(2002{\natexlab{b}})}]{Mostafazadeh:2001nr}%
  \BibitemOpen
  \bibfield  {author} {\bibinfo {author} {\bibfnamefont {A.}~\bibnamefont {Mostafazadeh}},\ }\bibfield  {title} {\bibinfo {title} {{Pseudo-Hermiticity versus $PT$ symmetry. II. A Complete characterization of non-Hermitian Hamiltonians with a real spectrum}},\ }\href {https://doi.org/10.1063/1.1461427} {\bibfield  {journal} {\bibinfo  {journal} {J. Math. Phys.}\ }\textbf {\bibinfo {volume} {43}},\ \bibinfo {pages} {2814} (\bibinfo {year} {2002}{\natexlab{b}})},\ \Eprint {https://arxiv.org/abs/math-ph/0110016} {arXiv:math-ph/0110016} \BibitemShut {NoStop}%
\bibitem [{\citenamefont {Mostafazadeh}(2002{\natexlab{c}})}]{Mostafazadeh:2002id}%
  \BibitemOpen
  \bibfield  {author} {\bibinfo {author} {\bibfnamefont {A.}~\bibnamefont {Mostafazadeh}},\ }\bibfield  {title} {\bibinfo {title} {{Pseudo-Hermiticity versus $PT$-symmetry III: Equivalence of pseudo-Hermiticity and the presence of antilinear symmetries}},\ }\href {https://doi.org/10.1063/1.1489072} {\bibfield  {journal} {\bibinfo  {journal} {J. Math. Phys.}\ }\textbf {\bibinfo {volume} {43}},\ \bibinfo {pages} {3944} (\bibinfo {year} {2002}{\natexlab{c}})},\ \Eprint {https://arxiv.org/abs/math-ph/0203005} {arXiv:math-ph/0203005} \BibitemShut {NoStop}%
\bibitem [{Note4()}]{Note4}%
  \BibitemOpen
  \bibinfo {note} {Formally, for the discriminant in Eq.~\protect \eqref {eq:eigenvalues}, there is another solution but located at much stronger field region at $eB_I^{\protect \mathrm {EP}} = {(m_{J/\psi } + m_{\eta _c})^2}/{2g_{J/\psi \eta _c\gamma }}$.}\BibitemShut {Stop}%
\bibitem [{\citenamefont {Barnes}\ \emph {et~al.}(2005)\citenamefont {Barnes}, \citenamefont {Godfrey},\ and\ \citenamefont {Swanson}}]{Barnes:2005pb}%
  \BibitemOpen
  \bibfield  {author} {\bibinfo {author} {\bibfnamefont {T.}~\bibnamefont {Barnes}}, \bibinfo {author} {\bibfnamefont {S.}~\bibnamefont {Godfrey}},\ and\ \bibinfo {author} {\bibfnamefont {E.~S.}\ \bibnamefont {Swanson}},\ }\bibfield  {title} {\bibinfo {title} {{Higher charmonia}},\ }\href {https://doi.org/10.1103/PhysRevD.72.054026} {\bibfield  {journal} {\bibinfo  {journal} {Phys. Rev. D}\ }\textbf {\bibinfo {volume} {72}},\ \bibinfo {pages} {054026} (\bibinfo {year} {2005})},\ \Eprint {https://arxiv.org/abs/hep-ph/0505002} {arXiv:hep-ph/0505002 [hep-ph]} \BibitemShut {NoStop}%
\bibitem [{\citenamefont {Bender}\ and\ \citenamefont {Boettcher}(1998)}]{Bender:1998ke}%
  \BibitemOpen
  \bibfield  {author} {\bibinfo {author} {\bibfnamefont {C.~M.}\ \bibnamefont {Bender}}\ and\ \bibinfo {author} {\bibfnamefont {S.}~\bibnamefont {Boettcher}},\ }\bibfield  {title} {\bibinfo {title} {{Real spectra in non-Hermitian Hamiltonians having $PT$ symmetry}},\ }\href {https://doi.org/10.1103/PhysRevLett.80.5243} {\bibfield  {journal} {\bibinfo  {journal} {Phys. Rev. Lett.}\ }\textbf {\bibinfo {volume} {80}},\ \bibinfo {pages} {5243} (\bibinfo {year} {1998})},\ \Eprint {https://arxiv.org/abs/physics/9712001} {arXiv:physics/9712001} \BibitemShut {NoStop}%
\bibitem [{\citenamefont {Bender}\ \emph {et~al.}(2002)\citenamefont {Bender}, \citenamefont {Brody},\ and\ \citenamefont {Jones}}]{Bender:2002vv}%
  \BibitemOpen
  \bibfield  {author} {\bibinfo {author} {\bibfnamefont {C.~M.}\ \bibnamefont {Bender}}, \bibinfo {author} {\bibfnamefont {D.~C.}\ \bibnamefont {Brody}},\ and\ \bibinfo {author} {\bibfnamefont {H.~F.}\ \bibnamefont {Jones}},\ }\bibfield  {title} {\bibinfo {title} {{Complex extension of quantum mechanics}},\ }\href {https://doi.org/10.1103/PhysRevLett.89.270401} {\bibfield  {journal} {\bibinfo  {journal} {Phys. Rev. Lett.}\ }\textbf {\bibinfo {volume} {89}},\ \bibinfo {pages} {270401} (\bibinfo {year} {2002})},\ \bibinfo {note} {[Erratum: Phys.Rev.Lett. 92, 119902 (2004)]},\ \Eprint {https://arxiv.org/abs/quant-ph/0208076} {arXiv:quant-ph/0208076} \BibitemShut {NoStop}%
\bibitem [{\citenamefont {Matsui}\ and\ \citenamefont {Satz}(1986)}]{Matsui:1986dk}%
  \BibitemOpen
  \bibfield  {author} {\bibinfo {author} {\bibfnamefont {T.}~\bibnamefont {Matsui}}\ and\ \bibinfo {author} {\bibfnamefont {H.}~\bibnamefont {Satz}},\ }\bibfield  {title} {\bibinfo {title} {{$J/\psi$ suppression by quark-gluon plasma formation}},\ }\href {https://doi.org/10.1016/0370-2693(86)91404-8} {\bibfield  {journal} {\bibinfo  {journal} {Phys. Lett. B}\ }\textbf {\bibinfo {volume} {178}},\ \bibinfo {pages} {416} (\bibinfo {year} {1986})}\BibitemShut {NoStop}%
\bibitem [{\citenamefont {Hashimoto}\ \emph {et~al.}(1986)\citenamefont {Hashimoto}, \citenamefont {Hirose}, \citenamefont {Kanki},\ and\ \citenamefont {Miyamura}}]{Hashimoto:1986nn}%
  \BibitemOpen
  \bibfield  {author} {\bibinfo {author} {\bibfnamefont {T.}~\bibnamefont {Hashimoto}}, \bibinfo {author} {\bibfnamefont {K.}~\bibnamefont {Hirose}}, \bibinfo {author} {\bibfnamefont {T.}~\bibnamefont {Kanki}},\ and\ \bibinfo {author} {\bibfnamefont {O.}~\bibnamefont {Miyamura}},\ }\bibfield  {title} {\bibinfo {title} {{Mass shift of charmonium near deconfining temperature and possible detection in lepton pair production}},\ }\href {https://doi.org/10.1103/PhysRevLett.57.2123} {\bibfield  {journal} {\bibinfo  {journal} {Phys. Rev. Lett.}\ }\textbf {\bibinfo {volume} {57}},\ \bibinfo {pages} {2123} (\bibinfo {year} {1986})}\BibitemShut {NoStop}%
\bibitem [{\citenamefont {Bender}\ \emph {et~al.}(2003)\citenamefont {Bender}, \citenamefont {Meisinger},\ and\ \citenamefont {Wang}}]{Bender:2003gu}%
  \BibitemOpen
  \bibfield  {author} {\bibinfo {author} {\bibfnamefont {C.~M.}\ \bibnamefont {Bender}}, \bibinfo {author} {\bibfnamefont {P.~N.}\ \bibnamefont {Meisinger}},\ and\ \bibinfo {author} {\bibfnamefont {Q.}~\bibnamefont {Wang}},\ }\bibfield  {title} {\bibinfo {title} {{Finite-dimensional $\mathcal{PT}$-symmetric Hamiltonians}},\ }\href {https://doi.org/10.1088/0305-4470/36/24/314} {\bibfield  {journal} {\bibinfo  {journal} {J. Phys. A}\ }\textbf {\bibinfo {volume} {36}},\ \bibinfo {pages} {6791} (\bibinfo {year} {2003})},\ \Eprint {https://arxiv.org/abs/quant-ph/0303174} {arXiv:quant-ph/0303174} \BibitemShut {NoStop}%
\bibitem [{\citenamefont {Mostafazadeh}(2003)}]{Mostafazadeh:2003gz}%
  \BibitemOpen
  \bibfield  {author} {\bibinfo {author} {\bibfnamefont {A.}~\bibnamefont {Mostafazadeh}},\ }\bibfield  {title} {\bibinfo {title} {{Exact $PT$-symmetry is equivalent to Hermiticity}},\ }\href {https://doi.org/10.1088/0305-4470/36/25/312} {\bibfield  {journal} {\bibinfo  {journal} {J. Phys. A}\ }\textbf {\bibinfo {volume} {36}},\ \bibinfo {pages} {7081} (\bibinfo {year} {2003})},\ \Eprint {https://arxiv.org/abs/quant-ph/0304080} {arXiv:quant-ph/0304080} \BibitemShut {NoStop}%
\bibitem [{\citenamefont {Gubler}\ \emph {et~al.}(2016)\citenamefont {Gubler}, \citenamefont {Hattori}, \citenamefont {Lee}, \citenamefont {Oka}, \citenamefont {Ozaki},\ and\ \citenamefont {Suzuki}}]{Gubler:2015qok}%
  \BibitemOpen
  \bibfield  {author} {\bibinfo {author} {\bibfnamefont {P.}~\bibnamefont {Gubler}}, \bibinfo {author} {\bibfnamefont {K.}~\bibnamefont {Hattori}}, \bibinfo {author} {\bibfnamefont {S.~H.}\ \bibnamefont {Lee}}, \bibinfo {author} {\bibfnamefont {M.}~\bibnamefont {Oka}}, \bibinfo {author} {\bibfnamefont {S.}~\bibnamefont {Ozaki}},\ and\ \bibinfo {author} {\bibfnamefont {K.}~\bibnamefont {Suzuki}},\ }\bibfield  {title} {\bibinfo {title} {{$D$ mesons in a magnetic field}},\ }\href {https://doi.org/10.1103/PhysRevD.93.054026} {\bibfield  {journal} {\bibinfo  {journal} {Phys. Rev. D}\ }\textbf {\bibinfo {volume} {93}},\ \bibinfo {pages} {054026} (\bibinfo {year} {2016})},\ \Eprint {https://arxiv.org/abs/1512.08864} {arXiv:1512.08864 [hep-ph]} \BibitemShut {NoStop}%
\bibitem [{\citenamefont {Choi}(2007)}]{Choi:2007se}%
  \BibitemOpen
  \bibfield  {author} {\bibinfo {author} {\bibfnamefont {H.-M.}\ \bibnamefont {Choi}},\ }\bibfield  {title} {\bibinfo {title} {{Decay constants and radiative decays of heavy mesons in light-front quark model}},\ }\href {https://doi.org/10.1103/PhysRevD.75.073016} {\bibfield  {journal} {\bibinfo  {journal} {Phys. Rev. D}\ }\textbf {\bibinfo {volume} {75}},\ \bibinfo {pages} {073016} (\bibinfo {year} {2007})},\ \Eprint {https://arxiv.org/abs/hep-ph/0701263} {arXiv:hep-ph/0701263} \BibitemShut {NoStop}%
\bibitem [{\citenamefont {Be{\v{c}}irevi{\'c}}\ \emph {et~al.}(2015)\citenamefont {Be{\v{c}}irevi{\'c}}, \citenamefont {Kruse},\ and\ \citenamefont {Sanfilippo}}]{Becirevic:2014rda}%
  \BibitemOpen
  \bibfield  {author} {\bibinfo {author} {\bibfnamefont {D.}~\bibnamefont {Be{\v{c}}irevi{\'c}}}, \bibinfo {author} {\bibfnamefont {M.}~\bibnamefont {Kruse}},\ and\ \bibinfo {author} {\bibfnamefont {F.}~\bibnamefont {Sanfilippo}},\ }\bibfield  {title} {\bibinfo {title} {{Lattice QCD estimate of the {\ensuremath{\eta}}$_{c}$(2S) {\textrightarrow} $J$/{\ensuremath{\psi}}{\ensuremath{\gamma}} decay rate}},\ }\href {https://doi.org/10.1007/JHEP05(2015)014} {\bibfield  {journal} {\bibinfo  {journal} {JHEP}\ }\textbf {\bibinfo {volume} {05}},\ \bibinfo {pages} {014}},\ \Eprint {https://arxiv.org/abs/1411.6426} {arXiv:1411.6426 [hep-lat]} \BibitemShut {NoStop}%
\bibitem [{\citenamefont {Ridwan}\ \emph {et~al.}(2025)\citenamefont {Ridwan}, \citenamefont {Arifi},\ and\ \citenamefont {Mart}}]{Ridwan:2024ngc}%
  \BibitemOpen
  \bibfield  {author} {\bibinfo {author} {\bibfnamefont {M.}~\bibnamefont {Ridwan}}, \bibinfo {author} {\bibfnamefont {A.~J.}\ \bibnamefont {Arifi}},\ and\ \bibinfo {author} {\bibfnamefont {T.}~\bibnamefont {Mart}},\ }\bibfield  {title} {\bibinfo {title} {{Self-consistent $M1$ radiative transitions of excited ${B}_{c}$ and heavy quarkonia with different polarizations in the light-front quark model}},\ }\href {https://doi.org/10.1103/PhysRevD.111.016011} {\bibfield  {journal} {\bibinfo  {journal} {Phys. Rev. D}\ }\textbf {\bibinfo {volume} {111}},\ \bibinfo {pages} {016011} (\bibinfo {year} {2025})},\ \Eprint {https://arxiv.org/abs/2409.13172} {arXiv:2409.13172 [hep-ph]} \BibitemShut {NoStop}%
\bibitem [{\citenamefont {Ghodrati}(2025)}]{Ghodrati:2025fah}%
  \BibitemOpen
  \bibfield  {author} {\bibinfo {author} {\bibfnamefont {M.}~\bibnamefont {Ghodrati}},\ }\bibfield  {title} {\bibinfo {title} {{Photonic Exceptional Points in Holography and QCD}},\ }\href@noop {} {\  (\bibinfo {year} {2025})},\ \Eprint {https://arxiv.org/abs/2510.15518} {arXiv:2510.15518 [hep-th]} \BibitemShut {NoStop}%
\bibitem [{\citenamefont {Shintani}\ \emph {et~al.}(2007)\citenamefont {Shintani}, \citenamefont {Aoki}, \citenamefont {Ishizuka}, \citenamefont {Kanaya}, \citenamefont {Kikukawa}, \citenamefont {Kuramashi}, \citenamefont {Okawa}, \citenamefont {Ukawa},\ and\ \citenamefont {Yoshie}}]{Shintani:2006xr}%
  \BibitemOpen
  \bibfield  {author} {\bibinfo {author} {\bibfnamefont {E.}~\bibnamefont {Shintani}}, \bibinfo {author} {\bibfnamefont {S.}~\bibnamefont {Aoki}}, \bibinfo {author} {\bibfnamefont {N.}~\bibnamefont {Ishizuka}}, \bibinfo {author} {\bibfnamefont {K.}~\bibnamefont {Kanaya}}, \bibinfo {author} {\bibfnamefont {Y.}~\bibnamefont {Kikukawa}}, \bibinfo {author} {\bibfnamefont {Y.}~\bibnamefont {Kuramashi}}, \bibinfo {author} {\bibfnamefont {M.}~\bibnamefont {Okawa}}, \bibinfo {author} {\bibfnamefont {A.}~\bibnamefont {Ukawa}},\ and\ \bibinfo {author} {\bibfnamefont {T.}~\bibnamefont {Yoshie}},\ }\bibfield  {title} {\bibinfo {title} {{Neutron electric dipole moment with external electric field method in lattice QCD}},\ }\href {https://doi.org/10.1103/PhysRevD.75.034507} {\bibfield  {journal} {\bibinfo  {journal} {Phys. Rev. D}\ }\textbf {\bibinfo {volume} {75}},\ \bibinfo {pages} {034507} (\bibinfo {year} {2007})},\ \Eprint {https://arxiv.org/abs/hep-lat/0611032} {arXiv:hep-lat/0611032} \BibitemShut {NoStop}%
\bibitem [{\citenamefont {Shintani}\ \emph {et~al.}(2008)\citenamefont {Shintani}, \citenamefont {Aoki},\ and\ \citenamefont {Kuramashi}}]{Shintani:2008nt}%
  \BibitemOpen
  \bibfield  {author} {\bibinfo {author} {\bibfnamefont {E.}~\bibnamefont {Shintani}}, \bibinfo {author} {\bibfnamefont {S.}~\bibnamefont {Aoki}},\ and\ \bibinfo {author} {\bibfnamefont {Y.}~\bibnamefont {Kuramashi}},\ }\bibfield  {title} {\bibinfo {title} {{Full QCD calculation of neutron electric dipole moment with the external electric field method}},\ }\href {https://doi.org/10.1103/PhysRevD.78.014503} {\bibfield  {journal} {\bibinfo  {journal} {Phys. Rev. D}\ }\textbf {\bibinfo {volume} {78}},\ \bibinfo {pages} {014503} (\bibinfo {year} {2008})},\ \Eprint {https://arxiv.org/abs/0803.0797} {arXiv:0803.0797 [hep-lat]} \BibitemShut {NoStop}%
\bibitem [{\citenamefont {Yamamoto}(2013)}]{Yamamoto:2012bd}%
  \BibitemOpen
  \bibfield  {author} {\bibinfo {author} {\bibfnamefont {A.}~\bibnamefont {Yamamoto}},\ }\bibfield  {title} {\bibinfo {title} {{Lattice QCD with strong external electric fields}},\ }\href {https://doi.org/10.1103/PhysRevLett.110.112001} {\bibfield  {journal} {\bibinfo  {journal} {Phys. Rev. Lett.}\ }\textbf {\bibinfo {volume} {110}},\ \bibinfo {pages} {112001} (\bibinfo {year} {2013})},\ \Eprint {https://arxiv.org/abs/1210.8250} {arXiv:1210.8250 [hep-lat]} \BibitemShut {NoStop}%
\bibitem [{\citenamefont {Yang}\ and\ \citenamefont {Lee}(1952)}]{Yang:1952be}%
  \BibitemOpen
  \bibfield  {author} {\bibinfo {author} {\bibfnamefont {C.-N.}\ \bibnamefont {Yang}}\ and\ \bibinfo {author} {\bibfnamefont {T.~D.}\ \bibnamefont {Lee}},\ }\bibfield  {title} {\bibinfo {title} {{Statistical theory of equations of state and phase transitions. I. Theory of condensation}},\ }\href {https://doi.org/10.1103/PhysRev.87.404} {\bibfield  {journal} {\bibinfo  {journal} {Phys. Rev.}\ }\textbf {\bibinfo {volume} {87}},\ \bibinfo {pages} {404} (\bibinfo {year} {1952})}\BibitemShut {NoStop}%
\bibitem [{\citenamefont {Lee}\ and\ \citenamefont {Yang}(1952)}]{Lee:1952ig}%
  \BibitemOpen
  \bibfield  {author} {\bibinfo {author} {\bibfnamefont {T.~D.}\ \bibnamefont {Lee}}\ and\ \bibinfo {author} {\bibfnamefont {C.-N.}\ \bibnamefont {Yang}},\ }\bibfield  {title} {\bibinfo {title} {{Statistical theory of equations of state and phase transitions. II. Lattice gas and Ising model}},\ }\href {https://doi.org/10.1103/PhysRev.87.410} {\bibfield  {journal} {\bibinfo  {journal} {Phys. Rev.}\ }\textbf {\bibinfo {volume} {87}},\ \bibinfo {pages} {410} (\bibinfo {year} {1952})}\BibitemShut {NoStop}%
\bibitem [{\citenamefont {Ozawa}\ and\ \citenamefont {Hayata}(2024)}]{Ozawa:2023oqc}%
  \BibitemOpen
  \bibfield  {author} {\bibinfo {author} {\bibfnamefont {T.}~\bibnamefont {Ozawa}}\ and\ \bibinfo {author} {\bibfnamefont {T.}~\bibnamefont {Hayata}},\ }\bibfield  {title} {\bibinfo {title} {{Two-dimensional lattice with an imaginary magnetic field}},\ }\href {https://doi.org/10.1103/PhysRevB.109.085113} {\bibfield  {journal} {\bibinfo  {journal} {Phys. Rev. B}\ }\textbf {\bibinfo {volume} {109}},\ \bibinfo {pages} {085113} (\bibinfo {year} {2024})},\ \Eprint {https://arxiv.org/abs/2307.14635} {arXiv:2307.14635 [cond-mat.mes-hall]} \BibitemShut {NoStop}%
\bibitem [{\citenamefont {Montag}\ and\ \citenamefont {Ozawa}(2026)}]{Montag:2026cqh}%
  \BibitemOpen
  \bibfield  {author} {\bibinfo {author} {\bibfnamefont {A.}~\bibnamefont {Montag}}\ and\ \bibinfo {author} {\bibfnamefont {T.}~\bibnamefont {Ozawa}},\ }\bibfield  {title} {\bibinfo {title} {{Non-Hermitian Landau Levels}},\ }\href@noop {} {\  (\bibinfo {year} {2026})},\ \Eprint {https://arxiv.org/abs/2605.23613} {arXiv:2605.23613 [quant-ph]} \BibitemShut {NoStop}%
\end{thebibliography}%

\end{document}